\documentclass[
 reprint,
 amsmath,amssymb,
 prx, superscriptaddress, nofootinbib]{revtex4-1}
\usepackage{xcolor}
\usepackage{ulem}
\usepackage{braket}
\usepackage{bbm}
\usepackage{graphicx}
\usepackage{dcolumn}
\usepackage{bm}
\usepackage{amsmath}
\usepackage{float}
\usepackage{lipsum}
\usepackage{enumerate}
\usepackage{mathtools}
\usepackage{tikz}
\usepackage[colorlinks]{hyperref}

\newcommand{\be}{\begin{equation}}
\newcommand{\ee}{\end{equation}}

\newcommand{\E}{\mathcal{E}}

\DeclareMathOperator{\Tr}{Tr}

\newcommand{\beq}{\begin{equation}}
\newcommand{\eeq}{\end{equation}}
\newcommand{\beqn}{\begin{eqnarray}}
\newcommand{\eeqn}{\end{eqnarray}}

\newcommand{\cE}{ {\cal E} }

\newcommand{\vect}[1]{{\bm{#1}}}

\newcommand{\hrho}{\hat{\rho}}

\makeatletter
\newsavebox{\@brx}
\newcommand{\llangle}[1][]{\savebox{\@brx}{\(\m@th{#1\langle}\)}%
  \mathopen{\copy\@brx\kern-0.5\wd\@brx\usebox{\@brx}}}
\newcommand{\rrangle}[1][]{\savebox{\@brx}{\(\m@th{#1\rangle}\)}%
  \mathclose{\copy\@brx\kern-0.5\wd\@brx\usebox{\@brx}}}
\makeatother

\newcommand{\Z}{{\mathbb{Z}}}

\usepackage{amsthm}
\theoremstyle{plain}
\newtheorem*{theorem*}{Theorem}

\newtheorem{definition}{Definition}

\begin{document}
\newcommand{\jianhao}[1]{ { \color{violet} \small (\textsf{JHZ}) \textsf{\textsl{#1}} }}
\newcommand{\zb}[1]{ { \color{orange} \small (\textsf{ZB}) \textsf{\textsl{#1}} }}
\newcommand{\yy}[1]{ { \color{red} \small (\textsf{YY}) \textsf{\textsl{#1}} }}
\newcommand{\shijun}[1]{ { \color{olive} \small (\textsf{SS}) \textsf{\textsl{#1}} }}

\title{Holographic View of Mixed-State Symmetry-Protected Topological Phases in Open Quantum Systems}
\author{Shijun Sun}
\thanks{These authors contributed equally to this work.}
\affiliation{School of Physics, Georgia Institute of Technology, Atlanta, Georgia 30332, USA}
\author{Jian-Hao Zhang}
\email{Sergio.Zhang@colorado.edu}
\thanks{These authors contributed equally to this work.}
\affiliation{Department of Physics and Center for Theory of Quantum Matter, University of Colorado, Boulder, Colorado 80309, USA}
\author{Zhen Bi}
\email{zjb5184@psu.edu}
\affiliation{Department of Physics, The Pennsylvania State University, University Park, Pennsylvania 16802, USA}
\author{Yizhi You}
\email{y.you@northeastern.edu}
\affiliation{Department of Physics, Northeastern University, Boston, Massachusetts 02115, USA}

\date{\today}
\begin{abstract}
We establish a holographic duality between \(d\)-dimensional mixed-state symmetry-protected topological phases (mSPTs) and \((d+1)\)-dimensional subsystem symmetry-protected topological states (SSPTs). Specifically, we show that the reduced density matrix of the boundary layer of a \((d+1)\)-dimensional SSPT with subsystem symmetry \(\mathcal{S}\) and global symmetry \(\mathcal{G}\) corresponds to a \(d\)-dimensional mSPT with strong \(\mathcal{S}\) and weak \(\mathcal{G}\) symmetries. Conversely, we demonstrate that the wavefunction of an SSPT can be constructed by replicating the density matrix of the corresponding lower-dimensional mSPT. This mapping links the density matrix in lower dimensions to the entanglement properties of higher-dimensional wavefunctions, providing an approach for analyzing nonlinear quantities and quantum information metrics in mixed-state systems. Our duality offers a new perspective for studying intrinsic mSPTs that are unique to open quantum systems, without pure state analogs. We show that strange correlators and twisted R\'enyi-$N$ correlators can diagnose these nontrivial phases and explore their connection to strange correlators in pure-state SSPTs. {\color{black} Furthermore, we discuss several implications of this holographic duality, including a method for preparing intrinsic mSPT states through the duality.}
\end{abstract}

\maketitle
\tableofcontents
\section{Introduction}

Topological phases of matter, characterized by long-range entanglement in many-body wavefunctions, have attracted considerable attention over the past few decades. Moreover, symmetry can significantly enhance the nontrivial topological properties of quantum many-body systems, even in the absence of long-range entanglement. A prominent example of this is the \textit{symmetry-protected topological} (SPT) phases \cite{XieChenScience, Chen:2011pg, cohomology, ZCGu2009, LevinGu, Ashvin2013, XieChen2014, Kapustin2015, Senthil_2015}. Previous research has focused on key properties of the ground states of topological phases in closed quantum many-body systems. However, the unavoidable effects of decoherence and dissipation, arising from the coupling between physical systems and their environments, necessitate extending the concept of topological phases to open quantum systems. As a result, there has been growing interest in exploring symmetry-protected topological orders in open quantum systems, including studies of systems driven by noisy quantum channels and ensembles of states generated by quenched disorder \cite{deGroot2022, MaWangASPT, LeeYouXu2022, ZhangQiBi2022, ma2024topological, Zhang_2023, ma2024symmetry, xue2024tensor, guo2024locally, chen2024separability, chen2024unconventional, chen2023symmetryenforced, lessa2024mixedstate, wang2024anomaly, prelim, zhang2024quantum, ellison2024towards,sohal2024noisy,chirame2024stable,sala2024decoherence,albert2014symmetries, zhang2024strong, zhang2024fluctuation}. These investigations have revealed a broad class of symmetry-protected topological phases in open systems, now referred to as \textit{mixed-state SPT} (mSPT) or \textit{average SPT} (ASPT).

Contrary to the intuition that decoherence always diminishes topological classification, it can also give rise to new mixed-state topological phases, known as intrinsic mSPTs. These phases are unique to open systems and cannot be realized in thermal equilibrium, such as in the ground state of a gapped Hamiltonian. Recent studies have uncovered a variety of intrinsic topological phases in mixed states, revealing quantum many-body topology exclusive to open quantum systems\footnote{\color{black}In Ref. \cite{ma2024topological}, the authors emphasized that there are two equivalent definitions of intrinsic mSPT mixed states: the one is the mSPT states that cannot be purified to pure state SPT states with the same symmetry class; the other is the elements of Atiyah-Hirzebruch spectral sequence that are obstructed by the last layer of differential maps.} \cite{lee2023quantum, Su_2024, sang2023mixed, sang2024stability, moharramipour2024symmetry, sang2024approximate, lessa2024strong, gu2024spontaneous, sala2024spontaneous, huang2024hydro, guo2024new}. 
It has been proposed that intrinsic mSPTs emerge from intrinsic gapless SPT states subjected to decoherence or disorder. While this connection is intriguing, intrinsic gapless SPTs are typically characterized by strong interactions and quasi-long-range correlations, making them challenging to realize in quantum circuits and programmable devices. A simpler purification of these intrinsic mixed-state SPTs remains elusive.

Symmetry plays a pivotal role in modern physics. In mixed-state density matrices, a unique feature is the distinction between strong (exact) symmetry and weak (average) symmetry. Strong symmetry refers to a symmetry operation that acts on one side of the density matrix and leaves it invariant (up to a global phase), namely:
\begin{align}
U(g)\rho=e^{i\theta}\rho.
\end{align}
Equivalently, this infers that every state in $\rho$ must remain symmetric under $U(g)$ with the same eigenvalue. In contrast, weak symmetry refers to a symmetry operation that preserves the density matrix only when applied simultaneously to both the left and right sides, i.e.,
\begin{align}
U(g)\rho\neq e^{i\theta}\rho,\quad U(g)\rho U(g)^\dag=\rho.
\end{align}
This means that the density matrix can always be diagonalized in a basis that respects the weak symmetry, although the symmetry charges may vary between the states. 

If we view the mixed-state density matrix (`system’) as the reduced density matrix of a wavefunction (dubbed `purified state’) in an enlarged Hilbert space, obtained by tracing out other degrees of freedom (`ancilla’), the weak symmetry \( G \) of the density matrix can naturally arise if the purified state (encompassing both the system and ancilla) remains invariant under \( G \). In contrast, strong symmetry 
$S$ requires that the system conserves $S$-charge on its own, without any exchange of symmetry charges with an ancilla. This charge conservation within each subsystem (either the system or ancilla) resembles the concept of `subsystem symmetry’ in the fracton literature \cite{Vijay2015-jj,Haah2011-ny,Chamon2005-fc,bravyi2010topological,Vijay2016-dr,yoshida2013exotic,pretko2020fracton,yoshida2015bosonic,ma2018fracton,you2019emergent,xu2007bond,paramekanti2002ring,tay2010possible,seiberg2020exotic,you2024quantum,nandkishore2019fractons,burnell2022anomaly}, where the symmetry operator acts only on the degrees of freedom of a specific subsystem (such as a plane or line), ensuring charge conservation within that subsystem.

When an entanglement cut is made on a 3d wavefunction exhibiting both subsystem and global symmetries, it naturally produces a reduced density matrix in a lower dimension with a strong symmetry—a descendant of the subsystem symmetry—as well as certain weak symmetries corresponding to the global symmetries. It has been demonstrated that the combination of subsystem symmetries and global symmetries can lead to various SPT phases, known as subsystem symmetry-protected topological (SSPT) phases \cite{you2018subsystem,you2018symmetric,you2019fractonic,you2019multipolar,devakul2018fractal,devakul2018strong,may2022interaction}. 
Motivated by this result, we establish a duality between \textit{mSPT in $d$ dimensions} and \textit{SSPT states in $d+1$ dimensions}. We show that tracing out the bulk degrees of freedom of an SSPT wavefunction yields an mSPT state in one lower dimension. Conversely, by replicating an mSPT density matrix, we can construct an SSPT wavefunction in one higher dimension, effectively serving as a purification of the mSPT state. This duality bridges the properties of mSPT density matrices with the entanglement characteristics of higher-dimensional SSPT wavefunctions, providing a framework to explore a broader class of mixed-state SPTs through dimensional elevation.
Notably, our findings indicate that intrinsic mSPTs, under this duality, are always mapped to higher-order SSPT states that exhibit gapless modes localized at hinges or corners of their surfaces. This holographic correspondence connects the nonlinear observables of $d$-dimensional mSPTs with the bulk correlation functions of the corresponding $d+1$-dimensional SSPT.
In particular, we introduce a new quantity dubbed \textit{twisted Rényi-$N$ correlator}, which measures long-range correlations in a mixed state along the `replica direction'. According to our duality, the twisted Rényi-$N$ correlator for a $d$-dimensional mSPT is dual to the bulk strange correlator of the corresponding $(d+1)$-dimensional SSPT, providing an alternative perspective on edge-bulk correspondence via holography, and can be treated as a new fingerprint for intrinsic mSPT.

\section{Generalities}
\subsection{A brief review of mixed-state SPT}

\begin{figure}
\centering
\includegraphics[width=0.48\textwidth]{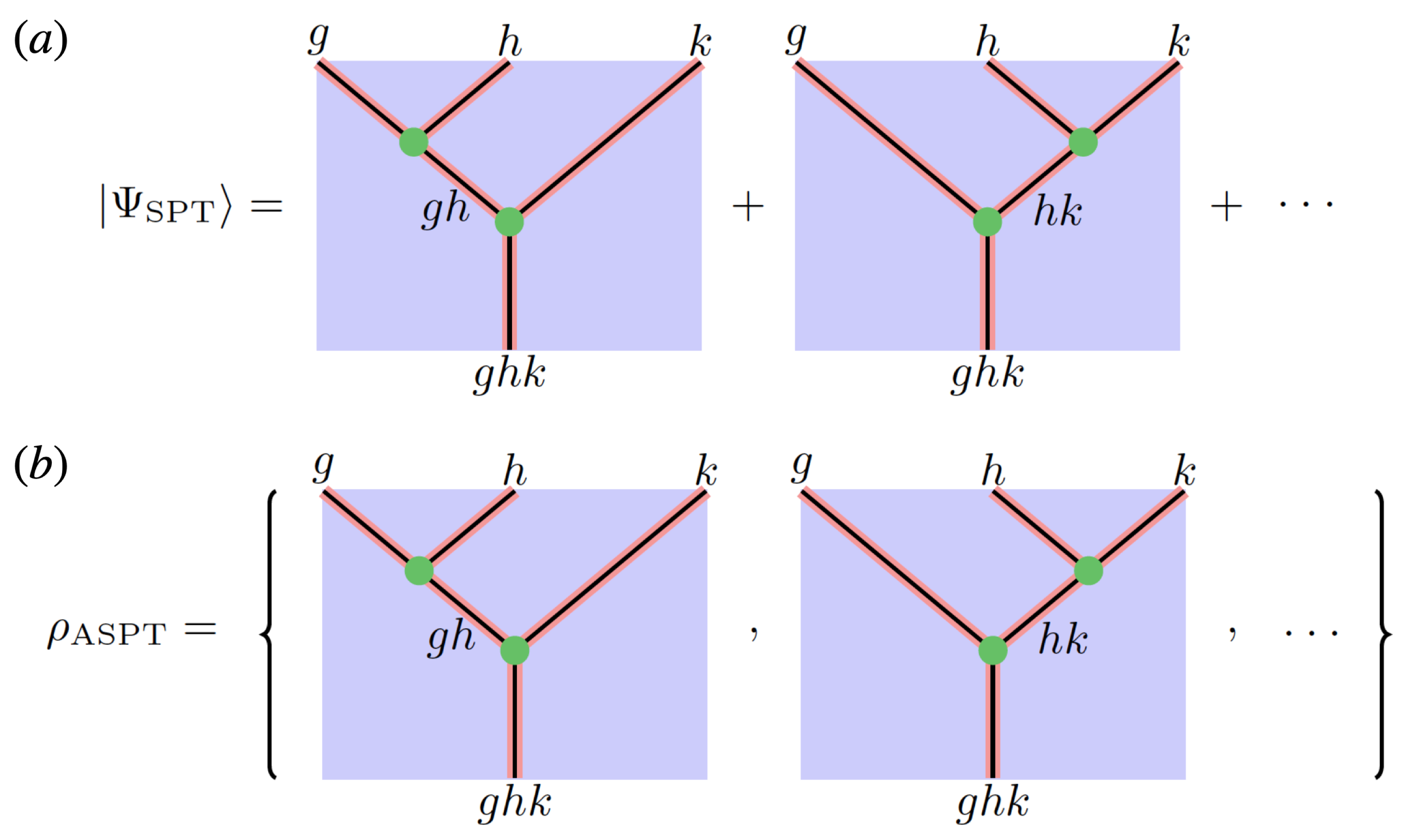}
\caption{Schematic illustration of decorated domain wall constructions of pure-state and mixed-state SPT phases.}
\label{Fig: DDW}
\end{figure}

We will first sketch the decorated domain wall construction for pure-state SPTs. For a generic symmetry group with the following extension,
\begin{align}
1\rightarrow K\rightarrow\tilde{G}\rightarrow G\rightarrow1,
\label{Eq: extension}
\end{align}
the decorated defect picture for pure-state SPT has the following paradigm [see Fig. \ref{Fig: DDW}(a)] \cite{XieChen2014}:
\begin{enumerate}[1.]
\item Consider a defect network of $G$ symmetry, decorated by $K$-symmetric short-range entangled (SRE) states;
\item Proliferate $G$ symmetry by quantum superpose all $G$ defect networks. 
\end{enumerate}
However, there might be some obstructions to the $G$ symmetry proliferation. To achieve a $G$-symmetric wavefunction, the following consistency conditions should be satisfied:
\begin{enumerate}[1.]
\item All $G$ defects should be decorated by $K$-symmetric gapped SRE states;
\item Local deformations of defects do not change the charge of $K$;
\item Deformations of $G$ defects will not acquire a trivial Berry phase.
\end{enumerate}

Suppose we consider open quantum systems where $G$ is a weak symmetry while $K$ remains a strong symmetry, the decorated domain wall paradigm is still well-defined in the sense that we decorated the $G$ defects by $K$-symmetric SRE states, and then ``classically'' or ``statistically'' proliferate $G$ defect networks to make a $G$ symmetric ensemble, i.e., put all possible $G$ defect networks to an ensemble rather than superposing them to a single wavefunction. The difference between SPT and mSPT is that there is no phase coherence in mixed states. A schematic of decorated domain wall construction of mSPT states is illustrated in Fig. \ref{Fig: DDW}(b), which demonstrates that the construction of pure-state and mixed-state SPTs are similar and the difference is one is a quantum superposition and the other one is a classical ensemble. 

Both pure-state and mixed-state SPT phases can be classified by generalized group cohomology, and computed by Atiyah-Hirzebruch spectral sequence (AHSS) \cite{Wang_2018Classification, Wang_2020, SpectralSequence}. Each decorated $G$ defect network can be described by an element of $E_2$ page of the AHSS, namely
\begin{align}
E_2^{p,q}=\mathcal{H}^p\left[G, h^q(K)\right],
\end{align}
where $h^q(K)$ is the classification of $K$-symmetric SRE states in $q$ spacetime dimension, and $p+q=d+1$. For pure-state SPT phases, all possible $G$ defect networks could be candidates for SPT wavefunctions, namely
\begin{align}
\bigoplus_{q=0}^{d+1}E_{2}^{d+1-q,q}=\bigoplus_{q=0}^{d+1}\mathcal{H}^{d+1-q}\left[G, h^q(K)\right],
\label{Eq: AHSS1}
\end{align}
while for mSPT phases, we should delete the layer of $q=0$ because of the phase incoherence, namely
\begin{align}
\bigoplus_{q=1}^{d+1}E_{2}^{d+1-q,q}=\bigoplus_{q=1}^{d+1}\mathcal{H}^{d+1-q}\left[G, h^q(K)\right].
\label{Eq: AHSS2}
\end{align}
Then the consistency/obstruction conditions we listed above can be formulated in terms of the differential maps in AHSS, namely
\begin{align}
\mathrm{d}_r:E_2^{p,q}\rightarrow E_{2}^{p+r,q-r+1},
\label{Eq: last layer}
\end{align}
and legitimate SPT states are in the kernel of all possible differential maps $\mathrm{d}_r$. In particular, the last layer of the differential map, namely $\mathrm{d}_{d+2-p}: E_2^{p,q}\rightarrow E_{2}^{d+2,0}$, characterizes the Berry phase consistency if we try to quantum superpose $G$ symmetry defects, which is not required if we want to construct an mSPT mixed state. 

{\color{black}

With fewer constraints, new SPT phases may emerge that are intrinsically mixed. These mSPT states cannot be purified into a pure-state SPT\footnote{\color{black}We note that there are two equivalent characters of intrinsic mSPT phases \cite{ma2024topological}: all elements in the $E_2$ page of AHSS \eqref{Eq: AHSS2} who are obstructed by the last layer of the differential \eqref{Eq: last layer}, and the mSPT density matrix who cannot be purified to a pure-state SPT wavefunction within the same symmetry class.}. Notably, as we will show in the following sections, the holographic mapping of bulk SSPT states to mSPT states is particularly useful for studying intrinsically mixed SPTs. In this framework, intrinsic mSPTs are mapped to the so-called higher-order SSPTs\cite{you2019multipolar, may2022interaction,Zhang_2023, zhang2023classification, prelim,you2024higher}. 
}

\begin{figure*}[ht!]
    \centering
\includegraphics[width=1\textwidth]{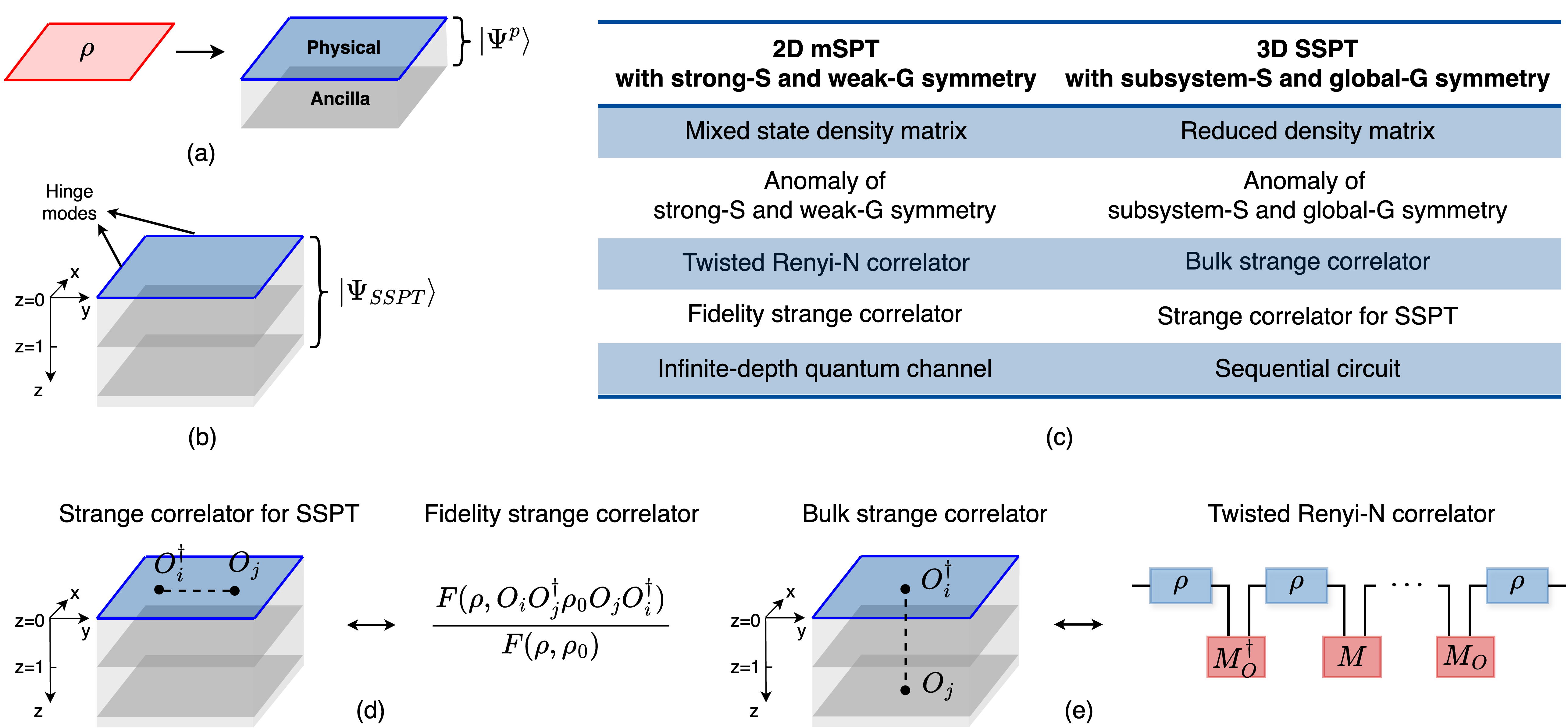}
    \caption{Summary of the mapping. a) Purification of the mixed state $\rho$. Tracing out the ancilla degrees of freedom (the bottom layers) of the purified wavefunction leads to the density matrix $\rho$. One can observe that the global and subsystem symmetries of the purified wavefunction correspond to the weak and strong symmetries of the mixed state. b) Illustration of the SSPT wavefunction. Tracing out the bottom layers $z>0$, one acquires a reduced density matrix of the $z=0$ layer, which describes an intrinsic mSPT in one lower dimension. c) Summary of the corresponding quantities. d) Illustration of the mapping between the strange correlator defined in Eq.~\eqref{eq:strangeSSPT} and the fidelity strange correlator of the mSPT. e) Illustration of the mapping between the strange correlator along the $z$-direction Eq.~\eqref{twsitedRenyi} and the twisted R\'enyi-$N$ correlator of the mSPT. }
    \label{overview}
\end{figure*}

\subsection{Holographic mapping between SSPT and mSPT}\label{sec:holography}

{
\color{black}
We highlight the motivations for introducing the subsystem symmetry:
\begin{enumerate}[1.]
\item Generically, a mixed-state density matrix \(\rho\) can be interpreted as the reduced density matrix of a pure-state wavefunction \(\ket{\psi}\) by tracing out part of the degrees of freedom(denote as ancilla). In this scenario, a global symmetry of the pure state is typically reduced to a weak symmetry upon tracing out those ancillae. Obtaining a reduced density matrix with a strong symmetry requires a more specialized conservation law in the corresponding pure-state wavefunction.
In this work, we focus on a subsystem-symmetric wavefunction \(\ket{\psi}\), where the charge is conserved within each individual layer or row. When we trace out certain layers (which can be viewed as ancilla degrees of freedom from the environment), the resulting reduced density matrix for the system of interest inherits a \textit{strong symmetry}. This occurs because the original subsystem symmetry—which prohibits charge fluctuations between layers—ensures that no charge exchange takes place between the system and the traced-out ancilla. As a result, the subsystem symmetry is promoted to a strong symmetry in the mixed state.

\item Previous studies \cite{ma2024topological} have shown that an intrinsic mixed-state symmetry-protected topological (mSPT) phase cannot be purified into a gapped SPT wavefunction within the same symmetry class and spatial dimension. Instead, most intrinsic mSPT phases are derived from so-called \textit{intrinsic gapless SPT states} by introducing decoherence or quenched disorder. In this work, we explore an alternative approach to purification using holographic correspondence. Specifically, we interpret the \(d\)-dimensional mixed state as the reduced density matrix of a \((d+1)\)-dimensional wavefunction exhibiting area-law entanglement. 
Within this framework, we map the strong symmetry of the mSPT state to the subsystem symmetry of the corresponding higher-dimensional subsystem symmetry-protected topological (SSPT) wavefunction. This embedding allows us to establish a direct connection between the mixed-state mSPT and the pure-state SSPT in one higher dimension.
\end{enumerate}
}

A mixed-state ensemble, described by the density matrix \(\rho\), can always be interpreted as the reduced density matrix of a wave function \(|\Psi^p\rangle\) (known as the purified state) in an enlarged Hilbert space.
This enlarged Hilbert space can be visualized as a multilayer system: the top layer contains the system degrees of freedom, while the other layers hold ancilla degrees of freedom from the environment, as shown in Fig.~\ref{overview}a. While the entire enlarged Hilbert space is in a pure state captured by the wavefunction \(|\Psi^p\rangle\), tracing out the ancilla qubits from the bottom layers yields the mixed state \(\rho\) for the system.

When a mixed state \(\rho\) exhibits weak symmetry \(\mathcal{G}\), its purified wavefunction \(|\Psi^p\rangle\) must also be \(\mathcal{G}\)-symmetric. If a mixed state \(\rho\) possesses strong symmetry \(\mathcal{S}\), all eigenvectors of the density matrix must carry the same \(\mathcal{S}\) charge. From the purified perspective, this means that the charge within the system must be conserved individually, with no \(\mathcal{S}\)-charge exchange between the system and ancilla layers. This idea parallels the concept of \textit{subsystem symmetry} in fracton literature, where the symmetry operator acts only on specific degrees of freedom within a particular subsystem, such as a single plane.

This leads to a natural question: \textit{Can the density matrix of an mSPT with strong symmetry \(\mathcal{S}\) be holographically mapped into a wave function that exhibits planar subsystem symmetry \(\mathcal{S}\)?} To gain some insight, we consider the 3d higher-order subsystem symmetric topological phases (HO-SSPT) discussed in Refs.~\cite{you2019multipolar, may2022interaction,Zhang_2023, zhang2023classification, prelim,you2024higher}.
These HO-SSPT states are protected by a global symmetry \( \mathcal{G} \) and a planar subsystem symmetry \( \mathcal{S} \), which acts on each \( xy \)-plane in 2d. The SSPT wave function shows short-range correlations in both the bulk and side surfaces but hosts gapless modes localized on the hinges (along the \( x \) or \( y \)-axis) as illustrated in Fig.~\ref{overview}b. These hinge modes display a \textit{mixed anomaly} between the global symmetry \( \mathcal{G} \) and subsystem symmetry \( \mathcal{S} \).

Consider this 3d SSPT wavefunction with an open boundary at \( z = 0 \). After tracing out all the bottom layers, the reduced density matrix \(\rho\) for the top layer at \( z = 0 \) exhibits weak \( \mathcal{G} \) symmetry and strong \( \mathcal{S} \) symmetry. While the 2d density matrix is short-range correlated in the bulk due to the gapped and short-range correlated side surfaces, its boundary—corresponding to the hinge of the original 3d SSPT state—exhibits a mixed anomaly between the weak \( \mathcal{G} \) and strong \( \mathcal{S} \) symmetries. This suggests that the reduced density matrix of the 3d SSPT wavefunction is reminiscent of a 2d mixed-state SPT.

Notably, such 2d mixed-state SPT is necessarily intrinsic in the sense that one cannot purify it into a pure-state SPT within the same dimensional Hilbert space. We prove this by a contradiction. We first assume the corresponding pure-state SPT exists. Since the pure state has the same symmetries as the corresponding mixed state, it possesses strong $\mathcal{G}$ and $\mathcal{S}$ symmetries. One can consider sticking such a 2d pure state to the bottom surface of the 3d SSPT state and obtain a 3d symmetric pure state. Then, the hinge modes of the 3d SSPT can be symmetrically gapped by a symmetric coupling to the edge modes of the 2d pure state. However, the stability of the hinge modes of the 3d SSPT prevents such processes. Therefore, the 2d pure state cannot be a non-trivial SPT state.

Motivated by this purification perspective, we will demonstrate the following duality:
\begin{center}
\textit{SSPT wavefunction in \((d+1)\)-dim }
$$\textit{holographic purification\ \ \ }\uparrow\downarrow\ \ \  \textit{tracing out the bulk}$$
\textit{mSPT in d-dim}
\end{center}

This duality establishes a holographic connection between the lower-dim mSPT and the reduced density matrix of a higher-dim SSPT, effectively linking the mixed state in the lower dimension to the entanglement properties of the higher-dimensional wavefunction.

In the remainder of this paper, we demonstrate the duality from both directions. From the dimensional reduction perspective, we start with the
fixed-point wavefunction of a 3d HO-SSPT that exhibits global symmetry \( \mathcal{G} \) and planar subsystem symmetry \( \mathcal{S} \) (acting on each \( xy \)-plane). These states have gapped side surfaces on the \( xy \)-plane, with gapless hinge modes localized at the \( x (y) \) hinges.
Next, we compactify the 3d SSPT state with \( L_z = 2 \) and trace out the bottom plane, resulting in the reduced density matrix \( \rho \) for the top layer. This \( \rho \) can be viewed as the density matrix of a 2d mixed ensemble that exhibits both weak \( \mathcal{G} \) symmetry and strong \( \mathcal{S} \) symmetry.
We will show that \( \rho \) exhibits characteristics of a 2d intrinsic mSPT.
The mixed anomaly between weak \( \mathcal{G} \) symmetry and strong \( \mathcal{S} \) symmetry at the edge of \( \rho \) is derived from the mixed anomaly between global \( \mathcal{G} \) symmetry and subsystem symmetry \( \mathcal{S} \) at the hinge of the 3d HO-SSPT. In this framework, the compactified 3d SSPT state with \( L_z = 2 \) acts as a purification of \( \rho \). A derivation of this duality from the dimensional reduction perspective will be presented in Sec.~\ref{sec:strangecor}.  From the dimensional extension perspective, we will show in Sec.~\ref{sec:dimext} that a 3d SSPT wavefunction can be constructed by replication of a 2d mSPT density matrix. Sec.~\ref{sec:2DSSPT} gives a concrete example of constructing a 2d HO-SSPT from replicating 1d intrinsic mSPT density matrices. After demonstrating the duality from both directions, in Sec.~\ref{sec:corner}-\ref{sec:dipole}, we will present a thorough examination of the duality using several concrete examples.

Our duality links the 2d mSPT density matrix with the entanglement properties of a 3d SSPT wavefunction, enabling the exploration of a broader class of mSPTs through dimensional reduction from higher-dimensional SSPT states. Notably, Ref.~\cite{zhang2023classification} provides a comprehensive analysis and classification of 3d HO-SSPT states with gapless hinge or corner modes. Building on our duality framework, we will show that the reduced density matrix of these 3d HO-SSPT wavefunctions can be interpreted as a 2d intrinsic mSPT.

\subsection{Dimension Extension from Tensor Network}\label{sec:dimext}

\begin{figure}[h!]
\centering
\includegraphics[width=0.48\textwidth]{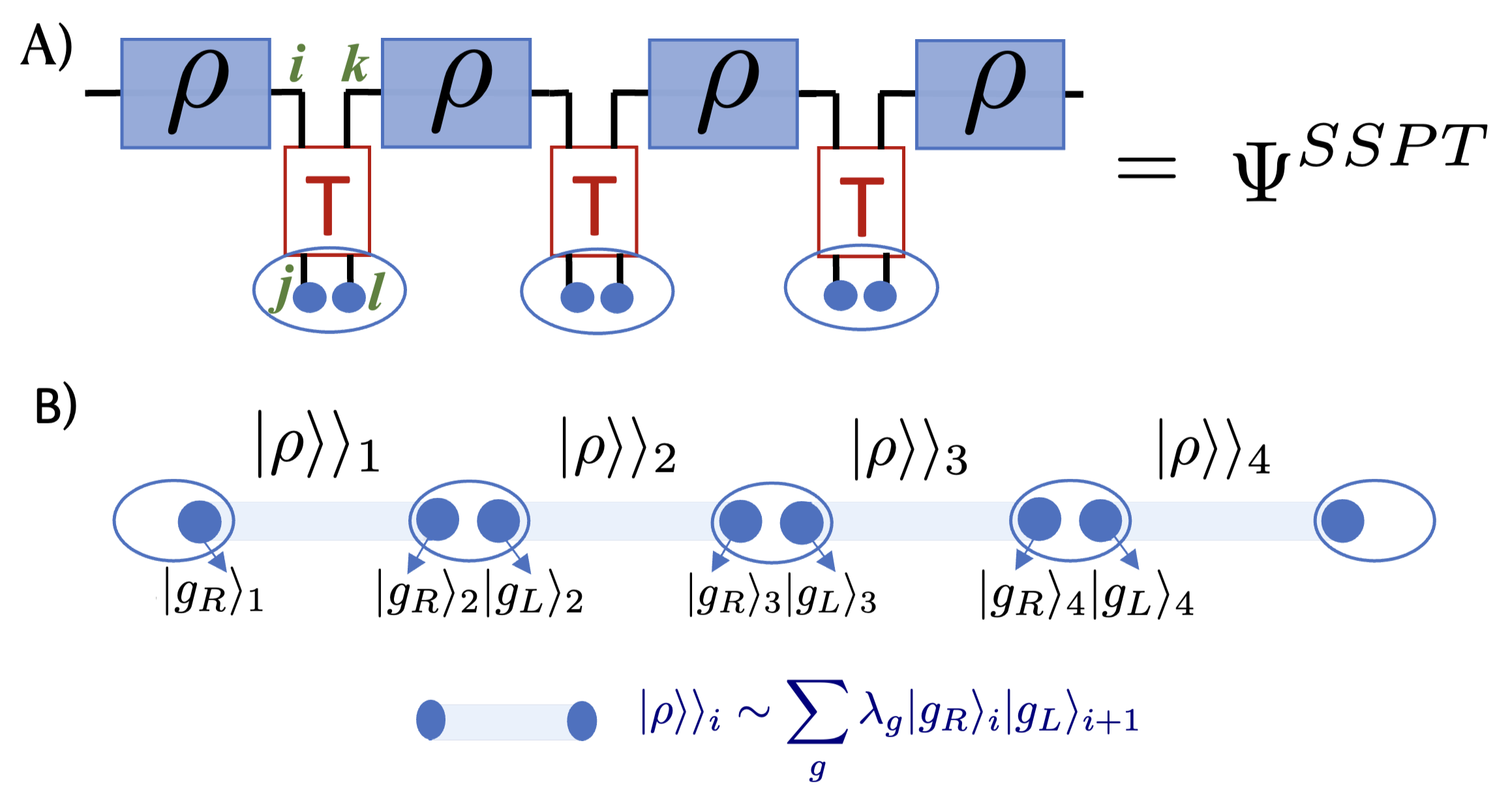}
\caption{A) Construct the 3d HO-SSPT wave function by replicating the 2d density matrix \(\rho\) and inserting a rank-4 \( T \)-tensor between each pair of replicas.
B) In this mapping, the bra space of the \(i\)-th density matrix and the ket space of the \(i+1\)-th density matrix together form a unit cell (represented by an oval). The ket and bra spaces of the \(i\)-th density matrix create a building block that includes entangled pairs between adjacent unit cells, depicted as blue lines.}
\label{wavefunc}
\end{figure} 

To illustrate the duality from the reversed perspective, we will demonstrate that the 3d HO-SSPT wavefunction (\(\Psi^{sspt}\)) can be constructed by replicating the 2d mSPT density matrix \(\rho\) and inserting the \(T\)-tensor between the replicated copies, as shown in Fig.~\ref{wavefunc}. This offers a dimensional extension view that maps a mixed-state density matrix to a higher-dimensional variational wavefunction.
For simplicity, we can choose \( T_{ijkl} = \delta_{ij} \delta_{kl} \) for the fixed-point wave function \(\Psi^{\text{sspt}}\). In this construction, the replica index corresponds to the \( z \)-layer index, effectively extending the 2d mixed state into a 3d wave function. Suppose we start with a 2d mixed state given by \(\rho = \lambda_g |g\rangle \langle g|\). The dual 3d state $\Psi^{sspt}$ is then constructed as a tensor product of entangled building blocks:
\begin{align}\label{eq:wfmap}
&|\Psi^{sspt}\rangle = \frac{1}{(\Tr(\rho^2))^N} \bigotimes_i (\sum_g \lambda_g |g_R\rangle_i |g_L\rangle_{i+1} )\nonumber\\
&\rho= \lambda_g |g\rangle \langle g| \rightarrow |\rho \rrangle_i \sim \sum_g \lambda_g |g_R\rangle_i |g_L\rangle_{i+1} 
\end{align}
The notation \(|g_{L/R}\rangle_i\) denotes the left/right element of the \(i\)-th layer. $|\rho \rrangle_i$ represents the entangled building block between the \(i\)-th and \(i+1\)-th layers.
Fig.~\ref{wavefunc} visually represents Eq.~\eqref{eq:wfmap}. 

In this mapping, each $x$-$y$ plane (referred to as a unit cell) contains two components: L (left) and R (right). The bra vector of the \(i\)-th \(\rho\) and the ket vector of the \((i+1)\)-th \(\rho\) together form a unit cell, composed of the L/R components within the same $z$-layer.
The R and L components of adjacent unit cells are entangled to form a building block, as illustrated in Fig.~\ref{wavefunc}. The 2$d$ density matrix, through the Choi–Jamiołkowski isomorphism, is mapped to a bilayer wavefunction $|\rho \rrangle_i \sim \sum_g \lambda_g |g_R\rangle_i |g_L\rangle_{i+1}$, which serves as an elementary building block by entangling the right component of the \(i\)-th layer with the left component of the \(i+1\)-th layer.
The noise channels (Kraus operators) that couple the bra and ket spaces of the density matrix are analogous to the interlayer coupling within each building block. 
\(\Psi^{\text{sspt}}\) is constructed as a tensor product of all these entangled building blocks, forming the fixed point wavefunction for the 3d SSPT. To deviate from this finely tuned state by adding interactions between the building blocks, one can adjust the rank-4 tensor \( T_{ijkl} \), introducing additional interactions between the L/R components within the same layer.
It's important to note that the R\'enyi-2 density matrix \(\frac{\rho^2}{\text{Tr}(\rho^2)}\) in Eq.~\eqref{eq:wfmap} accurately replicates the reduced density matrix for \( |\Psi^{sspt} \rangle\) when a spatial cut is made along the \( xy \)-plane. The von Neumann entropy of \(\rho^2\) corresponds directly to the entanglement entropy of \( |\Psi^{sspt} \rangle\) for this spatial bipartition.

The strong \(\mathcal{S}\) and weak \(\mathcal{G}\) symmetries act on $|\rho \rrangle_i$ as follows:
\begin{align}
& \mathcal{S}: U_S|\rho \rrangle_i= \sum_g \lambda_g (U_S|g_R\rangle_i)~ |g_L\rangle_{i+1} \nonumber\\
& \mathcal{G}: U_G|\rho \rrangle_i= \sum_g \lambda_g (U_G |g_R\rangle_i)~ (U^{\dagger}_G |g_L\rangle_{i+1})~ 
\end{align}
Here, the operators \( U^{\dagger}_G \) and \( U_G \) act on the L (left) and R (right) components in a Hermitian conjugate manner, originating from the bra vector of the density matrix. Since the fixed-point wavefunction in Eq.~\eqref{eq:wfmap} is a tensor product of entangled pairs, it is evident that \(\Psi^{\text{sspt}}\) exhibits both global \(\mathcal{G}\) and subsystem \(\mathcal{S}\) symmetries.
\begin{widetext}
 \begin{align}
& \mathcal{S}: U_S|\Psi^{sspt}\rangle= [\sum_g \lambda_g (U_S|g_R\rangle_a)~ |g_L\rangle_{a+1} ][\sum_g \lambda_g |g_R\rangle_{a-1}~ (U^{\dagger}_S|g_L\rangle_{a}) ][\prod_{i \neq a,a-1} \otimes \sum_g \lambda_g |g_R\rangle_i |g_L\rangle_{i+1}] \nonumber\\
& \mathcal{G}: U_G|\Psi^{sspt}\rangle= \bigotimes_i [\sum_g \lambda_g (U_G |g_R\rangle_i)~ (U^{\dagger}_G |g_L\rangle_{i+1})~ ]
\end{align}   
\end{widetext}
Here, the global symmetry \( U_G \) acts on all layers, while the subsystem symmetry \( U_S \) applies only to a specific layer at \( z = a \). The strong symmetry condition on \(\rho\) ensures the planar subsystem symmetry of the SSPT wave function\footnote{The fixed-point SSPT wavefunction, as a tensor product of entangled pairs, is finely tuned and contains local symmetries. However, introducing intra-layer interactions by adjusting the T-tensor would break the local symmetry, while the global and subsystem symmetries would remain intact.}.

Under this duality formalism, the R\'enyi-$N$ correlation function of $\rho$ can be mapped to the inner product between $\Psi^{sspt}$ and $\Psi^{trivial}$, with $M$ being chosen depending on the natural of the trivial state.
\begin{align}
&\langle \Psi^{trivial}  |\Psi^{sspt}\rangle  = \text{Tr}[\rho M \rho  M  \rho M\rho M...\rho M\rho M]
\end{align}
A typical choice for \( M \) is the identity matrix, which corresponds to a symmetric trivial state, represented by a direct product of unentangled layers. In this configuration, the L/R components within the same layer are coupled like a symmetric EPR state.
Notably, this inner product is essential for calculating the strange correlator of the SSPT state.

This duality connects the nonlinear observables of \(d\)-dimensional mSPTs with the strange correlator of the dual \(d+1\)-dimensional SSPT. Here, we introduce a new quantity dubbed \textit{twisted Rényi-$N$ correlator}, which measures the long-range correlation in a mixed state along the `replica direction'. According to our duality, the twisted Rényi-$N$ correlator for the 2d mSPT correlator is dual to the strange correlator\cite{you2014wave,lepori2023strange} of the 3d SSPT along the z-direction.
\begin{align}\label{twsitedRenyi}
&C^M(|i-j|)=\frac{\langle \Psi^{trivial}_M | O^{\dagger}_i O_j |\Psi^{sspt}\rangle}{\langle \Psi^{trivial}_M |\Psi^{sspt}\rangle}  \nonumber\\
& =\frac{ \text{Tr}[\rho \bm{M^{\dagger}_{O}}  ~(\rho M)^{i-j-1}\rho \bm{ M_{O}} ~(\rho M)^{i-j-1}]}{\text{Tr}[(\rho M)^{2(i-j)}]}
\end{align}
 The choice of the \( M \) matrix is determined by \(\Psi_M^{\text{trivial}}\), while \( M_O \) depends on the observable \( O \) being measured on the corresponding SSPT state. 
In most cases, the trivial state can be chosen as an onsite symmetric EPR state, simplifying the choice to \( M = I \), and the twisted Rényi-\(N\) operator reduces to:
\begin{align}\label{twsitedRenyisimple}
&C(|i-j|)=\frac{\langle \Psi^{trivial} | O^{\dagger}_i O_j |\Psi^{sspt}\rangle}{\langle \Psi^{trivial} |\Psi^{sspt}\rangle}  \nonumber\\
& =\frac{ \text{Tr}[ \bm{M^{\dagger}_{O}}  ~\rho^{i-j} \bm{ M_{O}} ~\rho^{i-j}]}{\text{Tr}[\rho ^{2i-2j}]}
\end{align}

Here, we insert the operators \(M_O\) and \(M_O^\dagger\) (which act on the same spatial point in a 2d mSPT) between the replica density matrices \(\rho^{i-j}\). Consequently, these operators are separated only in the replica direction. In the dual SSPT wavefunction, the replica direction is mapped to the \(z\)-axis, so \(O^\dagger_i O_j\) acts on the \(i\)-th and \(j\)-th layers, which are separated only along the \(z\)-axis.

We will demonstrate later that the twisted Rényi-\(N\) correlator displays (quasi) long-range order along the replica direction for intrinsic mSPTs, but only short-range order for conventional mSPTs. This difference serves as a key diagnostic for identifying intrinsic mSPTs. Detailed examples and further explanations of the twisted Rényi-\(N\) correlator will be presented in Secs.~\ref{sec:replica} and \ref{sec:twisted}.

\subsection{Dimensional reduction and fidelity strange correlator}
\label{sec:strangecor}

\begin{figure}[t!]
    \centering
    \includegraphics[width=0.48\textwidth]{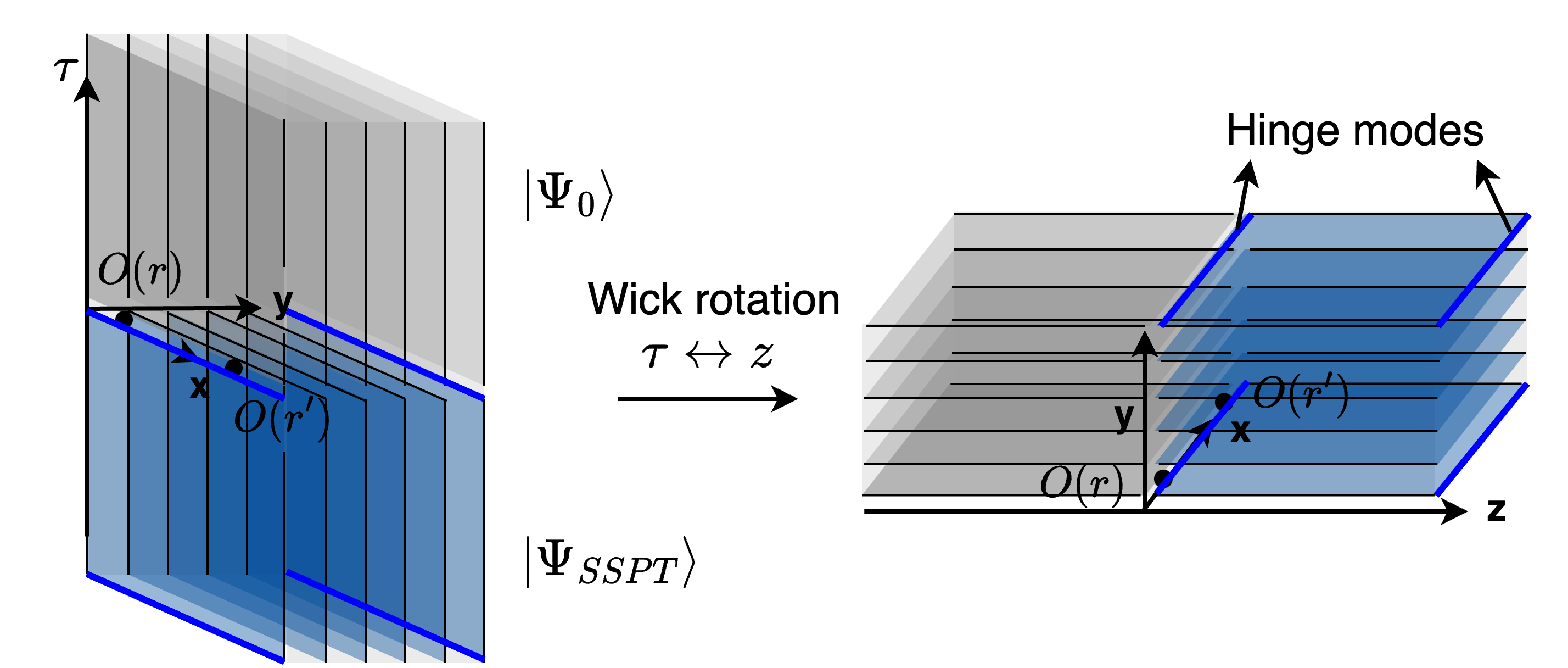}
    \caption{Strange correlator for diagnosing SSPT states. Both left and right graphs are $(3+1)$d, with the $z$ direction and the $\tau$ direction omitted in the left and the right graph, respectively. The right graph is obtained by a spacetime Wick rotation in the $\tau-z$ plane of the left graph}
    \label{fig:Strange_SSPT}
\end{figure}

This section derives the duality from a dimensional reduction perspective. We will demonstrate that tracing out the bulk of a $(d+1)$-dimensional pure state SSPT will necessarily produce a non-trivial mSPT mixed state in $d$-dimension if one of the subsystem symmetry is preserved by the trace-out procedure (i.e. if they become strong symmetries of the resulting mixed state), by deriving a correspondence of strange correlators of SSPT and fidelity correlators of mSPT. Specifically, we will show that the strange correlator of an SSPT state gives a lower bound for the fidelity strange correlator of the mixed state. Since the strange correlator of SSPT state is finite or algebraically decays with distance, our result in this section suggests that the fidelity strange correlator of the mixed state is necessarily finite or decays algebraically, which means the mixed state is a non-trivial mSPT state. 

Strange correlator~\cite{you2014wave,lepori2023strange,Scaffidi16} is a powerful tool for detecting SPT states. It is defined as 
\begin{align}
C(r,r^{\prime}) = \frac{\langle\Psi_{0}| O(r) O^{\dag}(r^{\prime})|\Psi\rangle}{\langle\Psi_{0}|\Psi\rangle},
\label{eq:strangeCorrelator}
\end{align}
where $|\Psi_0\rangle$ is the reference state, usually taken as a trivial product state, and $|\Psi\rangle$ is the state to be diagnosed, $O(r)$ is a local operator carrying nonzero charges of the symmetry group. Nevertheless, to diagnose the nontriviality of SSPT, one needs an alternative strange correlator because of the failure of Eq. \eqref{eq:strangeCorrelator}. A strange correlator can be understood as a correlation function at the temporal boundary between the product state $|{\Psi_0}\rangle$ and the state $|{\Psi}\rangle$ to be diagnosed, which is equivalent to a correlation function at a spatial interface under Wick rotation~\cite{you2014wave}. For SPT states, the interface between the trivial state and the SPT state cannot be short-range correlated due to the presence of the 't Hooft anomaly, and therefore, the strange correlator cannot be exponentially small. For SSPT states, however, the anomalous modes only live on a low-dimensional submanifold of the interface, with the rest of the interface fully gapped. Therefore, a legitimated strange correlator for SSPT phases needs to ensure both $r$ and $r'$ are located at the same submanifold that carries the anomaly of the bulk SSPT states.

We propose an alternative strange correlator for detecting SSPT states, namely
\begin{equation}
    C(r,r^{\prime}) = \frac{\langle\Psi_{0}| O(r) O^{\dag}(r^{\prime})|\Psi_{\text{SSPT}}\rangle}{\langle\Psi_{0}|\Psi_{\text{SSPT}}\rangle}\bigg|_{r,r^\prime \in \text{bottom layer}},
    \label{eq:strangeSSPT}
\end{equation}
where $|\Psi_{0}\rangle$ and $|\Psi_{\text{SSPT}}\rangle$ are the reference state and the state to be diagnosed, respectively, and $O(r)$ is a local operator carrying nonzero charges of the subsystem symmetry of the bottom layer. Here, we restricted the location of the two operators $r$ and $r^{\prime}$ to the bottom layer.
\begin{figure}[t!]
    \centering
    \includegraphics[width=0.4\textwidth]{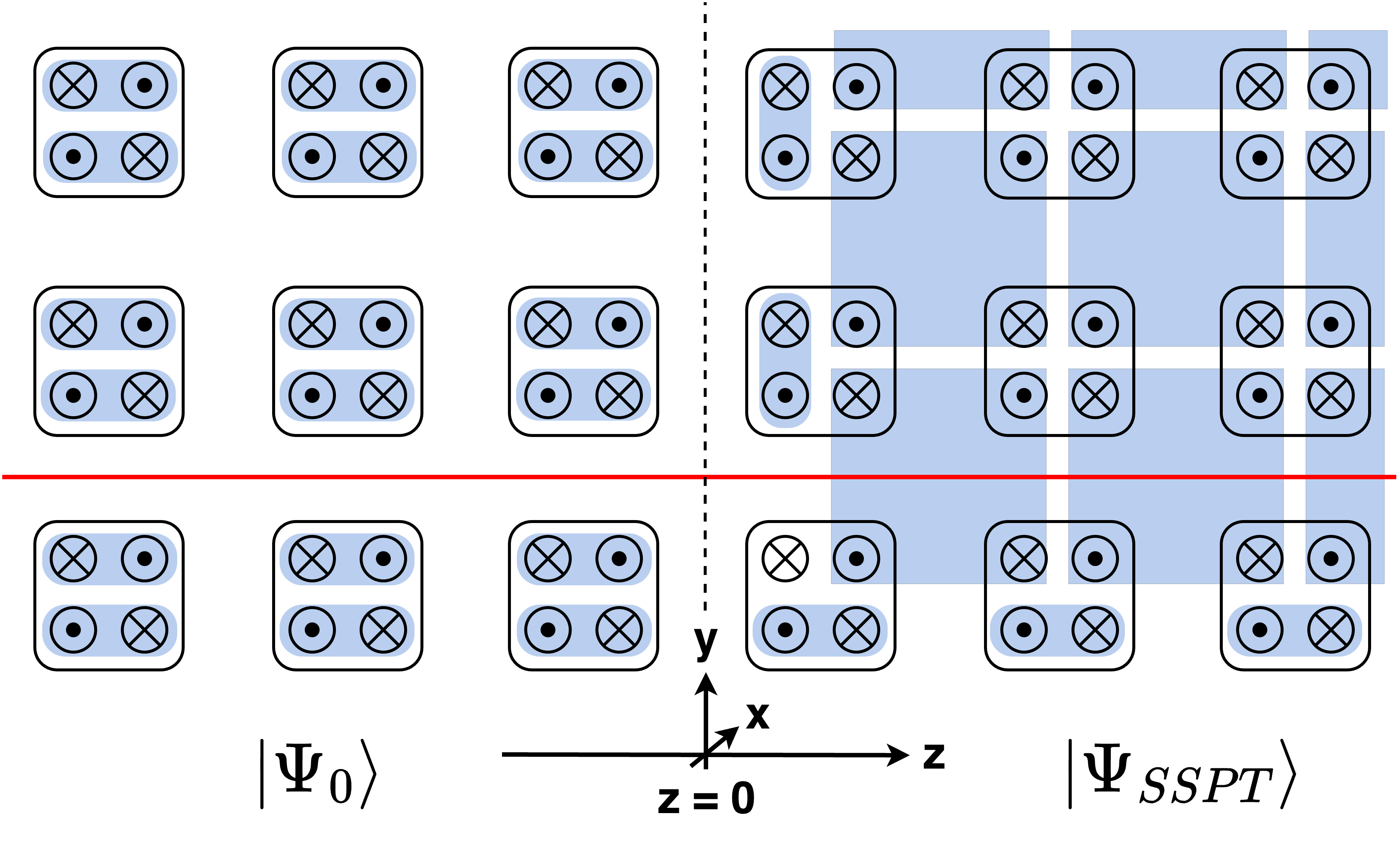}
    \caption{Coupled wire construction of the SSPT and the reference states. The reference state is chosen to be a trivial product state. The interface between the reference and the SSPT state is described by the vertical dashed line. }
    \label{fig:Strange_coupledwire}
\end{figure}

To see that the strange correlation Eq.~\eqref{eq:strangeSSPT} is not exponentially small for SSPT states, we utilize a coupled wire construction~\cite{zhang2023classification} for the states $|\Psi_{\text{SSPT}}\rangle$ and $|\Psi_0\rangle$. Here, we consider 3d bosonic SSPT with 2-foliated subsystem $\mathbb{Z}_{2}$ symmetry as an example; other cases can be demonstrated using a similar argument. The SSPT state and the reference state can be constructed as the ground state of the coupled wire model shown in Fig.~\ref{fig:Strange_coupledwire}. Each site contains four (1+1)d wires, with each wire described by a 2-component Luttinger liquid
\begin{equation}
    \mathcal{L}_{0} = \frac{1}{2\pi} \partial_{x}\phi_{1}\partial_{\tau}\phi_{2}+\frac{1}{4\pi}\sum_{\alpha,\beta=1,2}\partial_{x}\phi_{\alpha}V_{\alpha\beta}\partial_{x}\phi_{\beta}.
\end{equation}
Each wire has a $\mathbb{Z}_{2}$ symmetry under the transformation
\begin{equation}
    \phi_{1}\rightarrow\phi_{1}+\pi, \,\, \phi_{2}\rightarrow\phi_{2}+\pi.
\end{equation}
The system has subsystem $\mathbb{Z}_{2}$ symmetries along each $x$-$y$ and $x$-$z$ plane. The Luttinger liquids can be gapped out by interactions among the 4 wires in the blue plaquette and pairs of wires in the blue ellipses, which are the only interactions allowed by the subsystem symmetries. For the SSPT state, the bulk wires are gapped out by the inter-site couplings among wires in the blue plaquettes, and the wires at the surface are gapped out by the inner-site couplings between the pairs of wires in the blue ellipses. This leads to one dangling Luttinger liquid at each hinge. For the reference state, all wires are gapped out by inner-site couplings, which leads to a trivial product state. From this setup, we notice that the bottom edge of the interface between the SSPT and the reference state is long-range correlated due to the dangling (1+1)d wire carrying anomaly at the interface. 

Having defined the strange correlator for detecting SSPT states, we now turn to the corresponding $d$-dimensional mixed state and show its fidelity strange correlator is necessarily nontrivial if the strange correlator Eq.~\eqref{eq:strangeSSPT} is nontrivial and if the subsystem symmetry of the bottom layer is a strong symmetry of the mixed state. We label $\rho=\ket{\Psi_{\text{SSPT}}}\bra{\Psi_{\text{SSPT}}}$ and $\rho_0=\ket{\Psi_0}\bra{\Psi_0}$, and formulate the ``trace out the top layer'' by a local quantum channel from the whole Hilbert space $\mathcal{H}$ to the Hilbert space of the bottom layer $\mathcal{H}_b$, namely $\E:~{\mathcal{H}\rightarrow\mathcal{H}_b}$. Then, tracing out the top layer gives an mSPT density matrix $\E[\rho]$ and a trivial reference state $\E[\rho_0]$ of the bottom layer. We note that if the SSPT has subsystem symmetry along the horizontal plane, the quantum channel $\E$ maps these subsystem symmetries to a strong symmetry in $\mathcal{H}_b$, while all other symmetries are mapped to weak symmetries. The corresponding fidelity strange correlator \cite{ZhangQiBi2022} is given by
\begin{align}
C_F(r,r^{\prime})=\frac{F\left(\E[\rho], O_r O_{r^{\prime}}^\dag \E[\rho_0] O_{r^{\prime}} O_r^\dag\right)}{F\left(\E[\rho], \E[\rho_0]\right)},
\label{Eq: FSC}
\end{align}
where $F(\rho, \sigma) = \Tr\sqrt{\sqrt{\rho} \sigma \sqrt{\rho}}$ is the fidelity of two density matrices $\rho$ and $\sigma$. Utilizing the Uhlmann's theorem of fidelity \cite{uhlmann1976transition}, namely 
\begin{align}
F(\rho, \sigma)=\max_{\ket{\psi_\rho}, \ket{\phi_{\sigma}}}\left|\langle\psi_\rho|\phi_{\sigma}\rangle\right|.
\end{align}
the denominator of the fidelity strange correlator \eqref{Eq: FSC} has the following purification form
\begin{align}
F\left(\E[\rho], \E[\rho_0]\right)=\max_{\ket{\psi_{\E[\rho]}}, \ket{\phi_{\E[\rho_0]}}}\left|\langle\psi_{\E[\rho]}|\phi_{\E[\rho_0]}\rangle\right|. 
\label{Eq: fidelity}
\end{align}
Suppose the two wave functions $\ket{\tilde{\psi}}$ and $\ket{\tilde{\phi}}$ are the purifications of $\E[\rho]$ and $\E[\rho_0]$ that optimize the fidelity \eqref{Eq: fidelity}, then
\begin{align}
|\langle\Psi_{\text{SSPT}}|\Psi_{0}\rangle|\leq|\langle\tilde\psi|\tilde\phi\rangle|,
\end{align}
since by definition, $\ket{\Psi_{\text{SSPT}}}$ and $\ket{\Psi_0}$ are also purifications of $\E[\rho]$ and $\E[\rho_0]$. Since different purifications of the same density matrix can be related by an isometric operator $U_A$ that is only supported on the ancilla Hilbert spaces,
\begin{align}
\ket{\Psi_{\text{SSPT}}}=U_A\ket{\tilde{\psi}},\quad \ket{\Psi_0}=\tilde{U}_A\ket{\tilde{\phi}},
\end{align}
where both $U_A$ and $\tilde{U}_A$ are supported exclusively in ancilla Hilbert spaces. While the purification $\ket{\tilde{\phi}}$ should always be a trivial product state, due to the symmetry restrictions, there are only two possibilities for the purified state $\ket{\tilde{\psi}}$ of $\E[\rho]$, namely
\begin{enumerate}[1.]
\item If we purify the strong symmetry of $\E[\rho]$ to a horizontal subsystem symmetry, the purified state is an SSPT state.
\item If we purify the strong symmetry of $\E[\rho]$ to global symmetry, then due to the fact that every holographic constructed mSPT from SSPT pure states is intrinsically mixed, the purified state $\ket{\tilde{\psi}}$ must have (quasi) long-range correlation \cite{ma2024topological}, for instance, SSB or gapless pure state.
\end{enumerate}

For the numerator of the fidelity strange correlator \eqref{Eq: FSC}, the purifications $\ket{\tilde{\psi}}$ and $\ket{\tilde{\phi}}$ are not guaranteed to optimize the fidelity. Equivalently, we have the following inequality for the numerator,
\begin{align}
F\left(\E[\rho], O_r O_{r^{\prime}}^\dag \E[\rho_0] O_{r^{\prime}} O_r^\dag\right) \geq \left|\bra{\tilde{\psi}}O_r O_{r^\prime}^\dag \ket{\tilde{\phi}}\right|. 
\end{align}
Gathering everything together, we obtain a lower bound of the fidelity strange correlator \eqref{Eq: FSC}, namely
\begin{align}
C_F(r-r^{\prime})\geq\frac{\left|\bra{\tilde{\psi}}O_r O_{r^{\prime}}^\dag \ket{\tilde{\phi}}\right|}{|\langle\tilde\psi|\tilde\phi\rangle|}, 
\end{align}
where the right-hand side (RHS) hosts two possibilities:
\begin{enumerate}[1.]
\item If $\ket{\tilde{\psi}}$ is an SSPT state, then the RHS is the corresponding strange correlator;
\item If $\ket{\tilde{\psi}}$ is an SSB or gapless state, then it is obvious that the RHS cannot be short-ranged.
\end{enumerate}
In either of these two cases, the lower bound of the fidelity strange correlator is not exponentially decaying with distance. Therefore, the mixed state of the bottom layer $\E[\rho]$ is a non-trivial mSPT state.

\section{Generate 2d SSPT via 1d intrinsic mSPT}
\label{sec:2DSSPT}

The duality formalism introduced in Sec.~\ref{sec:dimext} offers a novel approach to exploring mSPT and SSPT phases. From the perspective of dimension reduction, it allows the derivation of intrinsic mSPT in 2d by considering the reduced density matrix of 3d SSPT, which we will elaborate in Sec.~\ref{sec:corner}-\ref{sec:dipole}. Conversely, from the dimension extension perspective, it provides a new method to generate $d$-dimensional SSPT states by replicating the density matrix of $d-1$ dimensional mSPT. Notably, controlling and manipulating SSPT enables the preparation of resource states for Measurement-Based Quantum Computing (MBQC) \cite{briegel2009measurement,raussendorf2003measurement}, where measurements on bulk qubits of a resource state facilitate universal quantum computation at the boundary.
Ref.~\cite{ma2024topological} explores a broad spectrum of intrinsic mixed-state SPTs, constructed by introducing quenched disorder to an intrinsic gapless SPT state \cite{wen2023bulk, thorngren2021intrinsically, li2023intrinsically}. In this section, we will demonstrate that the intrinsic mSPT proposed in Ref.~\cite{ma2024topological} can be dual to a higher-dimensional SSPT wavefunction.

\subsection{1d intrinsic mSPT on a spin chain} \label{sec:decoheredvsgapless}

We begin by reviewing a concrete example of a 1d intrinsic mSPT proposed in Ref.~\cite{ma2024topological}. The local Hilbert space of this model consists of Ising spins, denoted as \(\sigma\), on the sites and \(\tau\) on the links. The system has a weak $G=\Z_2$ symmetry and a strong $K=\Z_2$ symmetry, with nontrivial group extension \eqref{Eq: extension}. The $\Z_4$ symmetry operators defined on the lattice model have the following form:
\begin{equation}
U_g=\prod_j \sigma^x_j e^{i\frac{\pi}{4} (1-\tau_{j+1/2}^x)},~U_k=\prod_j \tau_{j+1/2}^x.
\end{equation}
Here $g$ and $k$ is the generator of $G$ and $K$, respectively. The unitaries are on-site and satisfy $U_k^2=1,~U_g^2=U_k$.

We define a projector as:
\begin{equation}
P=\prod_j P_j,~P_j= \frac{1+\sigma^z_j \tau_{j+1/2}^x\sigma^z_{j+1}}{2}.
\end{equation}
In the subspace where \( P = 1 \), an Ising domain wall, characterized by \(\sigma_j^z \sigma_{j+1}^z = -1\), is embedded with a charge \(\tau_{j+1/2}^x = -1\), known as a decorated domain wall pattern. When projecting the Hilbert space onto the subspace with \( P = 1 \), \( U_g \) takes the following form:
\begin{equation}\label{z2an}
U_g=\prod_j \sigma^x_j e^{i\frac{\pi}{4} (1-\sigma_{j}^z\sigma_{j+1}^z)},
\end{equation}
which takes the form of the anomalous $\Z_2$ symmetry of the Levin-Gu edge model.
In the low-energy subspace, it is evident that \( U_k \) reduces to the identity operator, at least in the bulk of the spin chain. Within this subspace, the effective theory corresponds to the edge Hamiltonian of the Levin-Gu model \cite{LevinGu}, as explored in Refs.~\cite{gaplessSPT, scaffidi2017gapless}. These works demonstrated that the system exhibits an intrinsic gapless SPT (igSPT) state characterized by an emergent \( Z_2 \) anomaly.





To obtain an intrinsic mSPT phase, we can add 
a random Ising disorder $-\sum_j h_j\sigma_j^z$ with $h_j=\pm 1$ to break $G=\mathbb{Z}_2$ symmetry. In the strong disorder limit, 
the mixed-state density matrix is an incoherent sum of the ground-state wavefunction for the disordered Hamiltonian,
\begin{equation}
\begin{gathered}\label{quench}
H_{\mathcal{D}}=\sum\limits_jP_j+h_j\sigma_j^z\\
|\Psi_{\mathcal{D}}\rangle=\bigotimes_j
|\sigma_j^z=h_j\rangle\otimes|\tau_{j+1/2}^x=h_jh_{j+1}\rangle
\end{gathered}.
\end{equation}
Thus we obtain a disordered ensemble $\{\ket{\Psi_{\cal D}}\}$ whose density matrix describes an intrinsic mSPT.
\begin{equation}\label{1dmsptd}
\rho =\frac{1}{2^N} \sum_{{\cal D}} |\Psi_{\mathcal{D}}\rangle \langle \Psi_{\mathcal{D}}|.
\end{equation}

\subsection{2d SSPT}
\begin{figure}
  \centering
\includegraphics[width=0.35\textwidth]{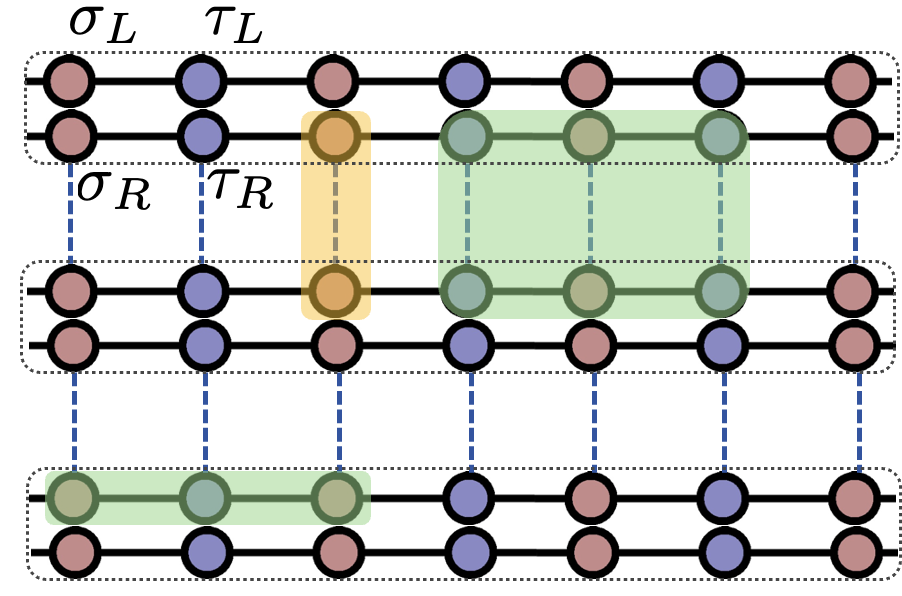}
\caption{Coupled Wire Construction for 2d SSPT: Each unit cell (within the dashed line) contains two sets of spin chains, labeled L and R. The building block involves the L and R spin chains across adjacent unit cells. The yellow/green shaded area highlights the cluster associated with stabilizer operators.} 
\label{1dchain}
\end{figure}

We now map the 1d mSPT in Eq.~\eqref{1dmsptd} into a 2d wavefunction, \(\Psi^{SSPT}\), using the replica trick proposed in Eq.~\eqref{eq:wfmap}. The replica index \( N \) serves as the row index along the y-axis. Each row, considered as a unit cell, contains two flavors of spin chains, \( \sigma_{L/R} \) and \( \tau_{L/R} \), as depicted in Fig.~\ref{1dchain}. The resulting 2d wavefunction \(\Psi^{SSPT}\) exhibits a subsystem \( \Z_2 \) symmetry along each x-row.
\begin{equation}
U_s(y)=\prod_{x} \tau_{R}^x(x,y)~ \tau_{L}^x(x,y)
\end{equation}
The subsystem symmetry arises from the strong \( \Z_2 \) symmetry in the 1d mixed state. Additionally, there is a global \( \Z_4 \) symmetry:
\begin{equation}
U_g=\prod_{x,y} \sigma^x_{L}(x,y) ~\sigma^x_{R}(x,y) e^{i\frac{\pi}{4}\sum_{x,y} [\tau_{L}^x(x,y))-\tau_{R}^x(x,y)]}
\end{equation}
This global \( \Z_4 \) symmetry stems from the weak \( \Z_4 \) symmetry in the 1d mixed state. Notably, \( (U_g)^2 \) represents a global \( \Z_2 \) symmetry that counts the total \(\tau^x\) charge, which is a subgroup of the subsystem \( \Z_2 \) symmetry.

A stabilizer Hamiltonian can be defined for the 2d SSPT wavefunction \(\Psi^{\mathrm{SSPT}}\):
\begin{widetext}
\begin{align}
    H=&-\sum_{\alpha=L/R}[\sigma^z_{\alpha}(x,y) \tau^x_{\alpha}(x,y) \sigma^z_{\alpha}(x+1,y)
    +J\sigma^z_{L}(x,y)\sigma^z_{R}(x,y+1)\nonumber\\
    &+\tau^z_{L}(x,y) \sigma^x_{L}(x+1,y) \tau^z_{L}(x+1,y) \tau^z_{R}(x,y+1) \sigma^x_{R}(x+1,y+1) \tau^z_{R}(x+1,y+1)]
\end{align}
\end{widetext}
The Hamiltonian can be decomposed into local stabilizer terms within each building block, illustrated as Fig.~\ref{1dchain}. Each building block consists of the L-spin chain at row \( y \) and the R-spin chain at row \( y+1 \). When the \( J \)-coupling is strong, it aligns the \(\sigma^z\) spins along the rung within each building block. Tracing out one of the spin chains in the building block yields a reduced density matrix that is diagonal in \(\sigma^z\), similar to the quenched disorder introduced in Eq.~\eqref{quench}. This Hamiltonian results in a gapped ground state \(\Psi^{SSPT}\) in the bulk.

In the presence of an open smooth boundary at $x=0$ (terminating at site $\sigma$), we can identify the following edge operators from each row:
\begin{align}\label{edgeoperator}
    &\tilde{Z}_{R/L}=\sigma^z_{R/L}(0,y),\nonumber\\
    &\tilde{X}_{R/L}=\sigma^x_{R/L}(0,y)\tau^z_{R/L}(0,y),\nonumber\\
    &\tilde{Y}_{R/L}=\sigma^y_{R/L}(0,y)\tau^z_{R/L}(0,y).   
\end{align}
At the edge site \( x=0 \), we turn off the coupling \(\sigma^z_{L}(0,y)\sigma^z_{R}(0,y+1)\), which does not impact the bulk physics. This ensures that all edge operators in Eq.~\eqref{edgeoperator} commute with the bulk stabilizers. Each y-row at the edge has two sets of dangling spin-\(\frac{1}{2}\) particles, labeled L and R. These can be entangled into an onsite singlet that is invariant under both the subsystem \( \Z_2 \) and global \( \Z_4 \) symmetries. Thus, a smooth boundary at $x=0$ can be gapped.

But what about the rough boundary? Suppose we have open boundaries at both \( x=0 \) and \( y=0 \). The edge at \( y=0 \) contains the R-spin chain that is decoupled from the bulk building block. We can introduce an edge transverse field \(\tau^x_R\) and \(\sigma^x_R\) at \( y=0 \), thereby gapping out the R-spin chain at this boundary. However, at the corner \( x=y=0 \), where these two edges intersect, there remain free Pauli operators \(\tilde{Z}_L\), \(\tilde{X}_L\), and \(\tilde{Y}_L\) that are decoupled from both the edge and the bulk. These spins form a projective representation under the subsystem \( \Z_2 \) and global \( \Z_4 \) symmetries, resulting in a corner mode. This concludes that the system represents a higher-order SSPT phase protected by subsystem \( \Z_2 \) and global \( \Z_4 \) symmetries.

{\color{black}

\subsubsection{Holographic duality from field theory picture}

We can also understand the holographic duality from a field theory argument to demonstrate the \(\Psi^{SSPT}\) state generated by replicating the 1d density matrix. Let us consider one building block by using the Choi–Jamiolkowski isomorphism. As outlined in Sec.~\ref{sec:decoheredvsgapless}, in the absence of decoherence or disorder, the building blocks exhibit a gapless theory anomalous under the \(\mathbb{Z}_2\) symmetry defined in Eq.~\eqref{z2an}. An effective field theory description of the $\mathbb{Z}_2$ anomaly can be written as the $O(4)$ non-linear sigma model as the following,
\begin{align}
\mathcal{L}_{O(4)}[\vec{m}]=\frac{1}{g}(\partial_{\mu} \vec{m})^2+\frac{i2\pi}{\Omega^3} \int_0^1 du\epsilon^{ijkl}  m_{i}\partial_z m_{j} \partial_t m_{k}\partial_u m_{l},
\label{O4}
\end{align}
where the $\mathbb{Z}_2$ symmetry acts as $\mathbb{Z}_2: \vec{m}\rightarrow -\vec{m}$. The anomaly is captured by the WZW term \cite{bi2015classification}. In the doubled space, the effective action is 
\begin{align}
&\mathcal{L}_{\text{doubled}}=\mathcal{L}_{O(4)}^L[\vec{m}_L]+\mathcal{L}_{O(4)}^{R*}[\vec{m}_R]-U\vec{m}_R\cdot \vec{m}_L 
\label{choidouble}
\end{align}
The effect of decoherence, which reduces strong symmetry to weak symmetry, is captured by the interaction in Eq.~\eqref{choidouble} that breaks strong symmetry while preserving weak symmetry. 


In the strong-coupling limit, the ket and bra vector bosons align in the same orientation, \(\vec{m}_{L} = \vec{m}_{R}\), the effect of the WZW terms cancels out. Consequently, the Choi-double state reduces to
\begin{align}
|\rho\rangle \,\sim\, \sum_{\vec{m}}\,|\vec{m}\rangle_L\,|\vec{m}\rangle_R.
\end{align}

Now consider the replica construction, the corresponding 2d wavefunction arises by stacking these Choi-double building blocks, in a manner akin to the coupled-wire construction. In this picture, each unit cell\footnote{This is not to be confused with the building block.} contains two WZW theories, denoted as $L_i$ and $R_i$ theories, where $i$ labels the unit cell. According to our construction as shown in Fig. \ref{wavefunc} and also in Fig. \ref{1dchain}, the interaction due to decoherence couples neighboring unit cells -- specifically, it couples $L_i$ and $R_{i+1}$ theories -- leading to a gapped bulk. However, it will leave a single copy of the WZW theory if the system has a boundary. In other words, the effective theory describing the 2d system will be a topological $\theta$-term at $\theta=2\pi$ \cite{bi2015classification}, which indicates the 2d bulk can be viewed as an SPT protected by the $\mathbb{Z}_2$ symmetry\footnote{Note that this $\mathbb{Z}_2$ symmetry is the weak symmetry of the original system. The subsystem symmetry sector is gapped (which is the strong symmetry) is not reflected in this effective WZW description.}.

This emergent SPT in bulk provides insight into the twisted Rényi-$N$ correlator. We have shown that the twisted Rényi-$N$ correlator for the 1d density matrix maps to the strange correlator of the 2d wavefunction. Due to the presence of the 2d SPT state, the strange correlator is nontrivial (long-range or power law), indicating nontrivial results for the twisted Rényi-$N$ correlator. Notably, we need to measure the twisted Rényi-$N$ correlator of the charge operators under weak $\mathcal{G}$ symmetries, as the SPT is protected by $\mathcal{G}$ symmetry. Moreover, the twisted Rényi-$N$ correlator will be trivial if the mixed-state density matrix is not an intrinsic mSPT. In this case, the argument based on anomaly and coupled-wire construction fails, as the system does not exhibit a low-energy anomaly in the first place. A more comprehensive examination of the twisted Rényi-\(N\) correlator is presented in Sec.~\ref{sec:twisted}.}

\begin{figure}
  \centering    \includegraphics[width=0.48\textwidth]{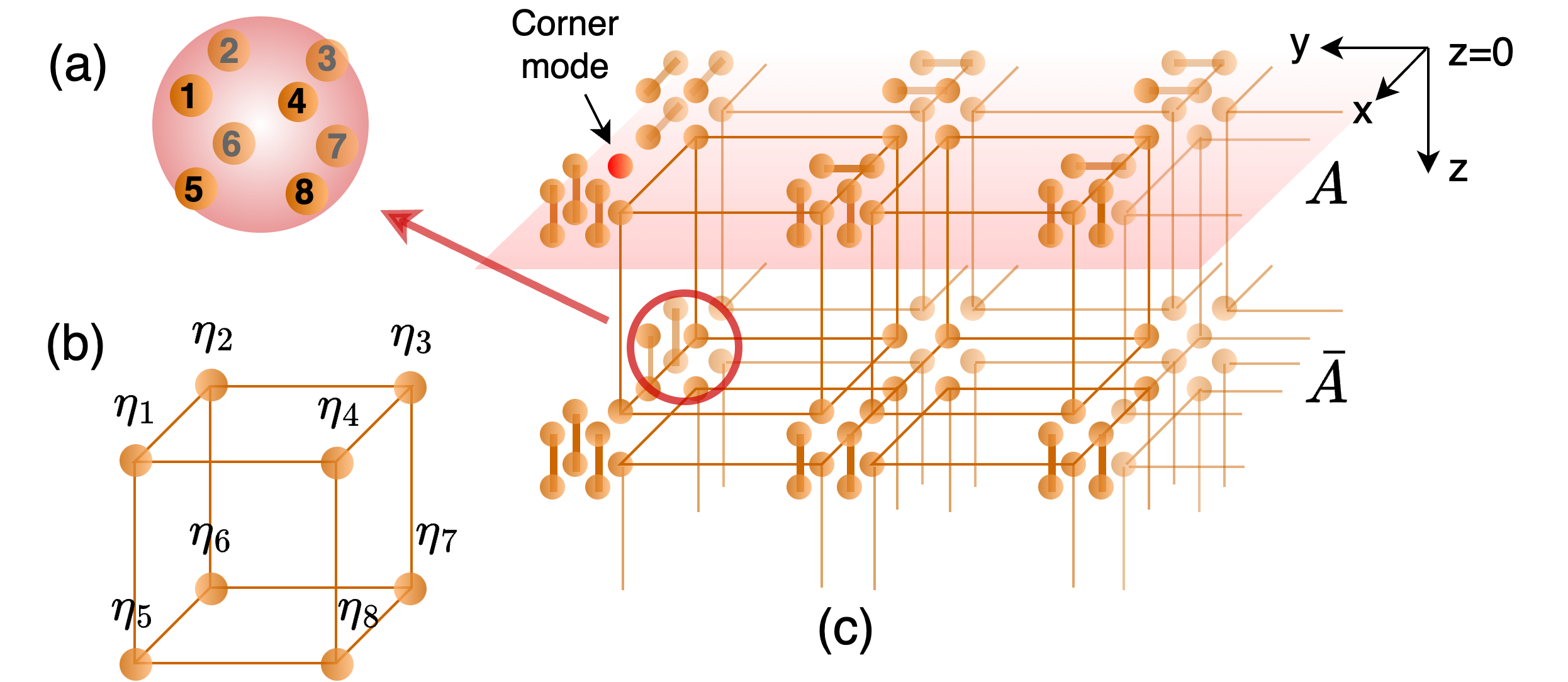}
  \caption{Schematic illustration of the lattice model of the 3D topological superconductor with Majorana corner modes. (a) Each site has eight Majoranas $\eta_1,\cdots,\eta_8$. (b-c) The eight Majorana zero modes living at the corner of each cube are projected into a unique ground state preserving coplanar fermion parity symmetry.} 
  \label{fermion}
\end{figure}

\section{Dimension reduction between 3d HOTSC and 2d intrisic mSPT with corner modes}\label{sec:corner}

\subsection{3d topological superconductor with Majorana corner modes protected by subsystem $\Z^{f}_2$}\label{sec:hotscbulk}

To start with, we first reboot our thinking by examining an exactly solvable higher-order topological superconductor (HOTSC) on a 3d cubic lattice protected by subsystem fermion parity symmetries proposed in Ref.~\cite{you2024higher}. This model exemplifies a `third-order topological superconductor' with Majorana modes localized at each corner, as illustrated in Fig.~\ref{fermion}. These corner modes are protected by the conservation of subsystem fermion parity on each $x$-$y$, $y$-$z$, and $x$-$z$ planes. Notably, such 3-foliated subsystem fermion parity symmetry prevents any fermion bilinear tunneling terms between sites, setting it apart from non-interacting systems described by standard band theory.

The lattice model consists of four complex fermions per site on the cubic lattice, as shown in Fig.~\ref{fermion}. These four fermions at each site can be decomposed into eight Majorana operators, labeled as $\eta_1, \dots, \eta_8$. The Hamiltonian can be decomposed into elementary building blocks $\mathcal{B}_r$, with each block consisting of a cube containing eight Majorana operators located at its eight corner sites:
\begin{widetext}
\begin{align} \mathcal{B}_r:~[ \eta_1(r), \eta_4(r-\hat{x}),\eta_2(r-\hat{y}),\eta_3(r-\hat{y}-\hat{x}), \eta_5(r+\hat{z}), \eta_8(r-\hat{x}+\hat{z}),\eta_6(r-\hat{y}+\hat{z}),\eta_7(r-\hat{y}-\hat{x}+\hat{z})] \label{eq:buildingblocks} . \end{align}
\end{widetext}
In this setting, each of the eight Majorana operators at a single site is uniquely associated with one of the eight adjacent cubes. This ensures that different building blocks contain non-overlapping Majorana degrees of freedom as illustrated in Fig.~\ref{fermion}.

The interaction between Majorana operators occurs only within the same building block $\mathcal{B}_r$. Thus, we can represent the Hamiltonian as a sum of local interactions for each building block $H^{\mathcal{B}_r}$. Due to the subsystem fermion parity symmetry on each $x$-$y$, $y$-$z$, and $x$-$z$ plane, any fermion bilinear term within the building block is prohibited so the system does not render any non-interacting analogy. To proceed, we first introduce a Majorana quartet interaction in each building block cube as:
\begin{align} 
H^{\mathcal{B}_r}_1=\eta_5\eta_6\eta_7\eta_8+\eta_1\eta_2\eta_3\eta_4.
\label{four}
\end{align}
Here, $\eta_1, \ldots, \eta_8$ refer to the Majorana operators within the building block ${\mathcal{B}_r}$ as defined in Eq.~\eqref{eq:buildingblocks}. For simplicity, their spatial coordinates are omitted.
Redefine these Majorana operators in terms of two sets of spinful fermions $\Psi_{\uparrow/\downarrow}$ and $\Psi'_{\uparrow/\downarrow}$. 
\begin{align} 
&\Psi_{\uparrow}=\eta_5+i\eta_6,\Psi_{\downarrow}=\eta_7+i\eta_8\nonumber\\
&\Psi'_{\uparrow}=\eta_1+i\eta_2,\Psi'_{\downarrow}=\eta_3+i\eta_4
\label{mapf}
\end{align}
This allows the quartet term in Eq.~\eqref{four} to be rewritten as:
\begin{align} 
H^{\mathcal{B}_r}_1=(n_{\Psi}-1)^2+(n_{\Psi'}-1)^2
\end{align}
The interaction term $H^{\mathcal{B}_r}_1$ favors the odd fermion parity state for both $\Psi$ and $\Psi'$. This allows us to map the ground state subspace of Eq.~\eqref{four} into two spin-1/2 degrees of freedom per building block:
\begin{align} \label{cp2}
&\vec{n}=\Psi^{\dagger} \vec{\sigma} \Psi, ~\vec{m}=\Psi'^{\dagger} \vec{\sigma} \Psi', 
\end{align}
The vector $\vec{n}$ or $\vec{m}$ characterizes a spin-1/2 degree of freedom in the $CP^1$ representation. In terms of these spin degrees of freedom, the second interaction within the building block is given by:
\begin{align} 
& H^{\mathcal{B}_r}_2=- m_x n_x-m_y n_y\nonumber\\
&= (\eta_5\eta_6-\eta_7\eta_8)(\eta_1\eta_2-\eta_3\eta_4)\nonumber\\
&+(\eta_5\eta_8-\eta_6\eta_7)(\eta_1\eta_4-\eta_2\eta_3)
\label{four2}
\end{align}
This denotes an XY interaction between the two spins in each building block, projecting them into an SU(2) singlet and resulting in a unique ground state. It is important to highlight that these interactions preserve the subsystem fermion parity symmetry on all i-j planes. By incorporating the cluster interactions described in Eqs.~\eqref{four}-\eqref{four2} in each building block, the bulk Hamiltonian becomes fully gapped and has a unique ground state.

Now let us examine the boundary in Fig.~\ref{fermion}. Each surface site contains four dangling Majoranas, and each hinge site contains two dangling Majoranas. These can be gapped out through onsite Majorana hybridization, which remains invariant under subsystem fermion parity symmetry.
The situation becomes tricky at the corners. Each corner site carries an odd number of dangling Majorana zero modes, which cannot be fully gapped out through onsite interactions. Additionally, due to the subsystem fermion parity symmetry, one cannot hybridize or gap out the Majorana zero modes from different corners via surface or hinge phase transitions without breaking the subsystem fermion parity on all \(i-j\) planes. Consequently, the corner host robust Majorana zero mode protected by the subsystem fermion parity symmetry.
Notably, the Majorana corner mode is an intrinsic feature emerging from the 3d bulk. This is evidenced by the impossibility of annihilating the corner modes through a gap closure that occurs solely on the hinge or surface. For example, if the gap closes on the top surface, it triggers long-range interactions among the four Majorana corner modes labeled \(\pi^1, \pi^2, \pi^3,\) and \(\pi^4\) on that surface. The only symmetry-allowed interaction term among these corner modes is \(\pi^1\pi^2\pi^3\pi^4\), which is insufficient to lift the degeneracy of all four corner modes.

\subsection{Reduce density matrix of the top layer} \label{sec:reduce}

The coupled building block Hamiltonian proposed in Sec.~\ref{sec:hotscbulk} yields a fixed-point ground state wave function with zero correlation. The eight Majorana modes in each building block form a highly entangled cube. As a result, the ground state wave function can be expressed as a tensor product of these entangled building blocks. 

Now, suppose we begin with the wave function $\Psi^{sspt}$ with open boundaries at $z=0$, illustrated in Fig.~\ref{fermion}. If we make a spatial cut parallel to the \(xy\)-plane, separating the top surface \(z=0\) (denoted as \(A\)) from all bottom parts \(z<0\) (denoted as \(\bar{A}\)), the reduced density matrix of the top layer, \(\rho^A\), can be treated as a mixed state ensemble in 2d.
The mixed state \(\rho^A\) exhibits a strong fermion parity symmetry (denoted as $Z^f_2$), stemming from the subsystem fermion parity conservation on the $x$-$y$ plane of \(\Psi^{sspt}\). Additionally, \(\rho^A\) displays a weak subsystem symmetry for fermion parity on the x-rows and y-columns (denoted as $Z^{f,x}_2, Z^{f,y}_2$), derived from the subsystem fermion parity conservation on the $x$-$z$ and $y$-$z$ planes of \(\Psi^{sspt}\).

Given that the wave function \(\Psi^{sspt}\) is a tensor product of entangled cubes, the entanglement cut between the top surface layer and the bottom layers separates \(\eta_1, \eta_2, \eta_3, \eta_4\) (in the lower half at \(z=1\)) from \(\eta_5, \eta_6, \eta_7, \eta_8\) (in the upper half at \(z=0\)) of each cube. The remaining dangling Majoranas at the top layer (\(z=0\)) are trivially entangled on-site and do not contribute to the entanglement with the other degrees of freedom, so we can ignore them for now.
To obtain the reduced density matrix \(\rho^A\), focus on a slab of building blocks that extends along the $x$-$y$ plane from \(z = 0\) to \(z = 1\). Tracing out the \(z = 1\) layer gives us the reduced density matrix \(\rho^A\), illustrated as Fig.~\ref{fermion}. 
Conversely, one can assert that the mixed ensemble \(\rho^A\) has a purification state, \(\Psi^{sspt}\), where the system qubits are located in the top layer at \(z = 0\). All bottom layers act as ancilla qubits that can be traced out.

The mixed ensemble \(\rho^A\) describes a mixed state SPT in 2d protected by strong fermion parity symmetry (\(\Z^f_2\)) and weak subsystem symmetry for fermion parity on the $x$-rows and $y$-columns (\(\Z^{f,x}_2, \Z^{f,y}_2\)). Given that the \(\Psi^{sspt}\) state has a finite correlation length in the bulk and side surfaces, \(\rho^A\) is a short-range entangled mixed state. When a rough boundary is present, such as a corner at \(y=0,~x=0\), a Majorana dangling mode appears, which cannot be eliminated through any interaction or symmetry-allowed quantum channels.

Notably, this Majorana corner mode does not exist in 2d SPT as a pure state in the presence of fermion subsystem symmetry. Therefore, the mSPT we define here is an \textit{intrinsic mSPT} that has no pure state analog under thermal equilibrium.
This is evidenced by the fact that if we shrink the 2d lattice into a tiny patch where four Majorana corner modes are close to each other by a lattice unit, it is impossible to gap these four Majorana corner modes while preserving subsystem fermion parity along the row and column. Alternatively, this also indicates that the 3d higher-order topological superconductor's surface with corner modes cannot be annihilated by adding 2d surface layers.

In the following paragraphs, we demonstrate that the nontriviality of this 2$d$ mSPT is reflected by the bipartite non-separability of the reduced density matrix of Majorana corner modes.

\begin{figure}
\begin{tikzpicture}[scale=0.88]
\tikzstyle{sergio}=[rectangle,draw=none]
\draw[ultra thick, color=red] (-2.5,0) -- (2.5,0);
\draw[ultra thick, color=red] (-2.5,-3) -- (2.5,-3);
\draw[ultra thick, color=red] (-1.5,0.5) -- (-1.5,-4);
\draw[ultra thick, color=red] (1.5,0.5) -- (1.5,-4);
\filldraw[fill=red!20, draw=red, thick] (-1.5,0)--(1.5,0)--(1.5,-3)--(-1.5,-3)--cycle;
\filldraw[fill=green!20, draw=black] (-2,0)circle (10pt);
\filldraw[fill=green!20, draw=black] (-1.5,-3.5)circle (10pt);
\filldraw[fill=green!20, draw=black] (1.5,-3.5)circle (10pt);
\filldraw[fill=green!20, draw=black] (-2,-3)circle (10pt);
\filldraw[fill=black, draw=black] (-1.5,-3)circle (3pt);
\filldraw[fill=black, draw=black] (1.5,-3)circle (3pt);
\filldraw[fill=black, draw=black] (-1.5,0)circle (3pt);
\filldraw[fill=black, draw=black] (1.5,0)circle (3pt);
\path (1.5,-3.5) node [style=sergio] {${P_f'}^y$};
\path (-1.5,-3.5) node [style=sergio] {$P_f^y$};
\path (-2,-3) node [style=sergio] {${P_f'}^x$};
\path (-2,0) node [style=sergio] {$P_f^x$};
\path (-1,-0.5) node [style=sergio] {\large $\gamma_1$};
\path (1,-0.5) node [style=sergio] {\large $\gamma_2$};
\path (-1,-2.5) node [style=sergio] {\large $\gamma_4$};
\path (1,-2.5) node [style=sergio] {\large $\gamma_3$};
\end{tikzpicture}
\caption{Majorana modes at corners.
}
\label{Majorana}
\end{figure}
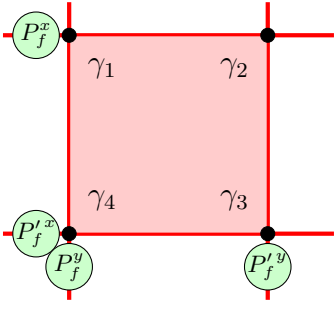

Consider the 2d subsystem on the topmost plane (obtained by tracing out all other layers), there are 4 dangling Majorana modes at the corner of the plane that are not coupled to other bulk and edge Majorana modes, as illustrated in Fig. \ref{fermion}, each layer is composed of two slices of Majorana zero modes, of which the Majorana zero modes included in the bottom layer were entangled with other bulk Majorana modes before tracing. In particular, the bulk wavefunction of the 3d SSPT model is composed of the tensor product of 8 Majorana modes forming the ground state of the local Hamiltonians \eqref{four} and \eqref{four2}. In particular, the ground state of $H_2$ is effectively a spin-singlet state of two spin-1/2 degrees of freedom,
\begin{align}
|\Psi_{ul}\rangle=\frac{1}{\sqrt{2}}\left(|\uparrow\rangle_u\otimes|\uparrow\rangle_l-|\downarrow\rangle_u\otimes|\downarrow\rangle_l\right),
\end{align}
where the subscript $u/l$ depicts the upper/lower half of the cubic (see Fig. \ref{fermion}). Then if we trace out the lower half, the resulting density matrix would be a maximally mixed state of the spin-1/2 degree of freedom, namely
\begin{align}
\rho_p=\Tr_l\left(|\Psi_{ul}\rangle\langle\Psi_{ul}|\right)=\frac{1}{2}\left(|\uparrow\rangle\langle\uparrow|_u+|\downarrow\rangle\langle\downarrow|_u\right).
\end{align}
Then we notice that these two effective spin-1/2 degrees of freedom are defined in the ground state subspace of the Hamiltonian \eqref{four}, and each spin-1/2 is formed by two complex fermions. In particular, the $S^z$ operator has the following form,
\begin{align}
S^z_u=\Psi_\uparrow^\dag\Psi_\uparrow-\Psi_\downarrow^\dag\Psi_\downarrow,
\end{align}
then the density matrix $\rho_p$ can be reformulated with the complex fermion basis, namely
\begin{align}
\rho_p=\frac{1}{2}(|01\rangle\langle01|+|10\rangle\langle10|),
\end{align}
where two digits depict the occupation number of fermions $\Psi_\uparrow$ and $\Psi_\downarrow$, respectively. 

Notice that all above discussions are irrelevant to the top slice of the topmost layer. Therefore, for each bulk site on the topmost layer, there are four additional Majorana zero modes that have been trivially gapped before tracing. Hence the overall bulk density matrix of the topmost layer is simply the tensor product of the atomic insulator $|00\rangle\langle00|$ in the top slice and the tensor product of $\rho_p$ for all different plaquettes in the bottom slice. 

Similarly, for each edge site on the topmost layer, there are six dangling Majorana zero modes that have been trivially gapped in the 3d SSPT model, and the ``tracing out'' procedure will not affect them. Therefore, the edge density matrix is simply the tensor product of the atomic insulator $|000\rangle\langle000|$. 

Finally, for each corner site on the topmost layer, there are seven dangling Majorana zero modes, six of which have been trivially gapped in the 3d SSPT model, and the ``tracing out'' procedure will not affect them. Therefore, we have four dangling Majorana modes on the topmost layer after tracing out all other layers, see Fig. \ref{Majorana}. In the subspace spanned by these 4 Majorana modes, the generator of the strong global fermion parity is
\begin{align}
P_f=-\gamma_1\gamma_2\gamma_3\gamma_4,
\end{align}
and the generators of weak subsystem fermion parities along horizontal and vertical directions are
\begin{align}
\begin{gathered}
P_f^x=i\gamma_1\gamma_2,~~{P_f'}^x=i\gamma_3\gamma_4\\
P_f^y=i\gamma_1\gamma_4,~~{P_f'}^y=i\gamma_2\gamma_3
\end{gathered}.
\end{align}
The reduced density matrix $\rho$ of these 4 Majorana modes should satisfy the following symmetry conditions,
\begin{align}
\begin{gathered}
P_f\rho=e^{i\theta}\rho\\
P_f^x\rho P_f^x=\rho,~{P_f'}^x\rho {P_f'}^x=\rho\\
P_f^y\rho P_f^x=\rho,~{P_f'}^x\rho {P_f'}^y=\rho
\end{gathered}.
\label{sym1}
\end{align}

Based on these, we further have some constraints on the density matrix as collieries, say
\begin{align}
\begin{gathered}
(i\gamma_1\gamma_2)\rho (i\gamma_3\gamma_4)=e^{i\theta}\rho\\
(i\gamma_1\gamma_3)\rho (i\gamma_2\gamma_4)=-e^{i\theta}\rho\\
(i\gamma_1\gamma_4)\rho (i\gamma_2\gamma_3)=e^{i\theta}\rho
\end{gathered}.
\label{sym2}
\end{align}

It is obvious that the corner density matrix $\rho$ must be tripartite non-separable, then we demonstrate that $\rho$ is actually bipartite non-separable. We choose the Fock space of two complex fermions defined through the 4 Majorana modes, namely
\begin{align}
c_{14}^\dag=\frac{1}{2}(\gamma_1+i\gamma_4),~c_{23}^\dag=\frac{1}{2}(\gamma_2+i\gamma_3),
\end{align}
then the corner Hilbert space is spanned by four states $\{|00\rangle,|01\rangle,|10\rangle,|11\rangle\}$, where two digits label the occupation of the complex fermions $c_{14}$ and $c_{23}$, respectively. Under the strong global fermion parity, these 4 states have the following symmetry properties,
\begin{align}
\begin{gathered}
P_f|00\rangle=|00\rangle,~P_f|11\rangle=|11\rangle\\
P_f|01\rangle=-|01\rangle,~P_f|10\rangle=-|10\rangle
\end{gathered},
\end{align}
then as a consequence, the states $\{|00\rangle,|11\rangle\}$ and $\{|01\rangle,|10\rangle\}$ cannot be mixed (neither coherent superposition nor convex sum) in the strongly symmetric corner density matrix $\rho$. 

We first consider the subspace spanned by $\{|00\rangle, |11\rangle\}$. Following the symmetry constraint $(i\gamma_1\gamma_2)\rho(i\gamma_3\gamma_4)=e^{i\theta}\rho$, we conclude that
\begin{align}
\begin{gathered}
(i\gamma_1\gamma_2)|00\rangle\langle00|(i\gamma_3\gamma_4)=-|11\rangle\langle11|\\
(i\gamma_1\gamma_2)|11\rangle\langle11|(i\gamma_3\gamma_4)=-|00\rangle\langle00|
\end{gathered},
\end{align}
hence if $\rho\propto|00\rangle\langle00|+|11\rangle\langle11|$, we get $(i\gamma_1\gamma_2)\rho(i\gamma_3\gamma_4)=-\rho$. Nevertheless, on the other hand, we see that $(i\gamma_1\gamma_2)\rho(i\gamma_3\gamma_4)=P_f^x\rho P_f^xP_f=\rho$. Therefore, $\rho$ cannot be the convex sum of $|00\rangle$ and $|11\rangle$. 

We conclude that if we require the symmetry constraints \eqref{sym1} and \eqref{sym2} within the subspace spanned by $|00\rangle$ and $|11\rangle$, the only allowed corner density matrix must be the convex sum of $(|00\rangle+i|11\rangle)(\langle00|-i\langle11|)$ and $(|00\rangle-i|11\rangle)(\langle00|+i\langle11|)$. We can see that $\rho$ has the form of a convex sum of two Bell states, hence it cannot be separable. For the subspace spanned by $|01\rangle$ and $|10\rangle$, following similar analysis, we conclude that only 
the convex sum of $(|01\rangle+i|10\rangle)(\langle01|-i\langle10|)$ and $(|01\rangle-i|10\rangle)(\langle01|+i\langle10|)$ is legitimated by the symmetry constraints \eqref{sym1} and \eqref{sym2}, which also has the same form of a convex sum of two Bell states. 

In principle, for discussing the separability problem of a density matrix, you should test all possible bases. For the present case, the calculations of separability of corner density matrix $\rho$ with different bases are the same. Therefore, we proved that the corner reduced density matrix $\rho$ must be bipartite non-separable, which implies that it carries a higher-order mixed anomaly of a weak fermion parity and two strong fermion parity symmetries.

\subsection{Prepare mSPT via Quantum channels}

Thus far, we have demonstrated that the reduced density matrix of the top surface of a HOTSC can be interpreted as a mixed-state SPT in 2d. In this section, we will show how this mixed-state SPT can be prepared using local quantum channels that preserve both weak and strong symmetries.
Consider a 2d square lattice, with each vertex hosting four Majoranas labeled \(\eta_1\), \(\eta_2\), \(\eta_3\), and \(\eta_4\). On each plaquette, we introduce a strong fermion quartet term that couples the four Majoranas within the plaquette:
\begin{align}
H_1=\eta_1(r)\eta_2(r+\hat{x})\eta_3(r+\hat{x}+\hat{y})\eta_4(r+\hat{y})
\label{2dh}
\end{align}
Such quartet coupling reduces the 4-dimensional Hilbert space spanned by the four Majoranas into a two-level system. Consequently, the ground state of Eq.~\eqref{2dh} contains \(2^{N_p}\) degenerate modes (with \(N_p\) being the number of plaquettes on the lattice), which span a degenerate ground state manifold \(\rho^{GS}\).
Now, we apply a set of local quantum channels to \(\rho^{GS}\),
\begin{align}
    \label{decohere1}
    & \hrho^D = \mathcal{E}[\rho^{GS}], \ \ \mathcal{E} = \prod_{\vect{r}} \cE^x_{\vect{r}} \cE^y_{\vect{r}}, \nonumber \\ 
    &  \cE^x_{\vect{r}}[\hrho^{GS}] =\frac{1}{2} \rho^{GS} + \frac{1}{2} \eta_1(r)\eta_2(r+\hat{x}) \rho^{GS} \eta_2(r+\hat{x})\eta_1(r), \nonumber \\ 
     &  \cE^x_{\vect{r}}[\hrho^{GS}] =\frac{1}{2} \rho^{GS} + \frac{1}{2} \eta_1(r)\eta_4(r+\hat{y}) \rho^{GS} \eta_4(r+\hat{y})\eta_1(r).
\end{align}
This quantum channel can be viewed as a pure measurement channel that measures the fermion bilinears \( i\eta_1(r)\eta_2(r+\hat{x}) \) and \( i\eta_1(r)\eta_4(r+\hat{y}) \) on all plaquettes, averaging over all possible outcomes. The quantum channel in Eq.~\eqref{decohere1}, together with the interaction defined in Eq.~\eqref{2dh}, respects both the strong fermion parity (\( \Z^f_2 \)) and the weak fermion parity along each row and column (\( \Z^{f,x}_2 \) and \( \Z^{f,y}_2 \), respectively). The density matrix obtained after applying the quantum channel exactly matches the reduced density matrix \(\rho^A\) described in Sec.~\ref{sec:reduce}.
Notably, implementing the measurement quantum channels in Eq.~\eqref{decohere1} is equivalent to introducing a unitary circuit that couples the system qubits with ancillas. Following our purification argument in Sec.~\ref{sec:reduce}, we can visualize the system and ancilla qubits as the top and bottom layers in the building block. Coupling them through the interaction proposed in Eq.~\eqref{four2} can be represented by a unitary operator that entangles these layers. When we trace out the ancilla (bottom layer), the unitary operator effectively becomes the quantum channel described in Eq.~\eqref{decohere1}.

\subsection{Generate 3d HOTSC via replica of 2d mSPT}\label{sec:replica}

In Sec.~\ref{sec:dimext}, we introduce a dimensional extension formalism to generate 3d SSPT wave functions (\(\Psi^{sspt}\)) by replicating the 2d mSPT density matrix \(\rho\). In this section, we adapt this protocol to generate the ground state wavefunction for the 3d HOTSC proposed in Sec.~\ref{sec:hotscbulk} using the 2d mSPT density matrix from Eq.~\eqref{decohere1}.

To establish this correspondence, we first examine a fundamental building block of the intrinsic 2d mSPT. Since the mixed-state density matrix is a tensor product of all such building blocks, this approach suffices to analyze the underlying physics.
This building block consists of a plaquette with four Majorana fermions, each originating from different sites located at the corners of the plaquette. The interaction between the four fermions, \(\gamma_1\gamma_2\gamma_3\gamma_4\), transforms each building block into a \textit{spin-1/2 degree of freedom}. After applying decoherence quantum channels, the resulting density matrix for each building block represents the thermal state of each spin-1/2 within the building block.
\begin{align}
    \rho=\frac{1}{2}(|\uparrow\rangle \langle\uparrow|+|\downarrow\rangle \langle\downarrow|)
\end{align}
Under the Choi-double mapping, the thermal matrix resembles a symmetric EPR pair, \( |\rho \rrangle \sim |\uparrow \uparrow \rangle + |\downarrow \downarrow \rangle \). The corresponding SSPT wavefunction along the z-tube is:
\begin{align}
 &|\Psi^{tube}\rangle \sim \bigotimes_i (|\uparrow_R \rangle_i |\uparrow_L \rangle_{i+1}+|\downarrow_R \rangle_i |\downarrow_L\rangle_{i+1} )
\end{align}
When we consider a single building block containing an entangled plaquette in the $x$-$y$ plane, the resulting \(\Psi^{tube}\) state forms a quasi-1d structure along the z-tube, made up of entangled cubes. This state is analogous to the AKLT chain, where each unit cell consists of two spins (denoted L/R), and each $z$-link includes a maximally entangled EPR pair. The strong fermion parity symmetry \(\Z_2^f\) does not affect the spin-1/2, while the two weak subsystem fermion parity symmetries $\Z_2^{f,x}, \Z_2^{f,y}$ (along the x-row and y-row) act as \(\pi\) rotations for \(S_x\) and \(S_y\).
\begin{align}
& Z_2^{f,x}: |v_R \rangle \rightarrow e^{i\pi S_x} |v_R \rangle,~|v_L \rangle \rightarrow e^{-i\pi S_x} |v_L \rangle \nonumber\\
& Z_2^{f,y}: |v_R \rangle \rightarrow e^{i\pi S_y} |v_R \rangle,~|v_L \rangle \rightarrow e^{-i\pi S_y} |v_L \rangle 
\end{align}
Notably, these symmetries act on the L/R components in a Hermitian conjugate manner because the L-component originates from the bra space of the density matrix. Notably,\(\pi\) rotations along the \(S_x\) and \(S_y\) axis renders a projective representation for spin 1/2 systems.

By extending \(\Psi^{tube}\) along the $x$-$y$ plane to encompass all building blocks, the 3d wavefunction \(\Psi^{sspt}\) is formed as a tensor product of all \(\Psi^{tube}\) states. Indeed, \(\Psi^{sspt}\) represents the ground state of the 3d HOTSC proposed in Sec.~\ref{sec:hotscbulk}. Notably, \(\Psi^{tube}\) exhibits a non-vanishing strange correlator along the z-direction, which can be expressed as the twisted R\'enyi-$N$ correlation of the mixed state, as described by the duality formalism in Sec.~ \ref{sec:dimext}:
\begin{align}\label{sc}
&\frac{\langle \Psi^{trivial} | O^{\dagger}_i O_j |\Psi^{tube}\rangle }{\langle \Psi^{trivial}  |\Psi^{tube}\rangle} \nonumber\\
&= \frac{\text{Tr}[\rho M \rho  ( M^{\dagger}_{O})  \rho M\rho M...\rho ( M_{O})\rho M]}{\text{Tr}[\rho M \rho  M  \rho M\rho M...\rho M\rho M]}=1 \nonumber\\
&|\Psi^{trivial}\rangle= \bigotimes_i (|\uparrow_L \rangle_i |\uparrow_R \rangle_{i}+|\downarrow_L \rangle_i |\downarrow_R\rangle_{i})\nonumber\\
&O^{\dagger}=S^+_L+S^-_R,~ M=I, ~M_{O}=\sigma^x
\end{align}
The M-matrix can be chosen as an identity operator, so the trivial state is a product state where the L/R components within the same unit cell are entangled as \(|\uparrow \uparrow \rangle + |\downarrow \downarrow \rangle\). The \(M_O\) matrix is chosen based on the operator \(O\) that is to be measured in the strange correlator. The long-range strange correlation in the dual 3d SSPT state guarantees the long-range order of the twisted R\'enyi-$N$ correlation of the intrinsic mSPT along the `replica direction'.

{\color{black}
In this example, the quasi-one-dimensional wavefunction \(\Psi^{tube}\) naturally admits an MPS representation along the tube. The crucial observation is that the twisted Rényi-\(N\) operator remains nonzero because the density matrix spectrum is degenerate. Specifically, each building block behaves like an effective spin-\(\tfrac{1}{2}\). The weak \(\mathbb{Z}_2^{f,x}\) and \(\mathbb{Z}_2^{f,y}\) symmetries act on this spin-\(\tfrac{1}{2}\) as \(\pi\) rotations along the \(S_x\) and \(S_y\) axes, respectively. As a result, the density matrix transforms projectively under these weak symmetries, giving rise to a twofold degeneracy in its eigenvalues. The corresponding eigenvectors can be chosen to be eigenstates of \(\mathbb{Z}_2^{f,y}\), which are related to each other by the \(\mathbb{Z}_2^{f,x}\) transformation.

In the large-\(N\) limit, \(\rho^N\) is always dominated by its two largest, degenerate eigenvalues, whose corresponding eigenvectors carry distinct \(\mathbb{Z}_2^{f,x}\) charges. When charged operators are inserted on both sides of \(\rho^N\), these degenerate eigenvectors are effectively exchanged under the symmetry action, leaving the density matrix invariant (\(\rho^N = \sigma^x \rho^N \sigma^x\)). As a result, the twisted Rényi-\(N\) operator remains nonvanishing. In this example, the projective representation of \(\rho^N\) under the weak symmetry \(G\) is crucial for generating long-range order (LRO) of the twisted Rényi-\(N\) operator in the replica direction.}

The duality mapping provides a numerical and experimental feasible approach to probe and detect intrinsic mSPT. Detecting non-linear quantities, such as those related to the twisted Rényi-$N$ density matrix, can be challenging in experimental settings. However, by connecting the twisted Rényi-$N$ correlator with the strange correlator in the dual wave function, it's possible to measure the Rényi-$N$ correlator on a quantum processor. Specifically, the SSPT wavefunction \(\Psi^{SSPT}\) can be generated using a shallow quantum circuit, and the strange correlator can be measured through qubit measurements with post-selection on digital quantum devices. This method offers a practical way to probe non-linear quantities in mixed states. We hope our duality formalism will pave the way for detecting unique quantum correlation patterns in non-equilibrium mixed ensembles. We will provide a more detail discussion on experimental probes of
twisted Rényi-$N$ correlator in Sec.~\ref{sec:detecttw}.

\section{Dimension reduction between 3d HOTSC and 2d intrisic mSPT with hinge modes}\label{sec:HOTSC2}

In our previous discussion, we demonstrated an intrinsic mSPT with a protected Majorana corner mode under weak subsystem fermion parity symmetries \(Z_2^{f,x}\) and \(Z_2^{f,y}\), along with strong fermion parity conservation. This mSPT can be understood as the surface state of a 3d HOTSC with three foliated subsystem symmetries on all i-j planes. Given this, a relevant question arises: What happens if we retain only one subsystem fermion parity symmetry, such as \(Z_2^{f,y}\)? In this section, we introduce an intrinsic mSPT protected by weak subsystem fermion parity symmetry \(Z_2^{f,y}\) and strong fermion parity conservation \(Z_2^{f}\). According to the duality formalism introduced in Sec.~\ref{sec:holography}, such a mixed state can arise as the surface density matrix of a 3d HOTSC, protected by two foliated subsystem symmetries on the $x$-$y$ and $y$-$z$ planes.

\subsection{Second Order Topological Superconductor with Chiral Hinge State}\label{sec:3dhinge}

\begin{figure}[h]
    \centering
    \includegraphics[width=0.4\textwidth]{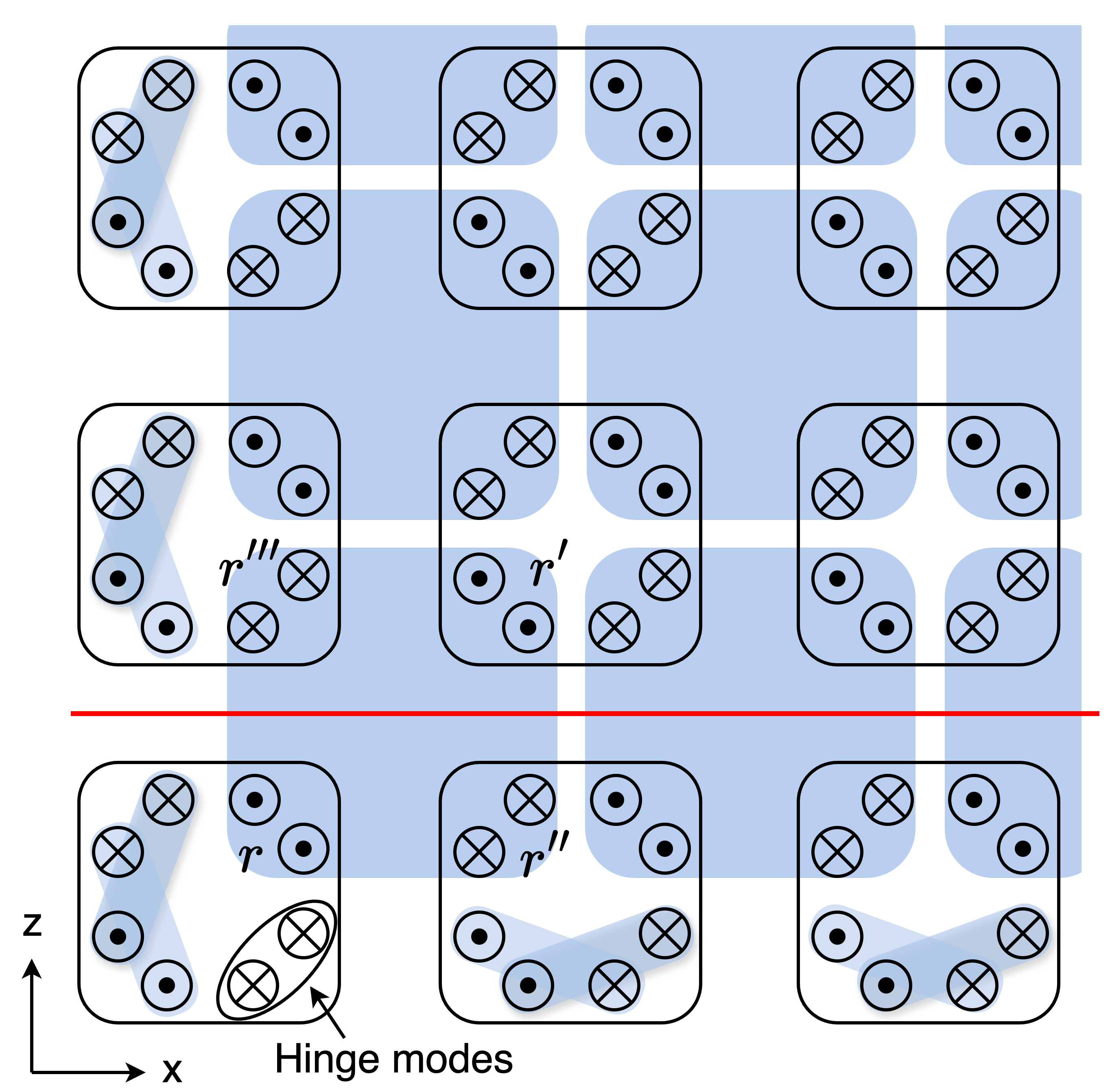}
    \caption{Illustration of 3d 2nd-order topological superconductor.  $\otimes$ and $\odot$ represent 1d Dirac fermions along the $+y$ and $-y$ directions, respectively. The blue plaquettes represent the four-body interactions Eq.~\eqref{eq:LatticeHamiltonianHOTSC} for the bulk fermions, and the blue ellipses represent two-body interactions along the edge. Two dangling chiral Dirac fermions remain at each y-hinge.}
    \label{chiral}
\end{figure}

To set the stage, we first review the 3d 2nd-order topological superconductor (HOTSC) proposed in Ref.~\cite{may2022interaction} with chiral hinge states that are protected by two-foliated subsystem symmetries. The concrete lattice model is built via the coupled wire construction, with a unit cell composed of four spinless $1$d Dirac fermions along the y-direction. These Dirac fermions can be written in terms of eight complex chiral fermions, $\chi^1_L$, $\chi^2_L$, $\chi^3_R$, $\chi^4_R$, $\chi^5_L$, $\chi^6_L$, $\chi^7_R$, $\chi^8_R$, where, as before, $R$ and $L$ indicate that the mode propagates along the $+y$ and $-y$-directions respectively. 

The Hamiltonian can be decomposed into a set of non-overlapping building blocks illustrated as Fig.~\ref{chiral}. Each building block consists of a tube along the y-direction, the interaction within the building blocks are:
\begin{equation}\begin{split}
    &\mathcal{H}_{\text{tube}}=\sum_{\bm{r}} \bm{\chi}^\dagger_{\bm{r}} i\partial_y \tau^{030}\bm{\chi}_{\bm{r}} + \mathcal{H}_{\text{int-SC}},\\
    &\mathcal{H}_{\text{int-SC}} = \sum_{\bm{r}} \Big[V_1 \chi^{1 \dagger}_{ L, \bm{r}}\chi^{2\dagger}_{ L,\bm{r}'}\chi^3_{ R,\bm{r}''}\chi^4_{ R,\bm{r}'''}\\
    &\phantom{=======} + V_2 \chi^{5 \dagger}_{L, \bm{r}}\chi^{6\dagger}_{L,\bm{r}'}\chi^7_{R,\bm{r}''}\chi^8_{R,\bm{r}'''}\\
    &\phantom{=======} + V_3 \chi^{1 \dagger}_{L, \bm{r}}\chi^3_{R,\bm{r}''}\chi^{ 5}_{L, \bm{r}}\chi^{7 \dagger}_{R,\bm{r}''}\\
    &\phantom{=======} + V_4 \chi^{1 \dagger}_{L, \bm{r}}\chi^4_{R,\bm{r}'''}\chi^{ 5 \dagger}_{L, \bm{r}}\psi^{8 }_{R,\bm{r}'''}\Big] + h.c., \\
\label{eq:LatticeHamiltonianHOTSC}\end{split}\end{equation}
where  $\bm{\chi} = (\chi^1_L, \chi^2_L,\chi^3_R,\chi^4_R,\chi^5_L,\chi^6_L,\chi^7_R,\chi^8_R)$, {\color{black}and $\tau^{030}=\mathbbm{1}_{2\times 2}\otimes\tau^z\otimes\mathbbm{1}_{2\times2}$ (here $\mathbbm{1}_{2\times2}$ is a $2\times2$ identity matrix, and $\tau^z$ is the Pauli-$z$ matrix).} The subsystem fermion parity symmetries are given by
\begin{equation}\begin{split}
\mathbb{Z}_2^{yz}: \bm{\chi}_{\bm{r}} \rightarrow e^{i \eta_{yz}(\bm{r}\cdot\hat{x})} \bm{\chi}_{\bm{r}},\\
\mathbb{Z}_2^{xy}: \bm{\chi}_{\bm{r}} \rightarrow e^{i \eta_{xy}(\bm{r}\cdot\hat{z})}\bm{\chi}_{\bm{r}},
\label{eq:SubSystemDef}\end{split}\end{equation}
where $\eta_{yz}$ and $\eta_{xy}$ are functions of $\bm{r}\cdot\hat{x} = n_x$ and $\bm{r}\cdot\hat{z} = n_z$ respectively, and are $\{0,\pi\}$ valued. 

The coupled wire construction consists of building blocks across four rows at $r,r',r'',r'''$ as Fig.~\ref{chiral} that can be treated as the hinge of an elementary tube.
The eight fermions $\chi^{ 1}_{L, \bm{r}}$, $\chi^{2}_{L,\bm{r}'}$, $\chi^3_{ R,\bm{r}''}$, $\chi^4_{R,\bm{r}'''}$, $\chi^{ 5}_{L, \bm{r}}$, $\chi^{6}_{L,\bm{r}'}$, $\chi^7_{R,\bm{r}''}$, and $\chi^8_{R,\bm{r}'''}$ only couple to one another in a bundle for fixed $\bm{r}$. To show that the interactions in Eq. \eqref{eq:LatticeHamiltonianHOTSC} gap out the bulk fermions, we use bosonization by expressing complex fermion modes in terms of vertex operators $\chi^j_{R/L} \sim e^{\mp i \varphi^j_{R/L}}$, where the $\mp$ are correlated to the $R/L$ subscript, and $j=1,\ldots 8.$
In terms of these bosonic fields, the interactions in Eq. \eqref{eq:LatticeHamiltonianHOTSC} become
\begin{equation}\begin{split}\label{inter}
\mathcal{H}_{\text{int-SC}} = &-g_1 \cos(\varphi^1_{L,\bm{r}} + \varphi^2_{L,\bm{r}'} + \varphi^3_{R,\bm{r}''} + \varphi^4_{R,\bm{r}'''})  \\ & -g_2 \cos(\varphi^5_{L,\bm{r}} + \varphi^6_{L,\bm{r}'} + \varphi^7_{R,\bm{r}''} + \varphi^8_{R,\bm{r}'''}) \\ & -g_3 \cos(\varphi^1_{L,\bm{r}} + \varphi^3_{R,\bm{r}''} -\varphi^5_{L,\bm{r}}-\varphi^7_{R,\bm{r}''})  \\ & -g_4 \cos(\varphi^1_{L,\bm{r}} + \varphi^4_{R,\bm{r}'''} + \varphi^5_{L,\bm{r}} + \varphi^8_{R,\bm{r}'''}).
\end{split}\end{equation}
These terms all commute with each other, and hence each wire bundle in the bulk is gapped at strong coupling. 

Having seen that the bulk is gapped we can consider surface boundaries normal to the $z$-directions. On such boundaries there exist gapless fermionic modes with vanishing chirality which can subsequently be gapped by turning on a surface coupling. Finally, on the y-hinges, the fermions $\chi^{ 1}_{L}$, $\chi^3_{ R}$, $\chi^4_{R}$, $\chi^{ 5}_{L}$, $\chi^7_{R}$, and $\chi^8_{R}$ are gapless. Hence, there are 4 chiral Majorana modes with ($c=2$ chiral central charge) at each y-hinge. 

In Ref. \cite{may2022interaction}, the authors demonstrated that a 2d TSC with a 1-foliated subsystem fermion parity symmetry along the $y$-row has a minimal central charge of \(c=4\) at the edge, consisting of eight chiral Majorana modes. Therefore, the surface theory with a chiral central charge of \(c=2\) at each $y$-hinge cannot be realized in 2d system thermal equilibrium.

\subsection{Reduced density matrix of the top layer}

The Hamiltonian in Eq.~\eqref{eq:LatticeHamiltonianHOTSC} defines a ground state wavefunction, \(\Psi^{sspt}\), with zero correlation length, representing a higher-order SSPT (HO-SSPT) state composed of a tensor product of entangled tubes. Each building block forms a highly entangled quasi-1d tube with coupled Majorana wires along the y-axis, and the overall SSPT wavefunction is a product of these tubes.

Consider the wave function \(\Psi^{sspt}\) with open boundaries at \(z=0\). If we make a spatial cut parallel to the \(xy\)-plane, separating the top surface at \(z=0\) (denoted as \(A\)) from the lower regions where \(z<0\) (denoted as \(\bar{A}\)), the reduced density matrix of the top layer, \(\rho^A\), becomes a mixed state ensemble in 2d.
This mixed state \(\rho^A\) manifests an intrinsic mSPT with strong fermion parity symmetry (\(\Z^f_2\)), originating from the subsystem fermion parity conservation in the \(xy\)-plane of \(\Psi^{sspt}\). Additionally, \(\rho^A\) exhibits a weak subsystem symmetry for fermion parity in the \(y\)-columns (\(\Z^{f,y}_2\)), due to subsystem fermion parity conservation in the \(yz\)-planes of \(\Psi^{sspt}\). In this construction, the 3d topological superconductor \(\Psi^{sspt}\) can be visualized as a purification of the mSPT, with short-range correlation in the bulk with anomaly edges.


{

}

\subsection{Prepare 2d mSPT from quenched disorder}

Moving forward, we will demonstrate that the reduced density matrix \(\rho^A\) at the top layer of the 3d HOSPT, as discussed in Sec.~\ref{sec:3dhinge}, can also be prepared in 2d by introducing quenched disorder.

\begin{figure}[h]
    \centering
\includegraphics[width=0.3\textwidth]{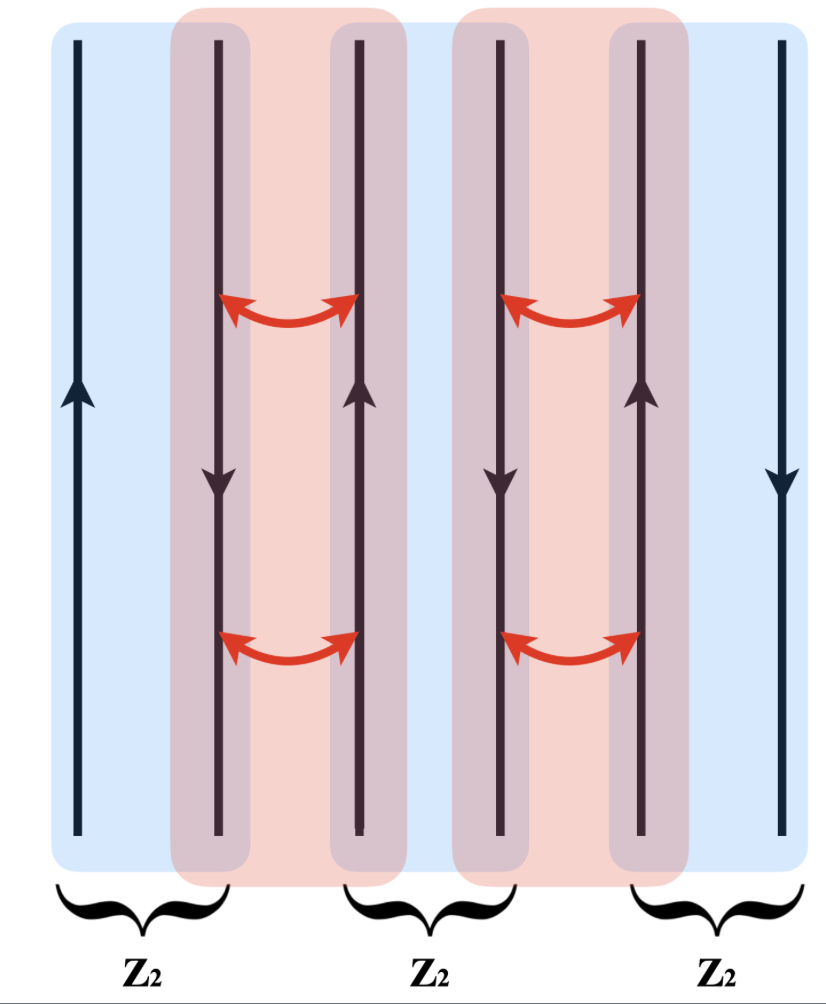}
    \caption{Coupled wire construction of a 2d TSC. Each blue block is a unit cell containing four copies of helical Majorana modes. The fermion parity of each unit cell is individually conserved. The basic building block consists of four left Majoranas at $y_i$ and four right Majoranas at $y_{i+1}$.}
    \label{2dTSC}
\end{figure}

We begin with the 2d coupled wire construction, where each wire at $y_i$ carries four left-moving and four right-moving Majorana modes. The elementary building block consists of four left Majoranas at $y_i$ and four right Majoranas at $y_{i+1}$, illustrated as Fig.~\ref{2dTSC}. Now and after, we will consider interactions and quenched disorders within each building block, and the 2d mixed state is a product of all building blocks.
The Hamiltonian of each building block is:
\begin{align}
    &H=\eta^T(k_y \sigma^{300})\eta \\
    & \mathbb{Z}_2^{f,y}: \eta \rightarrow \sigma^{3} \eta 
    \label{c1}
\end{align}

The subsystem fermion parity symmetry $\mathbb{Z}_2^{f,y}$ acts like a chiral symmetry that dresses the left and right moving majorana with different $\mathbb{Z}_2^{f,y}$ charges.
We can now add a disorder $O(4)$ mass vector $\vec{m}=(m_1,m_2,m_3,m_4)$ to the gapless wire in the building block.
\begin{align}
    &H=\eta^T(k_y \sigma^{300}+m_1 \sigma^{222}+m_2 \sigma^{102}+m_3 \sigma^{230}+m_4 \sigma^{210})\eta
    \label{dmass}
\end{align}
The $O(4)$ vector mass $\vec{m}$ is odd under subsystem fermion parity $\mathbb{Z}_2^{f,y}$, leading each mass pattern to break the conservation of subsystem fermion parity. However, if we consider a mixed ensemble $\rho$ as an incoherent sum of all possible quenched disorder mass patterns for $\vec{m}=(m_1, m_2, m_3, m_4)$, the resulting density matrix respects the weak subsystem fermion parity symmetry and exhibits short-range correlations. 

Before considering the disorder ensemble, we first treat O$(4)$ as a dynamical mass that fluctuates in space-time. Integrating out the fermion in the building block leads to an effective theory of the O$(4)$ rotor, described by a non-linear sigma model with a Wess-Zumino-Witten term:
\begin{equation}
\mathcal{L}_\uparrow=\frac{1}{g}(\partial_{\mu} \vec{m})^2+\frac{2\pi}{\Omega^3} \int_0^1 du\epsilon^{ijkl}  m_{i}\partial_y m_{j} \partial_t m_{k}\partial_u m_{l},
\label{eq:NLSM}\end{equation}
In this case, the fermionic excitation in the building block wires is gapped, and the gapless degree of freedom is purely bosonic, arising from the WZW term fluctuation of the O$(4)$ rotor.

Now, let us introduce quenched disorder into each building block by incorporating a randomly disordered mass vector \(\vec{m}(r,t)\) in Eq.~\eqref{dmass}. While any specific disorder mass configuration explicitly breaks subsystem fermion parity, the mixed ensemble of all possible disorder configurations results in a density matrix \(\rho\) that retains a weak \(\mathbb{Z}_2^{f,y}\) symmetry. The mixed-state density matrix for each building block can be written as: 
\begin{align}\label{mixed0}
    \rho \sim \sum_{\{\vec{m} \}} |\vec{m}\rangle \langle \vec{m} |
\end{align}
For each specific pattern \(\vec{m}\), the ket vector \(|\vec{m}\rangle\) represents the ground state of the building block in Eq.~\eqref{dmass}, where Majoranas couple to the \textit{static vector mass} \(\vec{m}\), resulting in a gapped, short-range correlated state. The density matrix in Eq.~\eqref{mixed0} is a convex sum of 1D gapped fermions in each building block with a disordered vector mass \(\vec{m} = (m_1, m_2, m_3, m_4)\). While the state \(|\vec{m}\rangle\) breaks the \(\mathbb{Z}_2^{f,y}\) symmetry, the incoherent sum over all possible \(|\vec{m}\rangle\) patterns retains a weak \(\mathbb{Z}_2^{f,y}\) symmetry. The WZW term in Eq.~\eqref{eq:NLSM} vanishes in the mixed state \(\rho\), as its effect is canceled by the opposing Berry phases from the bra and ket spaces. 

The 2d mixed state in Eq.~\eqref{mixed0} can also be obtained by adding quantum channels.
Following our duality argument in Sec.~\ref{sec:3dhinge}, we can envision the system and ancilla qubits in each building block tube as representing the top and bottom layers of the tube in Eq.~\eqref{eq:LatticeHamiltonianHOTSC}. The ancilla degree of freedom atop the building block adds another set of four helical modes, producing an additional $O(4)$ field, $\vec{m'}$, with an opposing WZW term \cite{you2014symmetry, xu2013wave, you2015bridging, you2016stripe}:
\begin{equation}
\mathcal{L}=\frac{1}{g}(\partial_{\mu} \vec{m}')^2-\frac{2\pi}{\Omega^3} \int_0^1 du\epsilon^{ijkl}  m'_{i}\partial_z m'_{j} \partial_t m'_{k}\partial_u m'_{l}\
\end{equation}
Notably, the $O(4)$ vector $\vec{m}'$ in the ancilla space undergoes the same transformation as $\vec{m}$ under the $\mathbb{Z}_2^{f,y}$ symmetry. 

In Eq.~\eqref{eq:LatticeHamiltonianHOTSC}, we coupled the system with the ancilla in each building block tube as follows:
\begin{equation}
H_{int}=-V \vec{m}' \cdot \vec{m} 
\end{equation}
When the coupling is strong, it forces the two $O(4)$ vectors, $\vec{m}$ and $\vec{m}'$, in the enlarged Hilbert space to align parallel, leading to the exact cancellation of the Wess-Zumino-Witten (WZW) term. The resulting theory simplifies to a trivial nonlinear sigma model (NLSM), with its fixed-point ground state described as follows:
\begin{align}\label{purified}
    |\Phi\rangle\ \sim \sum_{\{\vec{m} \}} |\vec{m}\rangle_s |\vec{m}\rangle_a
\end{align}
Here, \( |...\rangle_{s/a} \) represents the system of interest or the ancilla from the environment, while \(\sum_{\{\vec{m} \}}\) accounts for all possible disorder patterns. If the ancilla is traced out from Eq.~\eqref{purified}, the resulting reduced density matrix \( \rho \) resembles the mixed-state ensemble in Eq.~\eqref{mixed0}. Importantly, when the system and ancilla are entangled, their Schmidt decomposition can be expressed using the eigenvectors of \( \vec{m} \). Consequently, the reduced density matrix becomes an incoherent sum over all \( \vec{m} \) patterns, equivalent to introducing quenched disorder or quantum channels that average over all possible \( \vec{m} \) patterns.

\color{black}
One can detect the mSPT in Eq.~\eqref{mixed0} by looking into the twisted R\'enyi-$N$ operator that is charged under $\mathbb{Z}_2^{f,y}$ symmetry, which can be chosen to be $m_1(x,y)=\eta^T \sigma^{222}\eta$ defined in Eq.~\eqref{dmass};
\begin{align}\label{cn}
&C(n)= \frac{\text{Tr}[ \rho^n m_1(x,y) \rho^n m_1(x,y)]}{\text{Tr}[\rho^{2n}]}
\end{align}
The \(m\) operator acts at the same spatial point but is separated along the replica direction. Note that \(\eta^T \sigma^{222}\eta\) corresponds to the \(m_1\) component of the vector mass \(\vec{m}\). Since the density matrix of the mSPT in Eq.~\eqref{mixed0} is an incoherent sum over all possible patterns of \(\vec{m}\), Eq.~\eqref{cn} remains nonvanishing in the large-\(N\) limit. 


\section{Twisted R\'enyi-$N$ operator as a fingerprint for intrinsic mSPT}\label{sec:twisted}
{\color{black}

In this section, we provide a detailed discussion on how to detect and distinguish intrinsic mSPTs from conventional mSPTs via the twisted Rényi-\(N\) operator, with the key result summarized in Table I. As shown in Ref.~\cite{ma2024topological}, an intrinsic mSPT can be obtained by decohering an intrinsic gapless SPT protected by the symmetries \(\mathcal{G}\) and \(\mathcal{S}\). Notably, in the gapless IR theory, the action of \(\mathcal{G}\) is anomalous.
In open-system settings, while \(\mathcal{S}\) remains a strong symmetry, \(\mathcal{G}\) is reduced to a weak symmetry. As a result, the \(\mathcal{G}\)-anomaly becomes diminished in the mixed-state density matrix, and the resulting mixed state is short-range entangled.
From a purification perspective, the decohering quantum channel can be viewed as coupling the intrinsic gapless SPT to another intrinsic gapless SPT in the ancilla degrees of freedom, thereby canceling the \(\mathcal{G}\)-anomaly. This mechanism is analogous to the coupled-wire construction of higher-order SSPTs (HO-SSPTs), which exhibit protected hinge or corner modes.

\begin{table*}\label{table1}
\renewcommand\arraystretch{1.4}
\begin{tabular}{| l | l | l |}
\hline
Properties & intrinsic mSPT & conventional mSPT \\ \hline
fidelity strange correlator & (quasi) long-ranged & (quasi) long-ranged \\ \hline
twisted R\'enyi-$N$ correlator & (quasi) long-ranged &  short-ranged  \\ \hline
holographic dual & higher-order SSPT & weak SSPT (layer stacking) \\ \hline
strange correlator along z-axis & (quasi) long-ranged  & short-ranged\\ \hline
\end{tabular}
\caption{Summary Table: Differentiating Intrinsic mSPTs from Conventional mSPTs}
\end{table*}

Based on this framework, we expect the \textit{intrinsic mSPT} to exhibit the following properties:

i)\textit{ Non-vanishing fidelity strange correlator under the strong symmetry $\mathcal{S}$:}

   There exists an operator $O_s$ charged under the strong symmetry that displays a long-ranged (or power-law decay) fidelity strange correlator defined in Sec.~\ref{sec:strangecor}:
\begin{align}\label{eq:fidsc}
C_F(r,r^{\prime})=\frac{F\left(\E[\rho], O_s(r) O^\dag_s(r^{\prime}) \E[\rho_0] O_s(r^{\prime}) O^\dag_s(r)\right)}{F\left(\E[\rho], \E[\rho_0]\right)},
\end{align}

ii)\textit{ Non-vanishing twisted R\'enyi-$N$ correlator under the weak symmetry $\mathcal{G}$:}

   There exists an operator $O_g$ charged under the weak symmetry that exhibits a long-ranged (or power-law decay) twisted R\'enyi-$N$ correlator:
\begin{align}\label{eq:twry}
&C(n)= \frac{\text{Tr}[ \rho^n O^{\dagger}_g(r) \rho^n O_g(r)]}{\text{Tr}[\rho^{2n}]}
\end{align}

These two properties serve as key indicators for identifying and characterizing the presence of an intrisic mSPT phase. Now we elucidate these properties in detail, from the holographic duality picture. We emphasis the properties we describe above should be true for intrinsic mSPT for any dimensions. In the following, we will use the example of 3d.

As discussed in Sec.~\ref{sec:strangecor}, the fidelity strange correlator for a 2d mixed-state SPT can be traced back to the strange correlator acting on the top surface of the 3d SSPT wavefunction. In particular, as shown in Sec.~\ref{sec:strangecor}, measuring the strange correlator of a 3d SSPT can be viewed as projecting all qubits on the top surface (except those at sites \(r\) and \(r'\)) onto a trivial symmetric state. The resulting post-projection state then exhibits long-range order between the qubits at \(r\) and \(r'\) that are spatially separated in the xy-plane of the top surface. Once the lower layers of the 3d SSPT are traced out, the reduced density matrix of the top surface is effectively an mSPT. In this scenario, the subsystem symmetry of the original 3d SSPT becomes a strong symmetry of the resulting mixed state, and the strange correlator of the dual 3d SSPT provides a lower bound on the fidelity strange correlator of that mixed state. Notably, a related long-range strange correlator for mixed states was proposed in Refs.~\cite{zhang2022strange, LeeYouXu2022}.

While the fidelity strange correlator reveals the presence of an mSPT phase, it does not by itself confirm that the mSPT is \textit{intrinsic}. To establish the \textit{intrinsic} nature of an mSPT, one must additionally examine the twisted R\'enyi-\(N\) operator in (ii).}

{\color{black}
Physically, intrinsic mixed-state symmetry-protected topological phases (mSPTs) can often be obtained by decohering an intrinsic gapless SPT state that carries an emergent anomaly associated with symmetry \(\mathcal{G}\). Upon decoherence, the system transitions into a mixed state. The corresponding Choi-double state \(\lvert \rho \rrangle\) exhibits a cancellation of this anomaly due to the decoherence-induced coupling between the two layers of the anomalous gapless \(\mathcal{G}\)-states. This coupling effectively neutralizes the individual anomalies, resulting in a short-range-entangled (SRE) bilayer system.
Within the framework of holographic duality employed here, the dual wavefunction of the symmetry-protected topological phase (SSPT) emerges through a replication of the density matrix. This is achieved by stacking the coupled bilayer systems, which are referred to as `building blocks' in Sec.~\ref{sec:dimext}. This construction mirrors the methodology used in Ref.~\cite{zhang2023classification} for higher-order SSPTs (HO-SSPTs), where anomalous hinge or corner modes are generated by coupling lower-dimensional layers that individually carry anomalies.}

{\color{black}
Surprisingly, this stacking of building blocks construction suggests that the system hosts a 3d SPT phase protected by the \(\mathcal{G}\) symmetry. One can argue this from a field theory perspective. Assuming the system is bosonic, the relevant \(2d\) \(\mathcal{G}\)-anomalous gapless states used to construct the intrinsic mSPT can be described by certain non-linear sigma model with WZW terms. In each unit cell of the replicated theory, we have a pair of WZW theories which carries opposite anomaly. Decoherence that breaks the strong symmetry to a weak symmetry has the effect to couple these WZW theories between neighboring unit cells (which forms a building block). This will induce a topological \(\theta\)-term in \(3d\), protected by the \(\mathcal{G}\)-symmetry, indicating an SPT. Thus, when one chooses an operator \(O_g\) carrying \(\mathcal{G}\)-charge, its strange correlator should be nontrivial (long range or power-law) along the layered direction. In our holographic dual description, this long-range-ordered strange correlator corresponds precisely to the twisted R\'enyi-\(N\) operator and manifests as nontrivial correlation in the replica direction.}

It is worth mentioning that in the twisted R\'enyi-\(N\) operator, \(O_g\) is inserted at the same spatial point in \(2d\), separated only along the replica direction. Hence, it differs from the fidelity strange correlator or type-II strange correlator, which involves pairs of \textit{correlators} acting on both the ket and bra spaces of the density matrix, as Fig.~\ref{overview}-d. In contrast, the twisted R\'enyi-\(N\) operator includes only a pair of \textit{charged operators} acting on the ket and bra spaces of \(\rho^N\) (at the same spatial point), and examines how this correlation survives in the large-\(N\) limit. Because \(O_g\) carries \(\mathcal{G}\)-charge, it detects long-range order associated with the weak \(\mathcal{G}\) symmetry along the `replica direction'. In the holographic dual perspective, this weak \(\mathcal{G}\) symmetry becomes a global \(\mathcal{G}\) in \(3d\), with the replica direction identified as the \(z\)-axis (see Fig.~\ref{overview}-e). Consequently, the twisted R\'enyi-\(N\) operator corresponds to a strange correlator along the \(z\)-axis.

Such long-range order (LRO) in the twisted R\'enyi-\(N\) correlator is a fingerprint of intrinsic mSPTs, arising when the mSPT is obtained by decohering a gapless SPT wavefunction whose gapless spectrum exhibits an emergent \(\mathcal{G}\)-anomaly. In contrast, if one decoheres a gapped SPT wavefunction to form a conventional mSPT, the fidelity strange correlator (Eq.~\eqref{eq:fidsc}) discussed in (i) still maintains LRO, whereas the twisted R\'enyi-\(N\) correlator (Eq.~\eqref{eq:twry}) in (ii) vanishes in the large-\(N\) limit. This distinction is also evident from the holographic dual perspective: a conventional mSPT corresponds to an SSPT that is simply a stack of decoupled 2d SPT layers,\footnote{Such a stacking of 2d SPT layers is referred to as a weak SSPT in some literature.} which does not display strange correlations along the \(z\)-direction.

This result suggests that, under our holographic duality, a conventional mSPT corresponds to a weak SSPT \cite{devakul2018classification,devakul2020strong,tantivasadakarn2020searching} that is equivalent (up to local unitary transformations) to a trivial stacking of lower-dimensional SPT layers with \(\mathcal{S}\otimes \mathcal{G}\) symmetry along the \(z\)-axis. Here, the \(\mathcal{S}\) symmetry, conserved in each \(xy\)-plane, acts as a subsystem symmetry, while a layer-coupling term can exchange \(\mathcal{G}\) charge. As a result, although non-vanishing strange correlations persist along the \(xy\)-plane, the strange correlations along the \(z\)-axis remain short-ranged. In contrast, an intrinsic mSPT maps to a higher-order SSPT (HO-SSPT) with long-range strange correlators on both the \(xy\)-plane and the \(z\)-axis. In the mSPT language, these correspond to the fidelity strange correlator and the twisted Rényi-\(N\) operator, respectively.

{\color{black}
Within the framework of measurement-based quantum computation (MBQC), the nontrivial twisted Rényi-N correlator and its dual—the strange correlator—arise from the fact that the SSPT wavefunction serves as a nontrivial resource state. In this context, measuring qubits in the bulk of the SSPT state generates entanglement between the two boundaries along the $z$-direction. This directly translates to a nonzero strange correlator or twisted Rényi-N correlator, as overlapping with a trivial wavefunction is equivalent to performing a projective measurement in the bulk. By conducting measurements along the $z$-direction, quantum information can be effectively transmitted from the top to the bottom surface of the SSPT state.}

However, it is important to note that the presence of a long-range twisted Rényi-\(N\) correlator (as described in point (ii)) alone is insufficient to confirm the existence of an intrinsic mSPT phase. This is because such a correlator only involves weak symmetry. Therefore, to unambiguously identify an mSPT phase, both the twisted Rényi-\(N\) correlator (associated with the weak symmetry) and the fidelity strange correlator (associated with the strong symmetry, as described in point (i)) must be considered together. Only the combination of these two correlators provides a definitive signature of the intrinsic mSPT phase.


{
\color{black}
\section{Implication of Holographic duality}
So far we established a holographic duality between intrinsic mSPT states and SSPT pure states in one higher dimension. 
In this section, we aim to address the implication of our holographic duality on state preparation and experimental probes of intrinsic mSPT phases on quantum devices. As we will show, our holographic duality enables us to establish two effective methods to prepare intrinsic mSPT phases through finite depth quantum channel (FDQC) and provide an experimental probe for the mixed-state anomaly.

\subsection{Holographic preparation of intrinsic mSPTs}
The quantum state preparation is one of the central questions in quantum resource theory \cite{QRT_RMP}. Here we consider the efficient preparation of mSPT states in a quantum device. 

By definition, an mSPT state is short-range entangled, which means in principle it is preparable from a trivial product state by a finite-depth local quantum channel $\E$, namely
\begin{align}
\rho=\E[\ket{0}\bra{0}]=\Tr_{\mathcal{A}}\left[U|0\rangle\langle0|\otimes|0\rangle\langle0|_\mathcal{A}U^\dag\right],
\end{align}
where $U$ is a local unitary circuit in an enlarged Hilbert space $\mathcal{H}\otimes\mathcal{A}$ composed by the physical Hilbert space and an ancilla Hilbert space. For an mSPT state with a pure-state analog, finding such quantum channel $\E$ is straightforward: we first prepare the corresponding pure-state SPT by a finite-depth local unitary circuit, then decohere a part of degrees of freedom. The combination is a finite-depth local quantum channel. However, the state preparation of intrinsic mSPT states is much more challenging: as introduced in Ref. \cite{ma2024topological}, an intrinsic mSPT state can be purified to an intrinsic gapless SPT in the same dimension, which in general requires a linear-depth quantum circuit to prepare. 

In this subsection, we will emphasize that the holographic duality between mSPT and SSPT states serves as an efficient and systematic preparation paradigm for intrinsic mSPT states. We call it the \textit{holographic preparation} of intrinsic mSPT states. In particular, we note that the holographic preparation of intrinsic mSPT states can be done with a finite-depth local quantum channel, which is much more efficient in comparison with the route through intrinsic gapless SPT \cite{gaplessSPT} in the same dimension. 

It was proven in Ref. \cite{cohomology} that an arbitrary SRE pure state $\ket{\psi}$ is equivalent to a product state $\ket{0}$ via a quasi-adiabatic evolution $U$ \cite{hastings2005quasiadiabatic}, namely
\begin{align}
\ket{\psi}=U\ket{0}.
\end{align}
In particular, if $\ket{\psi}$ is an SPT wavefunction, then we call $U$ an ``entangler''. An SPT entangler is usually realized by a quantum cellular automaton (QCA), which is a locality-preserving unitary operator. i.e., for any operator $O$ supported in a ball of finite radius $r$, the conjugated operator $UOU^\dag$ is supported in a ball of finite radius $r+d$ for some constant $d$. 

Reconsider the 1D $\Z_4$ intrinsic mSPT example in Sec. \ref{sec:decoheredvsgapless}, which is dual to a 2D SSPT state $\ket{\Psi^{SSPT}}$ with a subsystem $\Z_2$ and a global $\Z_4$ symmetries. $\ket{\Psi^{SSPT}}$ can be prepared by a QCA $U_S$ from a product state, namely
\begin{align}
\ket{\Psi^{SSPT}}=U_S\ket{0}.
\end{align}
Then if we treat the bottom layer of the SSPT state is defined in the physical Hilbert space $\mathcal{H}$, and all other layers are defined in the ancilla Hilbert space $\mathcal{A}$, we can easily prepare the 1D $\Z_4$ intrinsic mSPT density matrix $\rho_{\Z_4}$, namely
\begin{align}
\rho_{\Z_4}=\Tr_{\mathcal{A}}\left[\ket{\Psi^{SSPT}}\bra{\Psi^{SSPT}}\right].
\end{align}
Therefore, we conclude that our holographic duality serves as an efficient and systematic paradigm of finite-depth preparation of intrinsic mSPT states. 

We note that to prepare an SSPT state, the entangler $U_S$ must locally break the symmetry. If the quantum device can only perform the symmetric finite-depth quantum circuits (FDQC), i.e., each local gate should be symmetric, we should design some other strategy to prepare the parent SSPT state of the desired intrinsic mSPT state. In Ref. \cite{stephen2024manybody}, the authors proposed a paradigm of symmetric preparation of SPT states by introducing a \textit{many-body quantum catalyst}. The basic idea of this proposal is that if a quantum state $\rho_a$ (either pure or mixed state) is invariant under the entangler $U_S$, then we can find a symmetric FDQC $\mathcal{V}$ to prepare the SPT state effectively, namely
\begin{align}
|0\rangle\langle0|\otimes|0\rangle\langle0|_{\mathcal{A}}\xrightarrow{\mathbbm{1}\otimes \E_a}|0\rangle\langle0|\otimes\rho_a\xrightarrow{\mathcal{V}}|\mathrm{SPT}\rangle\langle\mathrm{SPT}|\otimes\rho_a,
\label{Eq: catalyst}
\end{align}
where $\E_a$ is the efficient preparation channel of $\rho_a$: $\E_a[|0\rangle\langle0|]=\rho_a$. We note that because of the invariance of $\rho_a$ under $\mathcal{V}$, we can use the catalyst multiple times to efficiently prepare our desired state once we have it.

If $\rho_a$ is a pure state, then $U_a$ should be a unitary circuit. Nevertheless, it has been proven in Ref. \cite{stephen2024manybody} that a pure-state catalyst should carry the quantum anomaly of the corresponding symmetry class of the desired SPT. A legitimate pure-state quantum catalyst could be an SSB state, a quantum critical state, or a topological order. All of them can only be prepared by a deep quantum circuit. Therefore, although a catalyst can be repeatedly used infinite times, the preparation of the pure-state catalyst itself is still costly.

If $\rho_a$ is a mixed state, it has been proven that if the desired SPT state has a $G$ symmetry, then $\rho_a$ can be a mixed state with strong-to-weak spontaneous symmetry breaking (SWSSB) of $G$. We note that an SWSSB state can easily be prepared from a product state by a depth-1 local quantum channel \cite{lessa2024strong}. Therefore, the mixed-state catalyst has the privilege that the preparation is very straightforward and a catalyst can be repeatedly used by infinite times. 

With the help of mixed-state catalyst $\rho_a$, the holographic paradigm of preparing an intrinsic mSPT state is summarized as:
\begin{enumerate}[1.]
\item Consider two copies of a product state with a subsystem symmetry $K$ and a global symmetry $G$, $|0\rangle\langle0|\otimes|0\rangle\langle0|_{\mathcal{A}}$.
\item Prepare an SWSSB $\rho_a$ of $K$ and $G$ from $|0\rangle\langle0|_{\mathcal{A}}$ by a depth-1 local quantum channel. 
\item Apply a symmetric FDQC $\mathcal{V}$ to the combined state $|0\rangle\langle0|\otimes\rho_a$, to get the desired parent SSPT state as $|\Psi^{SSPT}\rangle\langle\Psi^{SSPT}|\otimes\rho_a$.
\item Trace out all layers but one. For the remaining layer, the previous subsystem symmetry $K$ becomes a strong symmetry and the global symmetry $G$ becomes a weak symmetry. 
\end{enumerate}
We note that all these procedures could be done in a quantum device with finite depth circuits. As a consequence, our holographic duality between intrinsic mSPT and SSPT states serves as an efficient \textit{holographic preparation} of the intrinsic mSPT states.

\subsection{Holographic duality between sequential quantum circuit and quantum channels}
\label{sec:sequential}

In this section, we briefly offer an outlook on how our holographic duality can shed light on the steady state of infinite-depth quantum channels. In particular, we show that if a \((d+1)\)-dimensional SSPT can be generated by a symmetric sequential circuit, then the corresponding \(d\)-dimensional mSPT can likewise be prepared as a steady state under repeated quantum channels.

Ref.~\cite{gopalakrishnan2023push} proposed that infinite-depth quantum channels in $d$ dimensions can be interpreted as sequential circuits in $d+1$ dimensions. In a sequential circuit\cite{chen2023sequential,chen2024sequential}, one applies a local unitary gate at each step that entangles the \(i\)-th qubits with the \((i+1)\)-th qubits, repeating this procedure step by step. As a result, the wavefunction produced by this process adheres to an area law of entanglement. Notably, such sequential circuits can generate nontrivial symmetry-protected topological phases while preserving the relevant symmetry at every unitary step.

\begin{figure*}
    \centering
\includegraphics[width=0.9\textwidth]{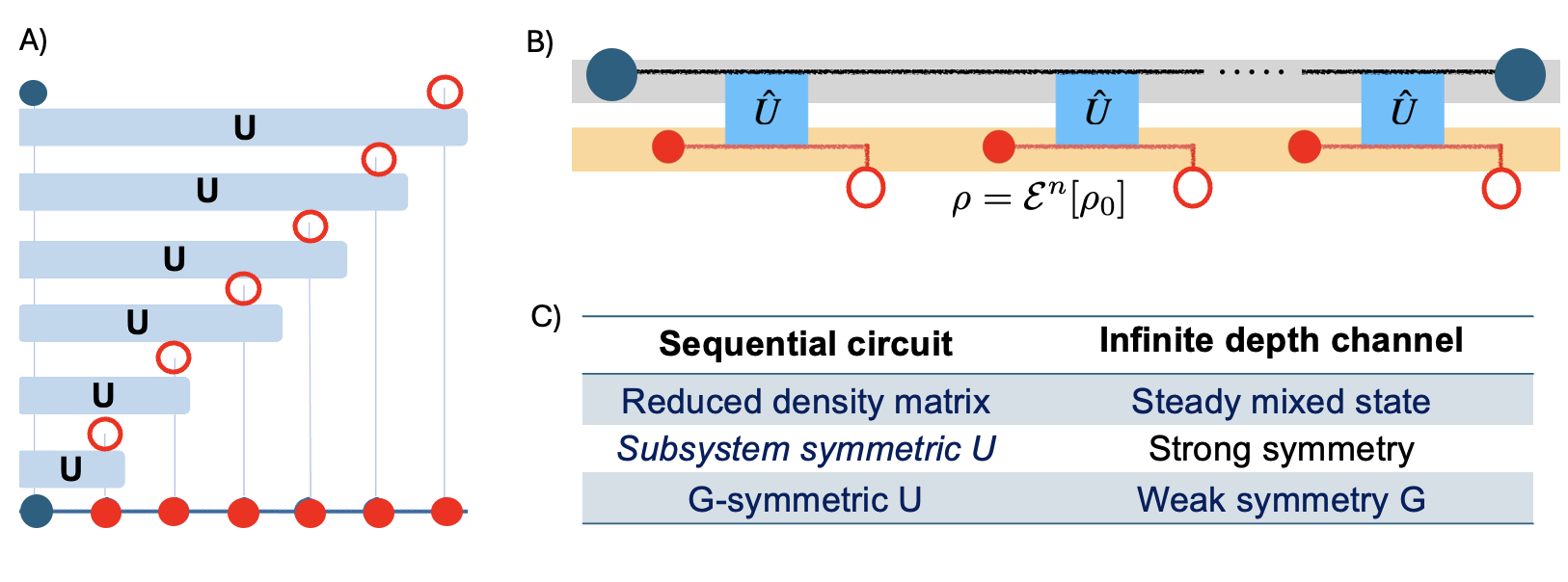}
    \caption{ A) The sequential circuit includes an additional SWAP gate at each step. The 1st qubit is entangled with the \((i+1)\)-th qubit during each step.
B) Dual to the quantum channels based on Ref.~\cite{gopalakrishnan2023push}, the blue dots represent the system qubits, while the red dots represent the ancilla qubits that become entangled with the system under a unitary gate at each step. The solid red dots indicate the input state, while the hollow red dots indicate the output state.
C) Summary table of this duality.}
    \label{uni3}
\end{figure*}

In the context of sequential circuit, each `qubits' can represent a single site or, more generally, the Hilbert space of a \(d\)-dimensional system. By introducing a SWAP gate into the sequential circuit, the circuit can be alternatively decomposed into the form \(U = \prod_i U_{1,i}\). In this decomposition, each unitary step \(U_{1,i}\) effectively entangles the first qubit with the \(i\)-th qubit at every stage of the process, as illustrated in Fig.~\ref{uni3}. 
If the first qubit is considered the system of interest, the \(i\)-th qubits are treated as ancillae for each step. In each unitary step, entangling the first qubit with the \(i\)-th qubit effectively feeds a quantum channel to the first qubit. By applying the sequential unitary \(U_{1,i}\) step by step, we are repeatedly applying a quantum channel to the first qubit.
If we assume that the information of all qubits is lost except for the first one, the sequential circuit essentially transforms into an infinite-depth quantum channel. In this scenario, the reduced density matrix of the first qubit represents the steady state of this channel.

Thus, our holographic duality offers a novel perspective on understanding $d$-dimensional quantum channels through the lens of $d+1$-dimensional wavefunctions, generated by sequential circuits.
On the one hand, this framework provides valuable insights into exploring exotic steady states of quantum channels \cite{chirame2024stable,sohal2024noisy} that exhibit mixed-state topological order or mixed-state symmetry-protected topological (mSPT) phases. These states can be studied by examining their dual 3d wavefunctions generated by sequential circuits. Since the steady mixed states are analogous to the reduced density matrix of a sequentially generated wavefunction, we anticipate that key properties of mixed states—including nonlinear quantities such as entanglement negativity and coherent information—can be mapped to specific features of the corresponding sequentially generated wavefunction.

Conversely, instead of seeking novel steady mixed states by iterating quantum channels, the problem can be approached by analyzing the entanglement properties of a sequentially generated wavefunction. Notably, strong (weak) symmetries in quantum channels are dual to subsystem (global) symmetries in sequential circuits. As such, exploring strongly symmetric quantum channels translates to investigating subsystem-symmetric quantum circuits. Similarly, the higher-order subsystem symmetry-protected topological (HO-SSPT) wavefunctions discussed here can, in principle, be generated via subsystem-symmetric sequential circuits. Once the unitary gates \( U_{1,i} \) for each step are identified, the corresponding quantum channels that generate the dual mSPT phases can be determined.

\subsubsection{Detection of Twisted R\'enyi-$N$ correlator from measurement-feedback protocol of sequential circuit}\label{sec:detecttw}

As described in Sec.~\ref{sec:twisted}, the twisted Rényi-\(N\) operator is mapped to the strange correlator of the dual SSPT state via a holographic mapping. Consequently, measuring the twisted Rényi-\(N\) operator is equivalent to detecting the strange correlator of the corresponding higher-dim SSPT state. In this section, we propose a feasible experimental scheme to probe the twisted Rényi-\(N\) correlator by instead measuring its dual strange correlator, using a combination of a measurement-feedback protocol and a sequential circuit. Both of these components can be realized on programmable quantum simulator platforms.

\begin{figure}[h]
    \centering
\includegraphics[width=0.48\textwidth]{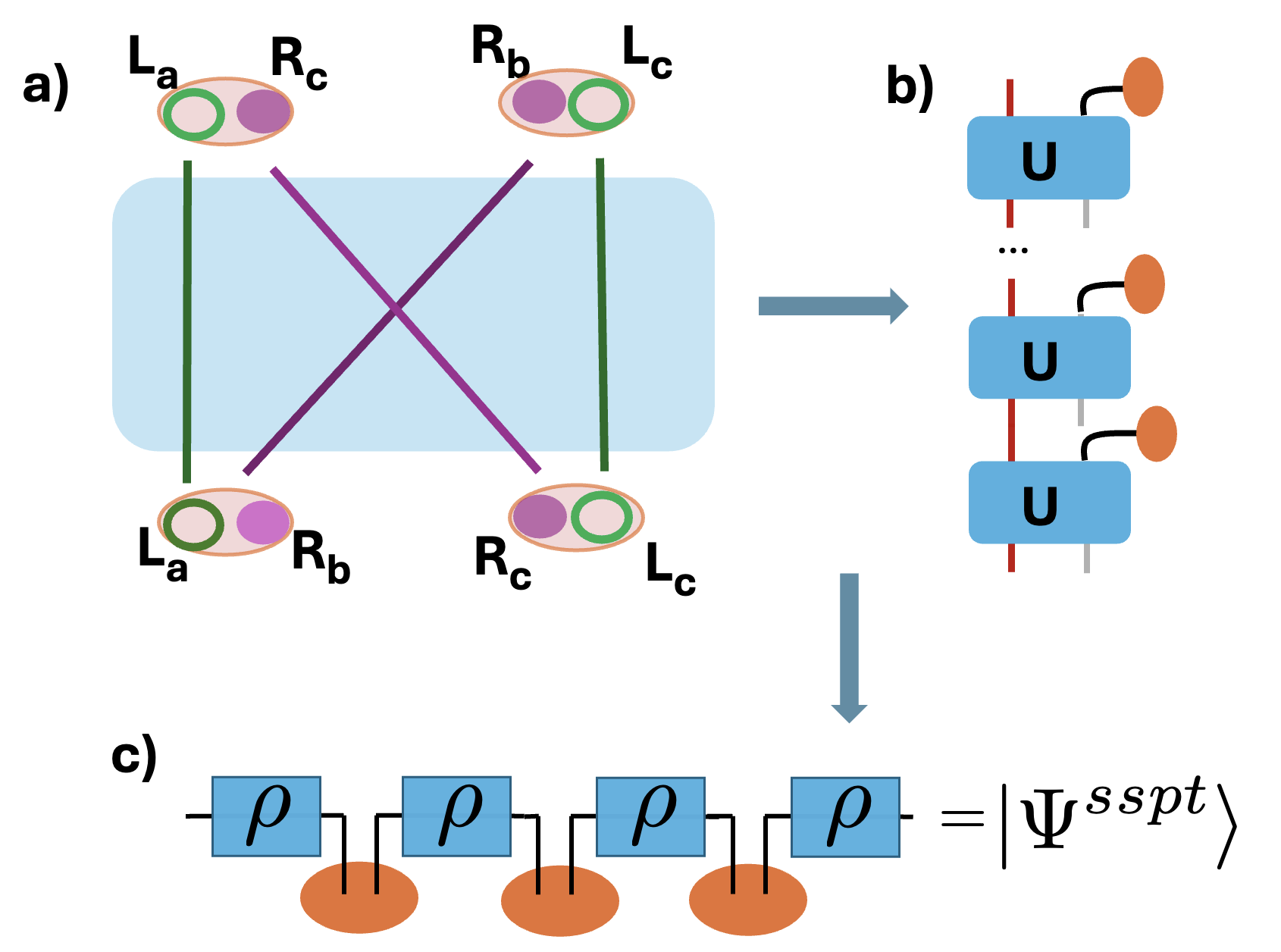}
    \caption{a) The green and purple dots in each oval  represent the left and right components of each row. The unitary gate is a partial SWAP gate that exchanges only the right component between adjacent rows.  
b) Repeating this partial SWAP gate generates the 3d SSPT state. The orange ovals represent output qubits. 
c) The sequentially generated wavefunction corresponds to the \(\Psi^{\text{SSPT}}\) obtained from the holographic mapping.}
    \label{seqmf}
\end{figure} 

To set the stage, we first describe a general scheme for realizing the SSPT wavefunction discussed in this work using a sequential circuit that employs a partial SWAP (PSWAP) gate, as depicted in Fig.~\ref{seqmf}. Specifically, suppose we begin with a trivial state composed of an on-site entangled state:
\[
|\phi_0\rangle \;\sim\; \prod_i \Bigl(\sum_a \lambda_a\, |L_a\rangle_i \,|R_a\rangle_i\Bigr).
\]
Next, we apply the partial SWAP gate step by step between the first row and the \(i\)th row, thus creating entanglement between them:
\[
\prod_{i} \mathrm{PSWAP}_{1,i}\,|\phi_0\rangle 
\;\sim\; \prod_{i} \Bigl(\sum_a \lambda_a\, |R_a\rangle_i \,|L_a\rangle_{i+1}\Bigr),
\]
This sequential unitary engenders the SSPT wavefunction obtained in Sec.~\ref{sec:holography} through replication of the mixed-state density matrix.

Measuring the strange correlator involves taking the inner product between the SSPT state and a trivial on-site EPR state. This requires performing measurements in the EPR basis on all sites. Without post-selection, the measurement outcomes are generated randomly.
To address the randomness of the measurement outcomes, we introduce a protocol for measuring the strange correlator that combines sequential unitary operations with measurement-based feedback at each step. Specifically, our procedure proceeds as follows: 

1. Apply \(\text{PSWAP}_{1,i}\) between the first row and the \(i\)-th row.  

2. Measure the \(i\)-th row qubits in the onsite EPR basis.  

3. Based on the measurement outcomes, apply a local unitary operator \(Q\) (on the first row) before proceeding to the next step.

For instance, if the measurement outcome on the \(i\)-th row corresponds to the symmetric EPR state \(\sum_a |L_a\rangle_i\,|R_a\rangle_i\), then, up to an overall normalization factor, the post-measurement state of the first row becomes
\begin{align}\label{postm}
  \sum_a \lambda_a^2 \,|R_a\rangle_1 \,|L_a\rangle_1.
\end{align}

If, instead, the measurement outcome on the \(i\)-th row corresponds to other EPR state, such as \(\sum_{a,b} U_{ab} \,|L_a\rangle_i\,|R_b\rangle_i\) (all EPR outcome is akin to a local unitary transformation acting on the symmetric EPR pair), then the post-measurement state of the first row becomes
\begin{align}
  \sum_{a,b} \lambda_a \,|R_a\rangle_1 \bigl(\lambda_b\,U_{ab}\bigr)\,|L_b\rangle_1.
\end{align}
By applying a local rotation \(Q_{ij} = \tfrac{\lambda_j}{\lambda_i}\, (U^\dagger)_{ij}\) on the left component, we can transform this state back into the desired form in Eq.~\eqref{postm}:
\begin{align}\label{eq:QU}
\sum_{a,b,c} \lambda_a & \,|R_a\rangle_1 \Bigl(\lambda_b\,U_{ab}\Bigr) Q_{bc}|L_c\rangle_1 
   \nonumber\\
  =\sum_{a,b,c} \lambda_a & \,|R_a\rangle_1 \Bigl(\lambda_b\,U_{ab}\Bigr) \Bigl(\tfrac{\lambda_c}{\lambda_b}\,U^\dagger_{bc}\Bigr)\,|L_c\rangle_1 
   \nonumber\\
   &\longrightarrow \sum_a \lambda_a^2\, |R_a\rangle_1\,|L_a\rangle_1.
\end{align}
Here, the vectors \( L/R \) are expressed in the eigenstate basis of the density matrix. The ranks of \( U \) and \( Q \) match that of the density matrix, so these transformations are defined within the subspace spanned by the nonzero eigenvectors of the density matrix.
Therefore, any measurement outcome can be corrected by applying an appropriate \(Q\) operator on the first row before we implement the next unitary gate as Fig.~\ref{pushfw}-a. This correction procedure is repeated at each step, ensuring that any random measurement outcome is compensated by a subsequent unitary feedback on the first row. 

To emphasize, in this protocol, rather than preparing the SSPT state in one step and then performing all measurements afterward, we interleave state preparation and measurement in a sequential manner. In each step, we apply \(\text{PSWAP}_{1,i}\) to entangle the first row with the \(i\)-th row, perform an EPR measurement on the \(i\)-th row, and use unitary feedback on the first row that depends on the measurement outcome (Fig.~\ref{pushfw}-a). Once this combined measurement and feedback operation is complete, we proceed to the next row by applying \(\text{PSWAP}_{1,i+1}\) and repeat this process. 
Overall, this protocol requires a sequential set of unitary measurement and feedback operations of length \(O(L)\).

\begin{figure}[h]
\centering
\includegraphics[width=0.48\textwidth]{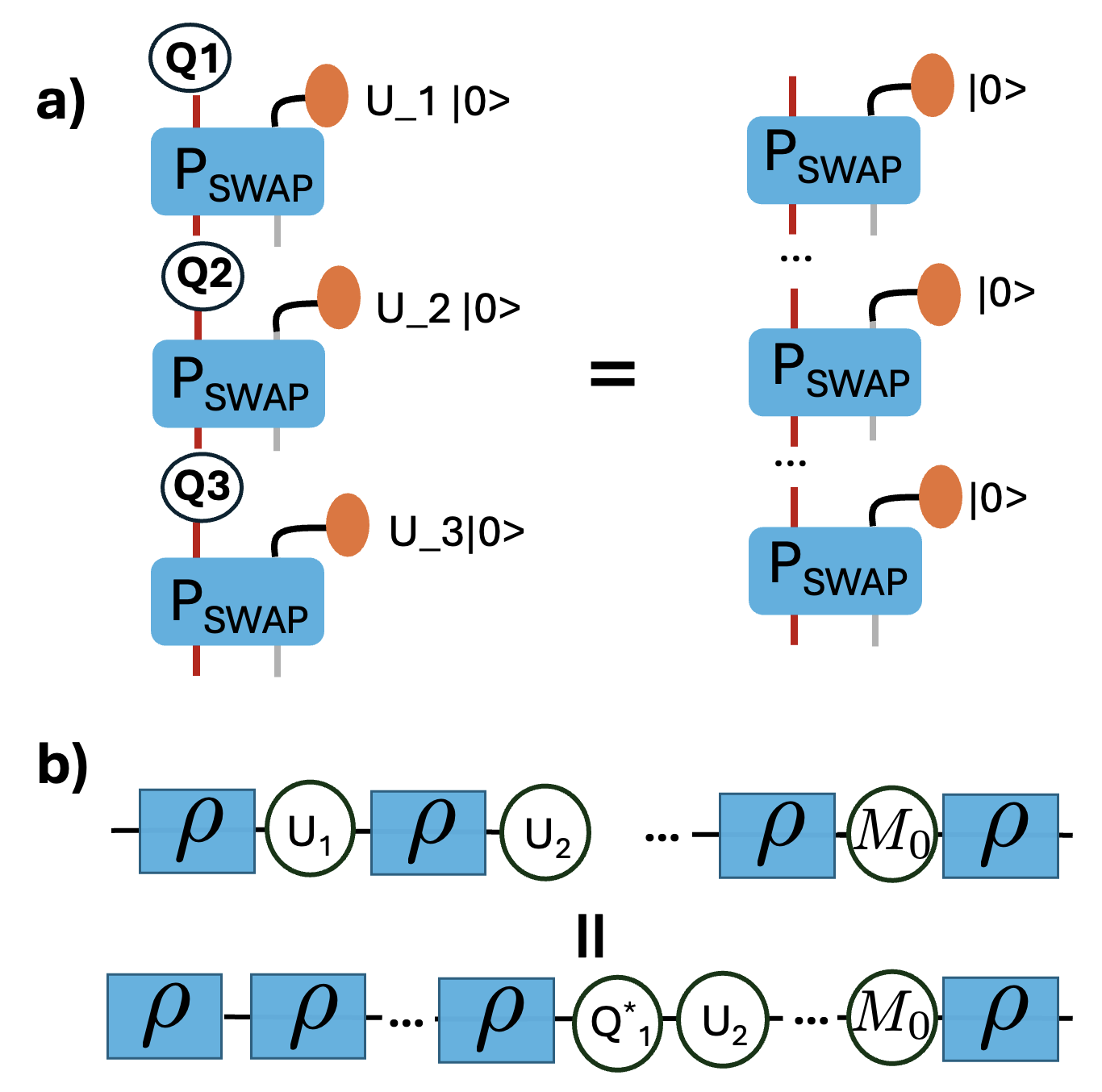}
\caption{a). Measure the qubits on the onsite EPR basis. Here $|0\rangle$ refers to measurement outcomes being the symmetric EPR state while $U|0\rangle$ denotes other EPR states. Based on the measurement outcomes, apply a local unitary operator \(Q\) (on the first row) before proceeding to the next step. 
b) The measurement outcome (in the EPR-basis) corresponds to inserting the U-operator between two density matrices in the twisted R\'enyi-\(N\) operator. One can always push the unitaries acting between copies of \( \rho \) toward either the left or right end.}
\label{pushfw}
\end{figure} 

Alternatively, one can prepare the SSPT wavefunction using a shallow circuit, then perform on-site EPR measurements in one step. 
Different measurement outcomes correspond to choosing different \(\Psi^{\text{trivial}}\) (product of on-site states) in the strange correlator. According to our holographic duality, this choice of trivial state is mapped to the \(M\)-matrix inserted between adjacent copies of the density matrix in the twisted R\'enyi-\(N\) operator defined in Eq.~\eqref{twsitedRenyi}.
As the bra vector of the \(i\)-th density matrix and the ket vector of the \((i+1)\)-th density matrix represent the left and right components in each row, projecting onto the on-site symmetric EPR state corresponds to multiple of two adjacent density matrices. Projecting onto any other on-site EPR basis instead corresponds to inserting a unitary operator between the bra vector of the \(i\)-th density matrix and the ket vector of the \((i+1)\)-th density matrix.
Thus, depending on the measurement outcomes, the strange correlator of $\Psi^{SSPT}$ is dual to the twisted R\'enyi-$N$ correlator  in an alternative form:
\[
\mathrm{Tr}\bigl[\,\bm{M}_O\,\rho\,U_1\,\rho\,U_2\,\dots\,\rho\,U_N\,\bm{M}^{\dagger}_O\,\rho\,U'_1\,\rho\,U'_2\,\dots\,\rho\,U'_N\bigr],
\]
The matrix \( U_i \), inserted between each copy of \( \rho \), are determined by the onsite EPR measurement outcomes. The operators \( \bm{M}_O \) and \( \bm{M}^{\dagger}_O \), which we aim to measure in the twisted Rényi-\(N\) correlator, are separated along the replica direction. From the density matrix perspective, one can show that any density matrix can be decomposed as follows:
\begin{align}
U\rho = \rho Q^{\dagger}
\end{align}
This is how we compensate for the measurement outcome using a subsequent unitary in Eq.~\eqref{eq:QU}. This approach allows us to `push' the unitaries acting between copies of \( \rho \) toward either the left or right boundary. Concretely, we obtain:
\begin{align}
& (\rho U_1 \rho U_2...\rho U_N)\bm{M}_O (\rho U'_1 \rho U'_2...\rho U'_N) \bm{M}^{\dagger}_O  \rightarrow  \rho^n \tilde{M}_O \rho^n \tilde{M}^{\dagger}_O \nonumber\\
&\tilde{M}_O=Q_1^{\dagger} U_2 Q_3^{\dagger}...U_N \bm{M}_O ,~\tilde{M}^{\dagger}_O=Q_1^{'\dagger} U'_2 Q_3^{'\dagger} ...U'_N \bm{M}^{\dagger}_O
\end{align}
In this approach, the outcome-dependent unitary operators are carried forward along the replica direction and effectively subsumed into the measurement operators \(\bm{M}^\dagger_O\) and \(\bm{M}_O\). A subtlety arises, however, because after \(n\) rounds of these local unitary transformations, the resulting operators \(\tilde{M}^\dagger_O\) and \(\tilde{M}_O\) may become nonlocal.

\subsection{Experimental probe of mSPT anomaly}

In addition to the efficient preparation of intrinsic mSPT states, we also propose a strategy to probe the boundary anomaly of mSPT states.

For closed systems with spatial dimensions higher than 1, the 't Hooft anomaly of a symmetry $G$ is reflected by (quasi) long-range two-point correlation function for arbitrary charged local operator $O_i$ on the boundary of the system, namely
\begin{align}
C(i,j)=\Tr(\rho O_iO_j^\dagger)-\Tr(\rho O_i)\Tr(\rho O_j^\dag).
\end{align}
Equivalently, the anomalous boundary state of a nontrivial SPT state cannot be a unique ground state of a symmetric gapped local Hamiltonian. 

Nevertheless, for open quantum systems, the generalization of 't Hooft quantum anomaly is not straightforward in the sense that there is no well-defined concept of Hamiltonian and corresponding energy gap for mixed states. Recently, to address this hardness of characterizing the mixed-state quantum anomaly, in Refs. \cite{ma2024topological, lessa2024mixedstate}, the authors systematically generalized ``energy gap'' to open quantum system by introducing the concept of \textit{gapped mixed states} or \textit{gapped Markovian mixed states} as the following.

\begin{figure}
\begin{tikzpicture}[scale=1.2]
\tikzstyle{sergio}=[rectangle,draw=none]
\filldraw[fill=green!20, draw=black, thick] (0,0)circle (2);
\filldraw[fill=violet!20, draw=black, thick] (0,0)circle (1);
\path (0,0) node [style=sergio,color=red] {\Large$A$};
\path (1.5,0.25) node [style=sergio,color=red] {\large $w(B)$};
\path (-1.075,1.075) node [style=sergio,color=red] {\Large$B$};
\path (-1.75,1.75) node [style=sergio,color=red] {\Large$C$};
\draw[thick, <->] (1,0) -- (2,0);
\end{tikzpicture}
\caption{A circular tripartition $A$, $B$, and $C$ of the system. Here $w(B)$ depicts the width of the buffer region $B$.}
\label{Fig: tripartition}
\end{figure}

\begin{definition}
Consider a tripartite separation of the system (see Fig. \ref{Fig: tripartition}), where $A$ is a disk regime, $B$ is a buffer regime surrounding $A$ with the width $w(B)$, and $C$ is the rest of the system. A mixed state $\rho$ is defined as ``gapped'' if for arbitrary tripartition, we have (here $S(A)$ is the von Neumann entropy of the regime $A$)
\begin{enumerate}[a.]
\item The mutual information (MI) is exponentially small:
\begin{align}
I_\rho(A:C)=S(A)+S(C)-S(AC)\sim e^{-w(B)/\xi_{\mathrm{MI}}},
\end{align}
with a finite correlation length $\xi_{\mathrm{MI}}$;
\item The conditional mutual information (CMI) is exponentially small:
\begin{align}
I_\rho(A:C|B)&=S(AB)+S(BC)-S(B)-S(ABC)\nonumber\\
&=e^{-w(B)/\xi_{\mathrm{CMI}}},
\end{align}
with a finite Markov length $\xi_{\mathrm{CMI}}$ \cite{sang2024stability}.
\end{enumerate}
\end{definition}

Then it has been proved in Ref. \cite{lessa2024mixedstate} that an anomalous edge state of an mSPT density matrix cannot be a gapped Markovian mixed state. 

Therefore, the ideal way to experimentally probe the mixed-state quantum anomaly requires the precise measurements of the von Neumann entropy. Nevertheless, due to the highly nonlinear nature of the von Neumann entropy (and any other information-theoretic quantities), a precise measurement of the von Neumann entropy requires an exponentially hard quantum state tomography (QST) to fully reproduce all the information of the density matrix. Therefore, the rigorous experimental probe of mixed-state quantum anomaly is exponentially hard with respect to the system size. 

Alternatively, in this section, we produce a reliable estimation of the mixed-state quantum anomaly which is experimentally accessible by classical shadow tomography \cite{Huang_2020, Rath_2021}. For open quantum systems, there is a heuristic argument that an mSPT density matrix corresponds to a pure-state SPT in doubled Hilbert space \cite{ma2024symmetry, xue2024tensor}, where a strong symmetry $K$ is mapped to a doubled symmetry $K_u\times K_l$ that are defined in upper and lower Hilbert spaces, respectively, and a weak symmetry $G$ is mapped to a diagonal symmetry $G_d$. Therefore, the 't Hooft anomaly of strong and weak symmetries in mixed states can be effectively captured by a (quasi) long-ranged two-point correlation function in doubled Hilbert space, namely
 \begin{align}
C^{(2)}(i,j)=\frac{\llangle\rho|O_i O_j^\dag|\rho\rrangle}{\llangle\rho|\rho\rrangle}\geq |i-j|^{-\gamma},
 \label{Eq: Renyi-2 correlator}
\end{align}
where $\gamma$ is a finite number. In particular, if we take $O_i=O_{i,u}O_{i,l}^\dag$ as a charged operator of $K_u\times K_l$, the (quasi) long-range two-point correlator in doubled Hilbert space is equivalent to a R\'enyi-2 correlator under the Choi–Jamiołkowski map, namely
\begin{align}
C^{(2)}(i,j)=\frac{\Tr\left(\rho O_i O_j^\dag \rho O_j O_i^\dag\right)}{\Tr(\rho^2)}\geq |i-j|^{-\gamma}.
\label{Eq: boundary Renyi-2}
\end{align}
Therefore, the quasi-long range boundary R\'enyi-2 correlation of the charged operators of strong symmetry gives a reliable indication of the mixed-state anomaly.

Following this fact, we design an experimental proposal for measuring the mixed-state quantum anomaly through quantum metrology. To be more specific, the quantum Fisher information (QFI) metric. Consider an mSPT density matrix $\rho$ with the open boundary, which is protected by a strong symmetry $K$ and a weak symmetry $G$. Define a local charge dephasing channel, namely
\begin{align}
\E=\circ_j\E_j,\quad\E_j[\rho]=(1-p)\rho+p O_j \rho O_j^\dag,
 \label{Eq: dephasing}
 \end{align}
 where $O_j$ is a local operator carrying charge of the strong symmetry $K$ and $p$ is the probability. Then the Bures metric $D_b^2(\E[\tilde\rho], \tilde\rho)$ of the twisted density matrix $\tilde{\rho}=\frac{1}{2}(\rho+O_i^\dag \rho O_i)$ is defined as
 \begin{align}
D_b^2(\E[\tilde\rho], \tilde\rho)=2-2F\left(\E[\tilde\rho], \tilde\rho\right),
\end{align}
where $F(\rho,\sigma)=\Tr\sqrt{\sqrt{\rho}\sigma\sqrt{\rho}}$ is the \textit{fidelity} between density matrices $\rho$ and $\sigma$. Accordingly, the QFI metric of $\tilde{\rho}$ is defined as the second derivative of the Bures metric with respect to the probability $p$, namely\footnote{We note that the QFI metric of $\rho$ can only give the information of strong symmetry $K$, and there is no difference between the systems with and without anomaly.}
 \begin{align}
\mathcal{F}_Q[\tilde{\rho}]=\left.2\frac{\partial^2D_b^2(\E[\tilde\rho], \tilde\rho)}{\partial p^2}\right|_{p=0}.
\label{Eq: QFI}
 \end{align}
\color{black}The Bures and QFI metrics of $\tilde{\rho}$ can be analytically calculated in terms of the following form of the series expansion, namely
\begin{widetext}
 \begin{align}
 \begin{gathered}
 D_b^2\left(\E[\tilde\rho],\tilde\rho\right)=\frac{Np}{2}+\frac{p^2}{4}\sum_{n=0}^\infty\sum_{m=0}^n\sum_{q=0}^m \binom{n}{m}\binom{m}{q}(-1)^m\sum_{j,r}\Tr\left[\rho^q O_jO_i^\dagger\rho O_iO_j^\dagger \rho^{m-q}O_rO_i^\dagger\rho O_i O_r^\dagger\right]+O(p^3)\\
 \mathcal{F}_Q[\tilde\rho]=\sum_{n=0}^{\alpha}\sum_{m=0}^n\sum_{q=0}^m \binom{n}{m}\binom{m}{q}(-1)^m\sum_{j,r}\Tr\left[\rho^q O_jO_i^\dagger\rho O_iO_j^\dagger \rho^{m-q}O_rO_i^\dagger\rho O_i O_r^\dagger\right]
 \end{gathered}.
 \label{Eq: Bures}
 \end{align}
 \end{widetext}
We note that each term in this expansion is a polynomial of the density matrix. In particular, the leading order of QFI metric is a sum of the denominator of the R\'enyi-2 correlator over the support of the local quantum channel \eqref{Eq: dephasing}, namely
\begin{align}
 \mathcal{F}_Q^{(0)}[\tilde{\rho}] = \sum_{j,r}\Tr\left(O_r^\dagger O_j\rho O_j^\dagger O_r\rho\right).
 \label{Eq: QFI 0th}
\end{align}
\color{black}

For an mSPT density matrix with a strong symmetry $K$ and a weak symmetry $G$, if we perform the local quantum channel \eqref{Eq: dephasing} on the open boundary, then the leading order of the QFI metric has the following scaling behavior, 
 \begin{align}
 \mathcal{F}_Q^{(0)}[\tilde\rho]\sim O(L^{2(d-1)-\gamma})\cdot\mathrm{Tr}(\rho^2),
 \end{align}
where $\gamma$ is defined in Eq. \eqref{Eq: boundary Renyi-2}, $L^{d-1}$ is the total number of particles on the boundary, and $d$ is the spatial dimension of the system. If we can directly measure the QFI for the state with a boundary perturbation, then from the scaling of the QFI with the system size, we can read off the power law exponent of the R\'enyi-2 correlation. 

However, if we do not have precise control and can only perturb the whole system rather than just on the boundary, assuming the R\'enyi-2 correlators follow Eq. \eqref{Eq: boundary Renyi-2}, the leading order of the QFI metric has the following scaling behavior,
 \begin{align}
 \mathcal{F}_Q^{(0)}[\tilde\rho]\sim O(L^{d})\cdot\Tr(\rho^2) + O(L^{2(d-1)-\gamma})\cdot\mathrm{Tr}(\rho^2),
 \end{align}
where the first term is from the exponentially small R\'enyi-2 correlator if one of the sites $j$ and $r$ is located in bulk, and the anomaly is characterized by the subleading term of the QFI metric. In this case, we have to do a careful scaling analysis and take out the leading term to extract the power law exponent. This is also in principle possible. 

Therefore, we are able to experimentally identify the mixed-state quantum anomaly by measuring the QFI metric of a twisted density matrix $\tilde{\rho}$ under the local charge dephasing channel \eqref{Eq: dephasing}. The measurement can be done by classical shadow tomography (CST): the CST is proposed to efficiently characterize an unknown quantum state by relatively few measurements, as well as the expectation values of many physical observables. For example, consider a qubit system for simplicity, the procedures of CST for a density matrix $\rho$ are summarized as
\begin{enumerate}[1.]
\item Apply a random unitary circuit $U$ to the density matrix, to introduce the randomness of the following measurement.
\item Perform a measurement in the computational ($Z$) basis, with the measurement outcome as a bit string $\ket{\boldsymbol{s}}\bra{\boldsymbol{s}},~\boldsymbol{s}\in\{0,1\}^{\otimes N}$. The probability of the post-measurement state is given by the Born's rule 
\begin{align}
p_{U,\boldsymbol{s}}=\Tr(U\rho U^\dag\ket{\boldsymbol{s}}\bra{\boldsymbol{s}}).
\end{align}
\item Repeatedly perform the randomized measurements multiple times. For each measurement outcome, a classical shadow is constructed as
\begin{align}
\sigma=U^\dag \ket{\boldsymbol{s}}\bra{\boldsymbol{s}}U,
\end{align}
such that 
\begin{align}
\mathop{\mathbb{E}}\limits_{\sigma\sim p(\sigma|\rho)}\sigma=\mathcal{M}[\rho],
\end{align}
where $\mathbb{E}$ depicts the expectation value of various randomized measurement outcomes, which can be formulated as a quantum channel $\mathcal{M}$.
\end{enumerate}

For each term in Eq. \eqref{Eq: QFI 0th}, we use the randomized measurements, namely
\begin{align}
\mathop{\mathbb{E}}\limits_{\sigma\sim p(\sigma|\rho)}\sigma=\mathcal{M}[\rho],
 \end{align}
 with the Born's rule $p(\sigma|\rho)\propto\Tr(\sigma\rho)$ and $\mathbb{E}$ depicts the expectation value with respect to different measurement outcomes $\sigma$. Here we label the randomized measurements as a local quantum channel $\mathcal{M}$. The formal reconstruction of $\rho$ is formulated as following,
 \begin{align}
\rho\simeq\mathop{\mathbb{E}}\limits_{\sigma\sim p(\sigma|\rho)}\mathcal{M}^{-1}[\rho],
 \end{align}
and each term in Eq. \eqref{Eq: QFI 0th} can be estimated by
\begin{align}
\Tr\left(O_r^\dag O_j \rho O_j^\dag O_r \rho\right) = \mathop{\mathbb{E}}\limits_{\substack{\sigma\sim p(\sigma|\rho) \\ \sigma'\sim p(\sigma'|\hat{\rho})}}\Tr\left( \mathcal{M}^{-1}[\sigma] \mathcal{M}^{-1}[\sigma']\right),
\label{Eq: numerator}
\end{align}
where $\hat{\rho}=O_j^\dag O_r \rho O_r^\dag O_j$, and
\begin{align}
\Tr\left(\rho^2\right)=\mathop{\mathbb{E}}\limits_{\substack{\sigma\sim p(\sigma|\rho) \\ \sigma'\sim p(\sigma'|{\rho})}}=\Tr\left( \mathcal{M}^{-1}[\sigma] \mathcal{M}^{-1}[\sigma']\right).
\label{Eq: denominator}
\end{align}
We note that from Eqs. \eqref{Eq: numerator} and \eqref{Eq: denominator}, we can directly estimate the boundary R\'enyi-2 correlator \eqref{Eq: boundary Renyi-2} which is already the leading order of QFI [cf. Eq. \eqref{Eq: QFI 0th}] as a reliable indicator of mixed-state quantum anomaly. The purpose of introducing the QFI \eqref{Eq: QFI} is that to improve the estimation accuracy, we can expand more terms and measure each term by classical shadow tomography. 
}

\color{black}
\section{Summary and Outlook}
In this work, we demonstrate a holographic duality between $d$-dimensional intrinsic mSPT and $(d+1)$-dimensional SSPT. This duality offers a pathway to identify intrinsic mixed-state SPTs by examining the reduced density matrix of SSPT states in higher dimensions. Similarly, one can investigate SSPT wave functions by considering replicas of mSPT in lower dimensions. 

{\color{black}We also systematically discuss the implication of our holographic duality on quantum devices, including:
\begin{enumerate}[1.]
\item We construct an efficient \textit{holographic preparation} of the intrinsic mSPT states which can be accomplished on a quantum device within finite depth by our holographic duality.
\item We discuss the implication of holographic duality through the sequential quantum circuits of SSPT states.
\item We further propose the experimental probe of the boundary anomaly of mSPT states by classical shadow tomography.
\end{enumerate}
}

We conclude our discussion by outlining several future directions:

\begin{enumerate}[1.]
\item In our proposed duality, the twisted Rényi-$N$ correlator of the mixed-state SPT maps to the strange correlator of the SSPT state. Since the mSPT ensemble can be viewed as the reduced density matrix of the SSPT on the surface, our duality offers an alternative perspective on bulk-edge correspondence where bulk correlations can be derived from the properties of the boundary's reduced density matrix. It would be valuable to explore how this correspondence and duality can be extended to generic mixed states.

\item In Ref.~\cite{lessa2024mixedstate,chen2023separability}, the mSPT (or its boundary) was examined through the lens of entanglement properties. Ref.~\cite{lessa2024mixedstate} presented an $N$-partite separability criterion for mixed states with strong symmetry and a long-range correlation in the sense of mutual information or conditional mutual information criterion for mixed states carrying mixed anomaly of strong and weak symmetries, while Ref.~\cite{chen2023separability} analyzed the entanglement behavior of each eigenvector within the mixed state. Given that our mSPT can be dual to the reduced density matrix of a higher-dimensional SSPT, it would be enlightening to investigate the entanglement aspects from the dual SSPT perspective.

\item Recently, there has been growing interest in the broadly defined mixed-state topological order. Gauging the 3d SSPT yields a symmetry-enriched fracton gauge theory, while gauging the 2d mSPT leads to a quantum field theory describing open SPTs within the Keldysh formalism. In Appendix B, we provide a brief discussion on how our duality connects the 3d hybrid fracton gauge theory with the 2d QFT in the Keldysh formalism on specific examples.  
Exploring how these two concepts can be unified within the framework of our duality would be an intriguing direction for future research.
\end{enumerate}

\acknowledgments 
We thank Chong Wang, Yichen Xu, and Yi-Zhuang You for their helpful discussions and comments. 
This work was performed in part at the Aspen Center for Physics (YY), which is supported by the National Science Foundation grant PHY-2210452 and the Durand Fund (YY). This research was supported in part by grant NSF PHY-2309135 to the Kavli Institute for Theoretical Physics (KITP). JHZ is supported by the U.S. Department of Energy under the Award Number DE-SC0024324. ZB acknowledges support from NSF under award number DMR-2339319. YY acknowledges support from NSF under award number DMR--2439118.

\appendix

\section{More examples of $d$-dim intrinsic mSPT and $(d+1)$-dim HO-SSPT}
{\color{black}
So far, we have demonstrated a duality between $d$-dim mSPT and $(d+1)$-dim HO-SSPT. From the dimension reduction side of this duality, the surface-reduced density matrix of the $(d+1)$-dim HO-SSPT, with subsystem $\mathcal{S}$ and global symmetry $\mathcal{G}$, can be treated as an mSPT in $d$-dim, protected by weak $\mathcal{G}$ and strong $\mathcal{S}$ symmetries. Notably, with this dimensional reduction, the resulting mSPT is always an intrinsic SPT. Specifically, no gapped SPT state, which can serve as the ground state of a lattice model, can support a gapless edge with such mixed anomaly between the $\mathcal{S}$ and $\mathcal{G}$ symmetries.

The intrinsic nature of this mSPT can be traced back to the duality, as the hinge states in the dual $(d+1)$-dim HO-SSPT, with a mixed anomaly between subsystem $\mathcal{S}$ and global $\mathcal{G}$, originate from the nontrivial bulk topology. Such hinge states cannot be realized through surface decoration.
This is also evident from our building block construction for the 3-dim HO-SSPT, where each 2-dim layer in the building block is anomalous under global $\mathcal{G}$ symmetry, and the anomaly cancels when the top and bottom layers in the building block are coupled.
Our duality offers a practical method to identify mSPT by examining the reduced density matrix of HO-SSPT in higher dimensions. Conversely, by reversing this duality, we can generate a $(d+1)$-dim SSPT wavefunction through the replication of $d$-dim intrinsic mSPT, as discussed in Sec.~\ref{sec:dimext}.

Finally, we discuss the relationship between our results and the intrinsic mSPT explored in Ref.~\cite{ma2024topological}. Ref.~\cite{ma2024topological} outlines a universal protocol to construct intrinsic mSPT by decohering or introducing disorder into an intrinsic gapless SPT state. This gapless SPT state, characterized by both $\mathcal{G}$ and $\mathcal{S}$ symmetries with the following group structure,
\begin{align}
1\rightarrow\mathcal{S}\rightarrow\tilde{\mathcal{G}}\rightarrow\mathcal{G}\rightarrow1,
\end{align}
exhibits a $\mathcal{G}$-anomaly in its low-energy spectrum (IR effective theory) within the bulk. When a mixed state is generated by adding quenched disorder (or quantum channels) that preserve weak $\mathcal{G}$ and strong $\mathcal{S}$ symmetries, the bulk $\mathcal{G}$-anomaly can be canceled by the `ancilla', resulting in a purified state that is anomaly-free in bulk. The resulting density matrix thus exhibits an intrinsic mSPT. 

In our SSPT construction, the top and bottom layers in the building block function as system and ancilla qubits, respectively. In this setup, the top layer alone within each building block is anomalous under $\mathcal{G}$ symmetry, meaning it cannot be symmetrically gapped. However, when the top and bottom layers are coupled, the $\mathcal{G}$ anomaly is canceled, resulting in a short-range entangled building block. This process closely mirrors the universal protocol for preparing intrinsic mSPT through decohering an intrinsic gapless SPT. Moreover, it suggests that the twisted Rényi-$N$ correlator [cf. Eq.~\eqref{twsitedRenyi}] of the $d$-dimensional intrinsic mSPT maps to the bulk strange correlator in the $(d+1)$-dimensional HO-SSPT wavefunction, both of which exhibit long-range or quasi-long-range order. In this scenario, the twisted Rényi-$N$ correlator serves as a `new metric' for detecting intrinsic mSPT and distinguishing it from conventional mSPT. In the case of conventional mSPT, which is obtained by decohering an SPT pure state, 
the twisted Rényi-$N$ correlator typically exhibits short-range behavior because, in the clean limit, the SPT wavefunction is a pure state.}

In this subsection, we discuss three general examples of d-dimensional intrinsic mSPT phases constructed using the duality by tracing out ancilla from the d+1-dimensional bosonic SSPT phases discussed in Refs.~\cite{zhang2023classification,prelim}.

\subsection{3D bosonic SSPT with $\mathcal{S}=\mathbb{Z}_2$}

\begin{figure}[t]
    \centering
    \includegraphics[width=0.48\textwidth]{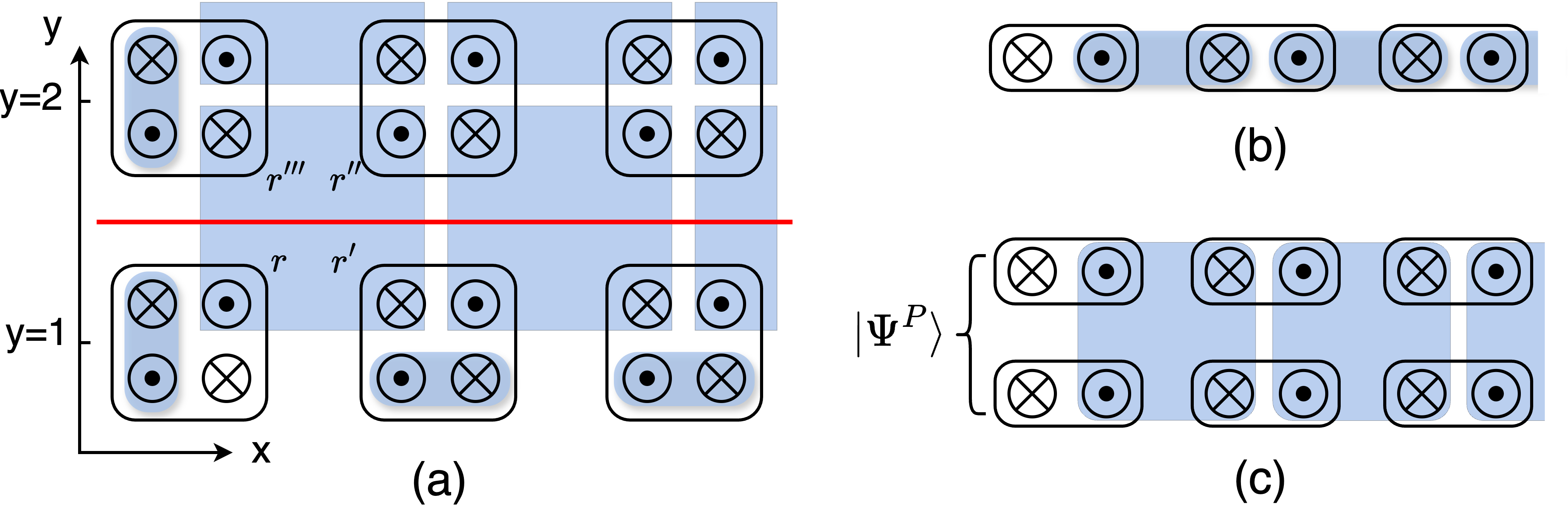}
    \caption{a) 3d SSPT with 2-foliated subsystem symmetry via coupled wire construction. b) 2d SSPT.  c) Double state of 2d mSPT. The top and bottom layers correspond to the left and right Hilbert spaces, respectively.}
    \label{fig:coupledwire}
\end{figure}

The first case we consider is a duality between (3+1)D $2^{\mathrm{nd}}$-order SSPT states with 2-foliated subsystem symmetries $\mathbb{Z}_{2}^{xz}$ and $\mathbb{Z}_{2}^{yz}$ on each $x$-$z$ and $y$-$z$ planes studied in Ref.~\cite{zhang2023classification} and a (2+1)D intrinsic mSPT. The coupled wire construction of the SSPT state is illustrated in Fig.~\ref{fig:coupledwire}a. Tracing out all top layers with $y>1$, one gets a mixed state for the bottom later at $y=1$. The mixed state has a strong global $\mathbb{Z}_{2}$ symmetry and weak subsystem symmetries $\mathbb{Z}_{2}^{z}$ on each z-column, which are the reminiscence of the 3D subsystem symmetries $\mathbb{Z}_{2}^{xz}$ and $\mathbb{Z}_{2}^{yz}$, respectively. In what immediately follows, we show that the mixed state of the bottom layer constructed this way is an intrinsic mSPT.

We start by reviewing the 3D 2nd order SSPT state~\cite{zhang2023classification}. Similar to Sec.~\ref{sec:3dhinge}, we use coupled wire construction for such SSPT states, as is illustrated in Fig.~\ref{fig:coupledwire}. Here, each site in the coupled wire construction of the 3D SSPT contains four (1+1)D wires, each of which can be described by a 2-component Luttinger theory
\begin{equation}
    \mathcal{L}_{0} = \frac{1}{2\pi} \partial_{z}\phi_{1}\partial_{\tau}\phi_{2}+\frac{1}{4\pi}\sum_{\alpha,\beta=1,2}\partial_{z}\phi_{\alpha}V_{\alpha\beta}\partial_{z}\phi_{\beta}.
    \label{eq:Luttinger_wire}
\end{equation}
These bosonic fields connect with the fermionic fields studied in the previous sections via a bosonization $\chi_{j} = e^{ i\phi_{j}}$. The $\mathbb{Z}_{2}$ symmetry transforms
\begin{equation}
    \phi_{1}\rightarrow\phi_{1}+\pi, \,\, \phi_{2}\rightarrow\phi_{2}+\pi.
    \label{eq:Z2Luttinger}
\end{equation}
The subsystem symmetries $\mathbb{Z}_{2}^{xz}$ and $\mathbb{Z}_{2}^{yz}$ correspond to applying the symmetry transformations Eq.~\eqref{eq:Z2Luttinger} to all wires in sites along $x$- or $y$-columns. In bulk, these symmetries allow for backscattering terms to gap out 4 wires in the blue plaquettes; on the boundary, they further allow for the pairs of wires in the blue ellipses, which are not included in the bulk interactions, to be gapped out. As a result, both the bulk and the surface are fully gapped. 

The system presents gapless hinge modes. Each hinge site has three dangling Luttinger liquids that are not included in the bulk plaquette interactions, among which two can be gapped out by the backscattering terms. This left one dangling Luttinger liquid as the gapless hinge mode of the system.

To demonstrate the correspondence between this SSPT state and a 2D mSPT state, we make a spatial cut along the red horizontal line (see Fig.~\ref{fig:coupledwire}) and trace out everything except the bottom layer. We demonstrate that by utilizing the Choi–Jamiołkowski isomorphism, the resulting mixed state of the bottom layer is an mSPT state by observing that the hinge modes on the bottom layer cannot be gapped out in the doubled Hilbert space. 


Since the bulk and the surface of the 3D SSPT state are fully gapped, after tracing out the top part of the system, the resulting density matrix will be a maximally mixed state of these degrees of freedom. Therefore, we can focus on the two hinge modes only. In the Choi-doubled space, a hinge mode can be written as a 4-component Luttinger liquid, namely
\begin{equation}
    \mathcal{L} = \frac{1}{4\pi}\partial_{z}\Phi^{T} K \partial_{\tau}\phi + \frac{1}{4\pi}\partial_{x}\Phi^{T}V\partial_{x}\Phi,
    \label{eq:Luttinger_Choidoubled}
\end{equation}
where $\Phi^{T} = \left(\phi_{1},\phi_{2},\phi_{3},\phi_{4}\right)$ is a 4-component bosonic field with the first 2 components in the left Hilbert space and the latter 2 components in the right Hilbert space, respectively, and $K=(\sigma^{x})\oplus(-\sigma^{x})$ is the $K$-matrix.

The symmetries of the bottom layer are strong global $\mathbb{Z}_{2}$ symmetry and weak subsystem $\mathbb{Z}_{2}^{z}$ symmetries along each $z$-column, which are mapped to a doubled symmetry $\Z_2^L\times\Z_2^R$ and a diagonal symmetry $\Z_2^z$ in the doubled Hilbert space. The symmetry actions on the bosonic field $\Phi$ are phrased as
\begin{subequations}
    \begin{align}
        \delta\Phi^{\mathbb{Z}_{2}^{L}} &=\pi(1,1,0,0)^{T}\label{eq:symPhi1}\\
        \delta\Phi^{\mathbb{Z}_{2}^{R}} &=\pi(0,0,1,1)^{T}\label{eq:symPhi2}\\
        \delta\Phi^{\mathbb{Z}_{2}^{z}} &=\pi(1,1,1,1)^{T}\label{eq:symPhi3}
    \end{align}
\end{subequations}
Here, \eqref{eq:symPhi1} and \eqref{eq:symPhi2} correspond to the strong symmetry transformations acting on the left and right Hilbert spaces, and \eqref{eq:symPhi3} is the weak $\mathbb{Z}_2^{z}$ symmetry, respectively.

The only symmetry-allowed backscattering term is
\begin{align}
\cos\left(\phi_1+\phi_2+\phi_3+\phi_4\right),
\end{align}
which is not enough to fully gap out the hinge. In fact, following the anomaly indicator of 1D Luttinger liquid with $\Z_2$ symmetry introduced in Ref. \cite{Heinrich_2018}, the anomaly indicator of $\Z_2^L$ and $\Z_2^R$ are non-vanishing, namely 
\begin{align}
\begin{gathered}
\nu_{L}\equiv1(\mathrm{mod}~2)\\
\nu_{R}\equiv1(\mathrm{mod}~2)
\end{gathered}
\end{align}
which implies that the corresponding 1D Luttinger liquid carries the 't Hooft anomaly of both $\Z_2^L$ and $\Z_2^R$ in the doubled Hilbert space. Equivalently, we conclude that in the physical Hilbert space, the hinge mode carries the 't Hooft anomaly of the strong $\Z_2$ symmetry, which implies a nontrivial (2+1)D mSPT state in the $y=1$ layer. 

Note that this mSPT state has no pure-state analog, since a 2D system with global $\mathbb{Z}_{2}$ symmetry and subsystem $\mathbb{Z}_{2}^{z}$ symmetry is anomalous. We illustrate this using a coupled wire construction shown in Fig.~\ref{fig:coupledwire}b. The gapless modes can be gapped out by onsite 2-body interactions, resulting in a trivial product state. However, inter-site interactions, for example, those depicted by the blue ellipses in Fig.~\ref{fig:coupledwire}b are forbidden by the subsystem $\mathbb{Z}_{2}^{z}$ symmetry. This should not be a surprise since if such 2D pure state SSPT exists, then the hinge modes of the aforementioned 3D bosonic SSPT are not stable, as they can be gapped out by adhering a 2D layer to surfaces of the 3D SSPT.  Therefore, this 2D system is either in a trivial product state or is anomalous in the bulk. On the other hand, in 2D mixed state with weak subsystem $\mathbb{Z}_{2}^{z}$ symmetry and strong global $\mathbb{Z}_{2}$ symmetry, the inter-site interactions depicted by blue plaquettes in Fig.~\ref{fig:coupledwire}c preserve the symmetries and can be included to gap out the Luttinger liquids in the bulk, giving a nontrivial mSPT with gapless hinge modes.

\subsection{3D bosonic SSPT with $\mathcal{S}=\mathbb{Z}_2$ and $\mathcal{G}=\mathbb{Z}_2^T$}

In this section, we obtain (2+1)D mSPT from (3+1)D second-order SSPT with 2-foliated subsystem $\mathbb{Z}_{2}^{xz}$ and $\mathbb{Z}_{2}^{yz}$ symmetry and a global $\mathbb{Z}_2^T$ symmetry studied in Ref.~\cite{zhang2023classification}. The SSPT state can be constructed via the coupled wire construction illustrated in Fig.~\ref{fig:coupledwire}a, where each wire can be described by a 2-component Luttinger theory Eq.~\eqref{eq:Luttinger_wire} with the following symmetry properties
\begin{subequations}
    \begin{align}
        &\mathbb{Z}_{2}: \,\, \phi_{1}\rightarrow\phi_{1}+\pi,\,\, \phi_{2}\rightarrow\phi_{2} \label{eq:Z2Luttinger_a}\\
        &\mathbb{Z}_{2}^{T}: \,\, \phi_{1}\rightarrow\phi_{1},\,\, \phi_{2}\rightarrow -\phi_{2}+\pi \label{eq:Z2TLuttinger_b}
    \end{align}
\end{subequations}

The subsystem symmetries corresponding to applying the symmetry transformation Eq.~\eqref{eq:Z2Luttinger_a} to all sites on each $x$-$z$ and $y$-$z$ planes; and a global $\mathbb{Z}_{2}^{T}$ symmetry corresponding to applying the transformation Eq.~\eqref{eq:Z2TLuttinger_b} to all wires in the system. With these symmetries, backscattering terms to gap out the wires in each blue plaquette and ellipses in Fig.~\ref{fig:coupledwire} are allowed. For each hinge site, only two among the three dangling Luttinger liquids can be gapped out, leaving one remaining Luttinger liquid to become the gapless hinge mode. 

If we make a spatial cut along the red horizontal line drawn in Fig.~\ref{fig:coupledwire} and trace out everything except for the bottom layer, the resulting mixed state of the bottom layer is a mSPT state. We show this by demonstrating the 't Hooft anomaly carried by the (1+1)D hinge mode. 

Using the Choi-Jamiolkowski isomorphism, we map the density matrix of the bottom layer to a pure state in the doubled Hilbert space. In the doubled space, each hinge mode can be described by the 1D Luttinger liquid with the Lagrangian Eq.~\eqref{eq:Luttinger_Choidoubled}.

The mixed state of the bottom layer obtained from tracing out the top part of the system has a strong global $\mathbb{Z}_{2}$ symmetry, weak subsystem symmetries $\mathbb{Z}_{2}^{z}$, and a weak global $\mathbb{Z}_{2}^
{T}$ symmetry. In doubled Hilbert space, all above symmetries are mapped to a doubled symmetry $\Z_2^L\times\Z_2^R$, a diagonal subsystem symmetry $\Z_2^z$ and a time reversal symmetry $\Z_2^T$. The symmetry actions on the 4-component bosonic field of each hinge are phrased as
\begin{subequations}
    \begin{align}
        \delta\Phi^{\mathbb{Z}_{2}^{L}} &=\pi(1,0,0,0)^{T}\\
        \delta\Phi^{\mathbb{Z}_{2}^{R}} &=\pi(0,0,1,0)^{T}\\
        \delta\Phi^{\mathbb{Z}_{2}^{z}} &=\pi(1,0,1,0)^{T}
    \end{align}
\end{subequations}
as well as the $\mathbb{Z}_{2}^
{T}$ symmetry acts as $\Phi\mapsto\left(\sigma^{z}\right)^{\oplus2}\Phi + \delta\Phi^{\mathbb{Z}_{2}^{T}}$ with $\delta\Phi^{\mathbb{Z}_{2}^{T}} = \pi(0,1,0,1)^{T}$. 

With these symmetries, the only symmetry-allowed interaction is
\begin{align}
\cos\left(\phi_{2}+\phi_{4}\right).
\end{align}
which is not enough to fully gap the hinge mode in the doubled Hilbert space. In fact, consider the generators of $Z_2^L$ and $Z_2^T$, $g^L$ and $\mathcal{T}$, the symmetry indicator \cite{Heinrich_2018} of the group element $g^L\mathcal{T}$ is non-vanishing, namely
\begin{align}
\nu_{g^L\mathcal{T}}\equiv1(\mathrm{mod}~2),
\end{align}
which implies that the hinge mode in the doubled Hilbert space carries the mixed anomaly of $\Z_2^L$ and $\Z_2^T$. Equivalently, the hinge mode in the physical Hilbert space carries the mixed anomaly between strong $\Z_2$ symmetry and weak $\Z_2^T$ symmetry.


It is worth noting that such mSPT constructed from the 3D SSPT is an intrinsic mSPT, with no pure state analog.  This is because a 2D pure state with subsystem $\mathbb{Z}_{2}^{z}$ symmetry and global $\mathbb{Z}_{2}$ and $\mathbb{Z}_{2}^
{T}$ symmetries is either a product state or is anomalous, i.e., the bulk cannot be gapped out by inter-site coupling shown in Fig.~\ref{fig:coupledwire}b as such couplings are not symmetric under the subsystem $\mathbb{Z}_{2}^{z}$ symmetry transformations. 

{\color{black}\subsection{3D fermionic SSPT with $\mathcal{S}=\mathbb{Z}_2^f$ and $\mathcal{G}=\mathbb{Z}_2 $}

In this section, we turn to fermionic systems, and obtain a (2+1)D mSPT from (3+1)D fermionic SSPT with 2-foliated subsystem fermion parity symmetries $\mathbb{Z}_{2}^{f,xz}$  and $\mathbb{Z}_{2}^{f,yz}$ on each $x$-$z$ and $y$-$z$ planes, and a global $\mathbb{Z}_2$ symmetry. This fermionic SSPT is constructed in Ref.~\cite{zhang2023classification} using the coupled wire construction. Each wire is described by the 2-component Luttinger theory Eq.~\eqref{eq:Luttinger_wire}. The fermion parity and the $\mathbb{Z}_2$ symmetry action on the 2-component field are defined as
\begin{subequations}
    \begin{align}
        &\mathbb{Z}_{2}: \,\, \phi_{1}\rightarrow\phi_{1},\,\, \phi_{2}\rightarrow-\phi_{2} \label{eq:Z2fermiona}\\
        &\mathbb{Z}_{2}^{f}: \,\, \phi_{1}\rightarrow\phi_{1}+\pi,\,\, \phi_{2}\rightarrow \phi_{2}+\pi \label{eq:Z2fermionb}
    \end{align}
\end{subequations}
The subsystem symmetry corresponds to applying the $\mathbb{Z}_{2}^{f}$ symmetry transformation \eqref{eq:Z2fermionb} to each wire on the $x$-$z$ and $y$-$z$ plane; and the global symmetry corresponds to applying the symmetry transformation \eqref{eq:Z2fermiona} to all wires in the system. With these symmetries present, the wires in the blue plaquette and the ellipses can be fully gapped, leaving one dangling Luttinger liquid on each of the four hinges. These gapless Luttinger liquids are the hinge modes of the constructed fermionic SSPT~\cite{zhang2023classification}.

Here, we trace out the top part of the aforementioned system (above the red cut in Fig.~\ref{fig:coupledwire}) to obtain a fermionic mSPT for the bottom layer at $y=1$. After tracing, the bottom layer has a strong global $\mathbb{Z}_{2}^{f}$ symmetry, weak subsystem symmetries $\mathbb{Z}_{2}^{f,z}$, and a weak global $\mathbb{Z}_2$ symmetry. Since the plaquettes and the ellipses are fully gapped in the 3D SSPT, after tracing, the resulting density matrix is a maximally mixed state of these degrees of freedom. Therefore, to show the bottom layer is in an mSPT state, it is sufficient to show that each hinge mode in the resulting mixed state carries some 't Hooft anomaly.

Using the Choi-Jamiolkowski isomorphism, we map the density matrix of the bottom layer to a pure state in the doubled Hilbert space, where each hinge mode can be described by the Luttinger liquid theory Eq.~\eqref{eq:Luttinger_Choidoubled}. The strong symmetry and the weak subsystem symmetry acting on the 4-component bosonic field $\Phi$ are given by
\begin{subequations}
    \begin{align}
        \delta\Phi^{\mathbb{Z}_{2}^{f L}} &=\pi(1,1,0,0)^{T}\\
        \delta\Phi^{\mathbb{Z}_{2}^{f R}} &=\pi(0,0,1,1)^{T}\\
        \delta\Phi^{\mathbb{Z}_{2}^{f,z}} &=\pi(1,1,1,1)^{T}
    \end{align}
\end{subequations}
and the weak global $\mathbb{Z}_2$ symmetry acts as $\Phi\mapsto\left(\sigma^{z}\right)^{\oplus2}\Phi$. In the presence of these symmetries, the only symmetry-allowed interaction is
\begin{align}
\cos\left(\phi_{1}+\phi_{2}+\phi_{3}+\phi_4\right),
\end{align}
which is not enough to fully gap out the hinge mode in the doubled Hilbert space. In fact, consider the generators of $\Z_2^{fL}$ and $\Z_2$, $P_f^L$ and $g$, its anomaly indicator \cite{Heinrich_2018} is non-vanishing, namely 
\begin{align}
\nu_{P_f^Lg}\equiv1(\mathrm{mod}~2),
\end{align}
which implies that the hinge mode in the doubled Hilbert space carries the mixed anomaly of $\Z_2^{fL}$ and $\Z_2$. Equivalently, the hinge mode in the physical Hilbert space carries the mixed anomaly of strong $\Z_2^f$ and weak $\Z_2$ symmetries.

Notably, this mSPT lacks pure state analog since a pure state with subsystem $\mathbb{Z}_{2}^{f,z}$ symmetry and global $\mathbb{Z}_{2}$ symmetry is either a product state or is anomalous in the bulk. Therefore, the mSPT we obtained in this section from tracing out the top layers of a fermionic SSPT is an intrinsic mSPT. 
}

\subsection{4D SSPT mapping to 3D intrinsic fractonic SPT in mixed state}

\begin{figure}[t]
    \centering
    \includegraphics[width=0.48\textwidth]{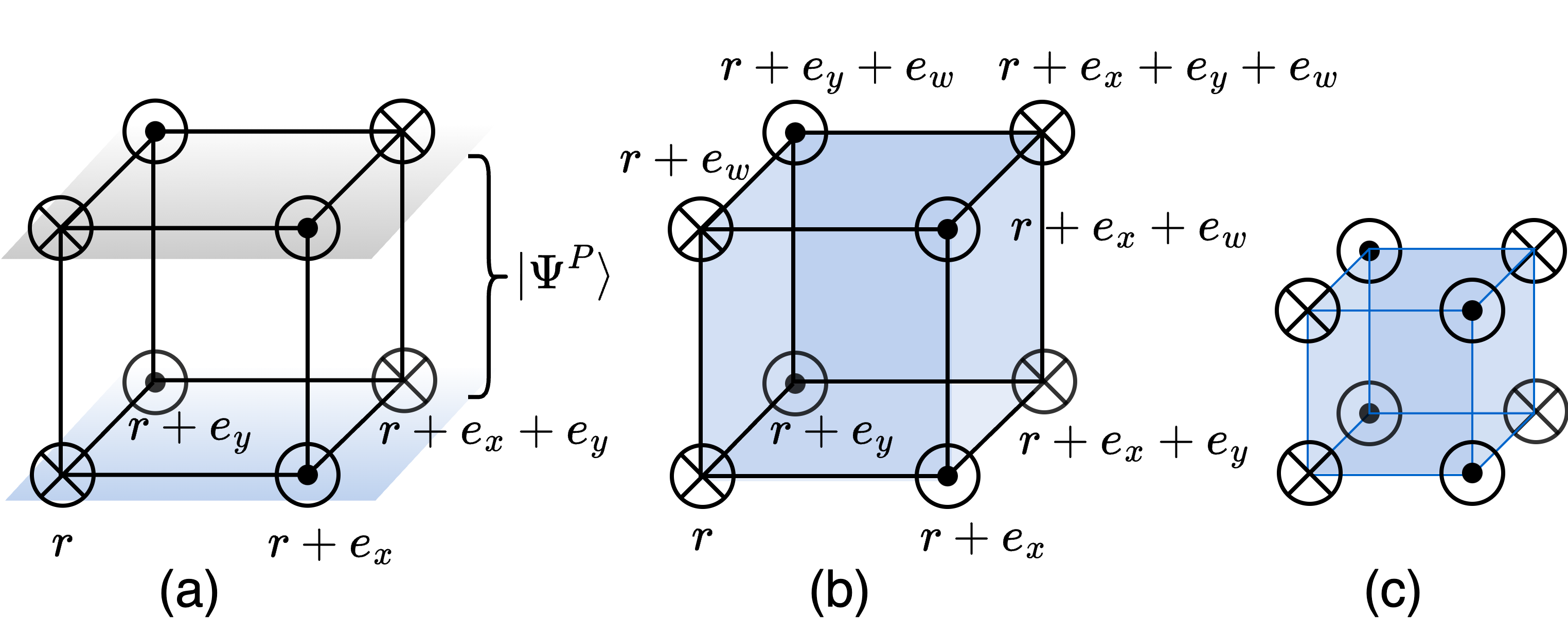}
    \caption{a) The building block of the bulk intersite coupling for 3D mSPT in the doubled Hilbert space. The bottom and top layers correspond to the left and right Hilbert spaces, respectively. b) Building block of the bulk intersite coupling for 4D SSPT. Here, the four spatial directions are labeled by $w,x,y,z$, and the wires moving along the $+z$ and $-z$ directions are represented by $\otimes$ and $\odot$, respectively. c) Each site in the 4D SSPT consists 8 wires.}
    \label{fig:4DSSPT}
\end{figure}

In this section, we present a 4D SSPT as the bulk theory of the 3D intrinsic fractonic mSPT with weak subsystem symmetries $U^{xz}(1)$ and $U^{yz}(1)$  and a strong global $U(1)$ symmetry studied in Ref.~\cite{prelim, Zhang_2023}. 

We start by reviewing the 3D intrinsic fractonic mSPT. Firstly, we examine the pure state analog of this mSPT state and illustrate that a 3D SSPT state cannot be constructed with subsystem symmetries $U^{xz}(1)$ and $U^{yz}(1)$ and a strong global $U(1)$ symmetry. We show this using the coupled wire construction illustrated in Fig.~\ref{fig:coupledwire}a, where each wire now represents a chiral boson field, labeled by $\phi_{L/R}^{1,2}(\mathbf{r})$, with the $L/R$ corresponds to left- and right-moving modes along the $z$-direction and $1,2$ labels the flavor of the bosons. Consider a blue plaquette in Fig.~\ref{fig:coupledwire}a, the overall Lagrangian of these four chiral Bosons are
\begin{equation}
    \mathcal{L} = -\partial_{\tau}\Phi^{T} K \partial_{z}\Phi + \partial_{z}\Phi^{T}V\partial_{z}\Phi,
\label{Eq: Luttinger}
\end{equation}
where $\Phi = (\phi_{L,\boldsymbol{r}}^{1},\phi_{R,\boldsymbol{r}^{\prime}}^{2},\phi_{L,\boldsymbol{r}^{\prime\prime}}^{2},\phi_{R,\boldsymbol{r}^{\prime\prime\prime}}^{1})$ is the 4-component field, and $K=(\sigma^{z})^{\otimes 2}$ is the $K$-matrix. The symmetries act on the boson field as follows:
\begin{subequations}
    \begin{align}
        \delta\Phi^{U(1)^{xz}} &=\alpha(1,1,0,0)^{T}\\
        \delta\Phi^{U(1)^{yz}} &=\beta(1,0,0,1)^{T}\\
        \delta\Phi^{U(1)} &=\theta(1,1,1,1)^{T},
    \end{align}
\end{subequations}
where $\alpha$, $\beta$, and $\theta\in\mathbb{R}$ are arbitrary phase factors. The only symmetry-allowed backscattering term is
\begin{equation}
\cos\left(\phi_{L,\boldsymbol{r}}^{1}-\phi_{R,\boldsymbol{r}^{\prime}}^{2}+\phi_{L,\boldsymbol{r}^{\prime\prime}}^{2}-\phi_{R,\boldsymbol{r}^{\prime\prime\prime}}^{1}\right),
\label{Eq: backscattering}
\end{equation}
which cannot fully gap out the bulk. Therefore, a pure state in 3D with these symmetries is either a product state or anomalous in the bulk. In particular, we elucidate that the anomalous bulk carries the mixed anomaly of $U(1)^{xz}$ and $U(1)^{yz}$. If we break the $U(1)^{xz}$ symmetry, an additional backscattering term could be further introduced to fully gap out the bulk, namely
\begin{align}
\cos\left(\phi_{L,\boldsymbol{r}}^1-\phi_{R,\boldsymbol{r}'''}^2\right).
\end{align}
Similarly, if we break the $U(1)^{yz}$ symmetry, another additional backscattering term could be further introduced to fully gap out the bulk, namely
\begin{align}
\cos\left(\phi_{L,\boldsymbol{r}}^1-\phi_{R,\boldsymbol{r}'''}^1\right).
\end{align}
Nevertheless, we can see that if we break the global $U(1)$ symmetry, the remaining symmetries still forbid all backscattering terms other than Eq. \eqref{Eq: backscattering}. Therefore, we conclude that for the anomalous bulk, each plaquette carries the mixed anomaly of $U(1)^{xz}$ and $U(1)^{yz}$.

Subsequently, we turn to the open quantum systems at which the subsystem symmetries $U(1)^{xz}$ and $U(1)^{yz}$ are weak. An immediate consequence we get is that the above anomalous bulk for pure states becomes anomaly-free because of the relevant symmetries are all weak. 

Using the Choi-Jamiolkowski isomorphism, one can map the density matrix of the mixed state to a pure state in the doubled Hilbert space. The building block of the intersite coupling in the doubled Hilbert space is drawn in Fig.~\ref{fig:4DSSPT}a, where the 8 chiral bosons can be gapped out without breaking the weak subsystem $U(1)^{xz}$ and $U(1)^{yz}$ and the strong $U(1)$ symmetries. One can further include onsite interactions to gap out the surface modes and two of the three remaining hinge modes. This left each hinge with one symmetry-protected chiral boson. Through this construction, one gets a 3D fractonic mSPT without any pure-state analogue.

Using the dimensional extension procedure described in Sec.~\ref{sec:dimext}, we can construct a 4D SSPT by replicating the 3D mSPT wavefunction in the doubled Hilbert space and layer them in the $w$-direction. The constructed 4D SSPT is protected by 3-foliated subsystem symmetries $U(1)^{wxz}$, $U(1)^{wyz}$, and $U(1)^{xyz}$. The former two subsystem symmetries correspond to the weak subsystem symmetries $U(1)^{xz}$ and $U(1)^{yz}$ in the 2-foliated (3+1)D system, whereas the last one corresponds to the strong global $U(1)$ symmetry of the mixed state.  

For the holographic constructed (4+1)D SSPT model, each site includes 8 chiral bosons. In bulk, they are gapped out by intersite couplings as shown in Fig.~\ref{fig:4DSSPT}, which is basically two copies of Eq. \eqref{Eq: Luttinger}, and ought to be anomaly-free. Each site on the (3+1)D surface of the system has 4 remaining chiral bosons, and each site on the (2+1)D hinge has 6 remaining chiral bosons, all of which can be gapped by pairwise onsite couplings. However, each (1+1)D corner has 7 remaining chiral bosons, 6 among which can be gapped out, leaving a dangling chiral boson per hinge that cannot be gapped out without breaking symmetry. 

\section{3d HOTI with dipole symmetry and subsystem charge conservation}\label{sec:dipole}

In this section, we introduce a 3d higher-order topological insulator (HOTI) that conserves the total \(x\)-dipole moment\cite{lam2024topological,lam2024classification,han2024topological}, defined as \(\int x\rho \, dV\), while also maintaining subsystem charge conservation on each \(x\)-\(y\) plane. In this system, both the bulk and side surfaces are fully gapped, while a helical mode is hosted along the hinge in the \(y\)-direction. The gapless hinge mode displays a mixed anomaly between the dipole \(U^d(1)\) symmetry and the subsystem \(U(1)\) symmetry.

\begin{figure}[h!]
\includegraphics[width=0.4\textwidth]{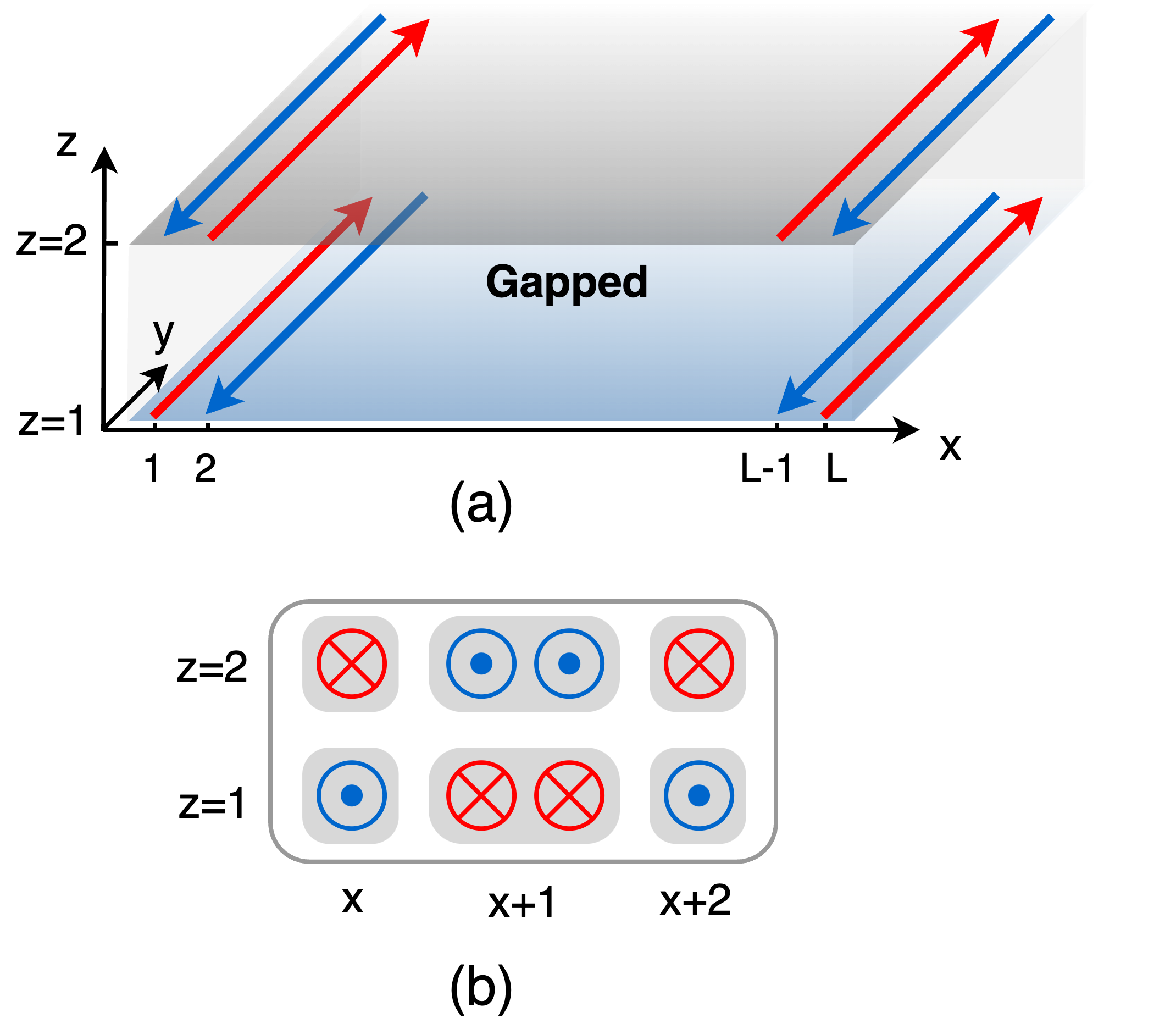}
\caption{Illustration of 3d HOTI with x-dipole symmetry. a) 3d HOTI in the thin slab limit. b) The elementary building blocks. $\otimes$ and $\odot$ represent 1d Dirac fermions along $+y$- and $-y$-directions, respectively.}
\label{qsh}
\end{figure}

A microscopic model for these anomalous hinge states can be constructed using the coupled wire approach. For simplicity, we develop a microscopic Hamiltonian with only two layers along the \(z\)-direction, effectively compactifying the 3d model into a \textit{thin slab limit}. Despite this reduction, the model retains the essential features of a higher-order topological insulator (HOTI) with subsystem and dipole symmetry in 3d. The full 3d model can then be realized by stacking these bilayer slabs.

We begin with a bilayer system extended on the \(x\)-\(y\) plane. Consider an array of 1d modes denoted as \(\phi^a_{L/R}(r)\) and \(\Theta^a_{L/R}(r)\) (\(a=1,2\) for the flavor index) extending along the \(y\)-direction. The \(\phi\)-modes correspond to the bottom layer at \(z=1\), while the \(\Theta\)-modes correspond to the top layer at \(z=2\). The elementary building blocks consist of these modes:
\begin{align} \label{bb}
&\phi^{1}_{L}(x,y), \phi^{1,2}_{R}(x+1,y), \phi^{2}_{L}(x+2,y),\nonumber\\
&\Theta^{1}_{R}(x,y), \Theta^{1,2}_{L}(x+1,y), \Theta^{2}_{R}(x+2,y). \nonumber \end{align} 


The \(x\)-dipole moment conservation \(D_x = \int x \rho (r) \, dV\) acts on these 1d wires as follows:
\begin{align} 
U^d(1): &\phi^a_{L/R}(r) \rightarrow \phi^a_{L/R}(r)+x\alpha,\nonumber\\
&\Theta^a_{L/R}(r) \rightarrow \Theta^a_{L/R}(r)+x\alpha,
\end{align}
The subsystem \(U(1)\) symmetry mandates that the charge on the top and bottom layers be conserved independently, and it operates on the 1d wires as follows:
\begin{align} \label{symmtryqsh}
U^A(1): &\phi^a_{L/R}(r) \rightarrow \phi^a_{L/R}(r)+\beta_1,\nonumber\\
U^B(1):&\Theta^a_{L/R}(r) \rightarrow \Theta^a_{L/R}(r)+\beta_2,
\end{align}
Here, \(U^A(1)\) and \(U^B(1)\) act only on the top (\(\phi\) mode) and bottom (\(\Theta\) mode) layers, respectively. The building block defined in Eq.~\eqref{bb} extends over three adjacent \(x\)-coordinates and includes four gapless boson modes in total. Four independent mass terms are required to gap out all the modes. To preserve the dipole and subsystem \(U(1)\) symmetries, we will couple the wires using quartic inter-wire interactions.
\begin{align}
\mathcal{H} (x) & = g  \cos(\phi^1_{L}(x) \! - \!\phi^2_{R}(x\!+\!1)
\!- \!\phi^1_{R}(x\!+\!1) \!+\! \phi^2_{L}(x\!+\!2))  \nonumber \\ 
&+g \cos( \Theta^1_{R}(x) \!-\! \Theta^2_{L}(x\!+\!1) \!-\! \Theta^2_{L} (x\!+\!1)\!+\!\Theta^1_{R}(x\!+\!2))  \nonumber \\ 
&+g \cos(\Theta^1_{R}(x) \!-\! \Theta^1_{L}(x\!+\!1) \!-\! \phi^1_{L}(x) \!+\! \phi^1_{R}(x\!+\!1)) \nonumber \\ 
&+g \cos(\Theta^1_{R}(x)\!-\!\Theta^2_{L}(x\!+\!1)\!-\!\phi^1_{L}(x)\!+\!\phi^2_{R}(x\!+\!1))
\label{eq:BosonLagrangian3}
\end{align}
The \(y\)-dependence of the fields is omitted for simplicity of notation. All cosine terms in Eq.~\eqref{eq:BosonLagrangian3} are independent and commute with each other. At sufficiently strong coupling \(g\), these terms can generate four independent mass terms, resulting in a fully gapped building block. As a result, the bulk degrees of freedom are gapped.

At the left boundary, the following chiral modes from the top and bottom layers remain gapless:
$$\phi^{1}_{L} (1),~ \phi^{1}_{R} (2),~ \Theta^{1}_{R} (1),~ \Theta^{1}_{L} (2).$$
To further gap out certain degrees of freedom through edge reconstruction, symmetry-allowed terms can be introduced to couple the edge modes as:
\begin{align}\label{interedge}
V =   v \cos(\Theta_{R} (1) - \Theta_{L}(2) - \phi_{L}(1) + \phi_{R}(2)) 
=v\cos(\Phi^v)
\end{align}
The flavor index has been dropped without loss of generality. Subsequently, only the following modes at the edge remain gapless:
\begin{align}
&2\Phi =-\Theta_{R}(1)+\Theta_{L}(2) -\phi_{L}(1) +\phi_{R}(2),\nonumber\\
&2\Tilde{\Phi} =-\Theta_{R}(1) -\Theta_{L}(2)+\phi_{L}(1) +\phi_{R}(2)
\end{align}
One can show $[\Phi (y) , \tilde{\Phi} (y') ] = i \pi  {\rm sgn} (y-y')$. The edge theory of this bilayer system can be described as a helical Luttinger liquid:
\begin{align}
&\mathcal{L} = \partial_t \Phi \partial_y \Tilde{\Phi} +\frac{v_1}{2} (\partial_y \Tilde{\Phi})^2+ \frac{v_2}{2} (\partial_y \Phi)^2
\label{qshedge}
\end{align}
Based on the symmetry assignment in Eq.~\eqref{symmtryqsh}, the $\Phi$ is charge neutral and transforms under dipole symmetry $U^{d}(1)$ as:
 \begin{align}
U^{d}(1): \Phi \rightarrow \Phi+ \alpha
\end{align}
while the $\Tilde{\Phi}$ is dipole neutral, it carries subsystem \(U(1)\) charge for both layers. This symmetry assignment on the edge theory reveals a mixed anomaly between the dipole \(U^d(1)\) and subsystem \(U(1)\) (in the bilayer setting, it is the charge difference between the two layers), which consequently prevents the gapping out of the helical mode in Eq.~\eqref{qshedge}. By introducing a subsystem \(U(1)\) flux insertion (on either the first or second layer only), the system facilitates the transfer of a dipole moment from the left to the right boundary.

This mixed anomaly at the boundary can be observed through the correlation function of operators that carry either dipole or subsystem \(U(1)\) charge on the boundaries of the bilayer system:
 \begin{align}\label{coredge}
\langle e^{i \Phi(y)} e^{-i \Phi(y')} \rangle=\frac{1}{(y-y')^{K}},\nonumber\\
\langle e^{i \Tilde{\Phi}(y)} e^{-i \Tilde{\Phi}(y')}\rangle=\frac{1}{(y-y')^{\frac{2\pi}{K}}} 
\end{align}
Both of these correlation functions exhibit quasi-long-range order, akin to the behavior observed in a 1d gapless system with a `t Hooft anomaly.

\subsection{Tracing out the ancillae to obtain mSPT}

If we consider the qubits in the top layer as ancilla degrees of freedom from the environment and the bottom layer as the system of interest, tracing out the top layer yields a mixed-state density matrix, denoted as \(\rho\). This mixed-state ensemble exhibits a strong \(U(1)\) symmetry since the purified wave function of the bilayer conserves charge in each layer. Additionally, it supports a weak dipole symmetry \(U^d(1)\), as the \(x\)-dipole moment can fluctuate between layers. The density matrix \(\rho\) characterizes an intrinsic mSPT, as originally proposed in Ref.~\cite{prelim}, and is protected by both strong \(U(1)\) and weak dipole symmetries. Introducing a dipole flux into the system, analogous to adding twisted boundary conditions, via a gauge potential \(A_y = \frac{2\pi x}{L_y}\), induces charge pumping with respect to the strong \(U(1)\) symmetry between the edges, rendering the boundary of the mixed state anomalous.

How can we detect such a mixed anomaly? In the purified state, where the system and the ancilla layer together form the ground state of a local Hamiltonian as described in Eq.~\eqref{eq:BosonLagrangian3}, one can measure the correlation of the charged operator, as shown in Eq.~\eqref{coredge}, to demonstrate that the boundary exhibits (quasi) long-range correlation. A challenge arises because the charged operators \(e^{i \Tilde{\Phi}}\) (and \(e^{i\Phi}\)) are bound states involving both the system's qubits and the ancilla. With the ancilla inaccessible, tracing them out results in the loss of information from these operators. This raises the question: How can we identify the mixed state anomaly by measuring only the system's qubits?

We start by examining the correlation of the operator \(e^{i \Phi^d} = e^{i(\Theta_{R}(1) - \Theta_{L}(2))}\) near the boundary of the system layer. This operator is neutral under the strong \(U(1)\) symmetry but carries a charge under the weak dipole symmetry.
 \begin{align}\label{coredgemixed}
\langle e^{i \Phi^d (y)} e^{-i \Phi^d(y')} \rangle \sim \langle e^{i \Phi (y)} e^{-i \Phi(y')} \rangle \langle e^{-i \Phi^v (y)} e^{i \Phi^v(y')} \rangle
\end{align}
\(\Phi^v\) is the operator defined in Eq.~\eqref{interedge}, which is pinned by the edge potential term \(\cos{\Phi^v}\). Thus, we can assume \(\Phi^v = 0\), and the correlator reduces to:
\begin{align}\label{coredgemixed}
\langle e^{i \Phi^d (y)} e^{-i \Phi^d(y')} \rangle=c\langle e^{i \Phi (y)} e^{-i \Phi(y')} \rangle  \sim  \frac{1}{r^K}
\end{align}
Here, \(\langle \rangle\) denotes the average value taken over the mixed state density matrix, represented as \(\Tr [e^{i \Phi^d (y)} e^{-i \Phi^d(y')} \rho]\).

Now, let us consider the operator \(e^{i \Phi^e} = e^{i(\Theta_{R}(1) + \Theta_{L}(2))}\), which is charged under the strong \(U(1)\) symmetry but remains neutral under the weak dipole symmetry. As noted in Ref.~\cite{prelim,xu2024average}, this operator exhibits short-range correlation. However, when we examine the Rényi-2 correlator of \(e^{i \Phi^e}\), it exhibits quasi-long-range order.
 \begin{align}\label{Renyi-2correlator}
\frac{\Tr [e^{i \Phi^e (y)} e^{-i \Phi^e(y')}\rho e^{-i \Phi^e (y)} e^{i \Phi^e(y')}\rho ]}{\Tr [\rho^2]}\sim 1/r^a
\end{align}
Note that the operator \(e^{i \Phi^e}\) acts on both the ket and bra spaces of the density matrix \(\rho\), making it charged under the strong \(U(1)\) symmetry while remaining neutral under the weak dipole symmetry. As elucidated in Ref.~\cite{sala2024spontaneous,prelim}, this operator is equivalent to doubling the purified wave function of the bilayer system, projecting the two ancillas onto a symmetric EPR pair, and then measuring the four-point correlator (acting on the first and second copies) of the post-projection wave function.

Finally, we want to emphasize that the mixed anomaly between dipole and charge symmetry in the mixed state \(\rho\) is unique to 2d open systems and has no counterpart in pure states. In a 2d pure state with both \(x\)-dipole moment and charge conservation, it is impossible to manifest a mixed anomaly between dipole and charge at the boundary. The reasoning is as follows: if the edge exhibited a mixed anomaly where a dipole flux insertion altered the charge density at the boundary, it would trigger charge pumping between the left and right boundaries. However, this charge pumping would inevitably change the bulk dipole moment, implying that a global dipole flux insertion would modify the total dipole moment in 2d. 
This scenario would make the 2d theory anomalous and incompatible with a lattice model with local interactions. This phenomenon is known as the hierarchy structure of anomaly in multipole-conserved systems \cite{lam2024topological,prelim,sala2024exotic}, where a charge anomaly at the boundary can induce a change in the charge multipole moment in the bulk, which makes the bulk anomalous.
In an open system with weak dipole symmetry, the dipole anomaly in the bulk, acting on both the ket and bra spaces, would cancel out, leaving the mixed-state density matrix anomaly-free. The absence of a mixed anomaly in pure states also has significant implications for the 3d HOTI described in Eq.~\eqref{eq:BosonLagrangian3}. While the surface of the 3d HOTI is gapped, with gapless modes localized on the hinge, such a surface state cannot be realized in a pure 2d theory under the same symmetry constraints.

\subsection{Effective theory}

In this section, we examine the implications of our holographic duality from SSPT to mSPT from the perspective of effective field theory. As a concrete example, we focus on a 3d higher-order topological insulator (HOTI) with both dipole symmetry and subsystem charge (see Appendix~\ref{sec:dipole}); however, the same approach can be extended to a broader class of systems in future studies.

In Appendix~\ref{sec:dipole}, we demonstrate that the surface state of this 3D HOTI, which conserves a total \(x\)-dipole moment \cite{lam2024topological,lam2024classification,han2024topological}, is dual to an intrinsic mSPT phase featuring a strong \(U(1)\) symmetry and a weak dipole symmetry. The effective theory of the 3D HOTI can be analyzed by compactifying the z-direction, taking \(L_z = 2\). Crucially, the anomaly in this HOTI arises from the interplay between the subsystem \(U(1)\) symmetry and a global dipole symmetry so this compact limit is fairly enough to capture the underlying physics. To make this explicit, we couple the theory to background gauge fields for the dipole symmetry as follows:
\begin{align}
    (A_t,A_{xx},A_y)\rightarrow (A_t \!+\! \partial_t\lambda,A_{xx}\! +\! \partial_x^2\lambda, A_y \!+\!\partial_y\lambda)
\end{align}
as well as the subsystem charge (that only acts on the top layer)
\begin{align}
    (a_t,a_x,a_y)\rightarrow (a_t +\partial_t\chi,a_x+\partial_x\chi,a_y+\partial_y\chi).
\end{align}
Here $\chi(x,y,0)$ is a function that is localized on the top layer.
The response theory can be expressed as:
\begin{align}\label{mcs}
    \mathcal{L}=&\,\frac{l_0}{2\pi}a_t (\partial_x^2 A_y-\partial_y A_{xx})\nonumber
    +\frac{l_0}{2\pi}a_x\partial_x(\partial_y A_{t}-\partial_t A_y)\nonumber
    \\
    &\,+\frac{l_0}{2\pi}a_y(\partial_t A_{xx}-\partial_x^2 A_t),
\end{align}
where \( l_0 \) denotes the lattice spacing along the \( x \)-direction. The system can be reformulated as a mutual Chern-Simons theory involving the subsystem gauge fields \( a_\mu = (a_t, a_x, a_y) \) and the dipole gauge fields \( \mathcal{A}_\mu = (l_0 \partial_x A_t, l_0 A_{xx}, l_0 \partial_x A_y) \). This formulation closely resembles the dipole spin Hall theory proposed in Ref.~\cite{lam2024topological}. 

This effective theory suggests that a \( U(1) \) Hall current localized on the top layer can be induced by a dipole electric field in the transverse direction.
\begin{align}
    J_y(x,y,0)=\frac{\delta\mathcal{L}}{\delta a_y}=\frac{l_0}{2\pi}(\partial_t A_{xx}-\partial_x^2 A_t)
\end{align}
This result is consistent with the fact that the SSPT exhibits a mixed anomaly between the subsystem charge and dipole moments at the hinge. A gauge transformation of the dipole symmetry leads to the pumping of a subsystem-symmetric charge (on the top layer) between the two hinges.  

Integrating out the top surface yields the mSPT, whose effective theory is described by a mutual Chern-Simons theory defined on the Choi double space.
\begin{align}
[\rho]&=\int D\bm{a^L} ~ D\bm{a^R}~D\bm{A}~\exp(-\mathcal{L}_{ket}+\mathcal{L}_{bra})\nonumber\\
    \mathcal{L}_{ket}=&\,\frac{l_0}{2\pi}a^L_t (\partial_x^2 A_y-\partial_y A_{xx})\nonumber
    +\frac{l_0}{2\pi}a^L_x\partial_x(\partial_y A_{t}-\partial_t A_y)\nonumber\\
    &\,+\frac{l_0}{2\pi}a^L_y(\partial_t A_{xx}-\partial_x^2 A_t)\nonumber\\
       \mathcal{L}_{bra}=&\,\frac{l_0}{2\pi}a^R_t (\partial_x^2 A_y-\partial_y A_{xx})\nonumber
    +\frac{l_0}{2\pi}a^R_x\partial_x(\partial_y A_{t}-\partial_t A_y)\nonumber
    \\
    &\,+\frac{l_0}{2\pi}a^R_y(\partial_t A_{xx}-\partial_x^2 A_t),
\end{align}
Here, \( a^{L/R} \) arises from the strong \( U(1) \) symmetry, which acts independently in the ket and bra spaces.

\bibliography{Refs.bib}

\providecommand{\noopsort}[1]{}\providecommand{\singleletter}[1]{#1}%
\begin{thebibliography}{107}%
\makeatletter
\providecommand \@ifxundefined [1]{%
 \@ifx{#1\undefined}
}%
\providecommand \@ifnum [1]{%
 \ifnum #1\expandafter \@firstoftwo
 \else \expandafter \@secondoftwo
 \fi
}%
\providecommand \@ifx [1]{%
 \ifx #1\expandafter \@firstoftwo
 \else \expandafter \@secondoftwo
 \fi
}%
\providecommand \natexlab [1]{#1}%
\providecommand \enquote  [1]{``#1''}%
\providecommand \bibnamefont  [1]{#1}%
\providecommand \bibfnamefont [1]{#1}%
\providecommand \citenamefont [1]{#1}%
\providecommand \href@noop [0]{\@secondoftwo}%
\providecommand \href [0]{\begingroup \@sanitize@url \@href}%
\providecommand \@href[1]{\@@startlink{#1}\@@href}%
\providecommand \@@href[1]{\endgroup#1\@@endlink}%
\providecommand \@sanitize@url [0]{\catcode `\\12\catcode `\$12\catcode
  `\&12\catcode `\#12\catcode `\^12\catcode `\_12\catcode `\%12\relax}%
\providecommand \@@startlink[1]{}%
\providecommand \@@endlink[0]{}%
\providecommand \url  [0]{\begingroup\@sanitize@url \@url }%
\providecommand \@url [1]{\endgroup\@href {#1}{\urlprefix }}%
\providecommand \urlprefix  [0]{URL }%
\providecommand \Eprint [0]{\href }%
\providecommand \doibase [0]{http://dx.doi.org/}%
\providecommand \selectlanguage [0]{\@gobble}%
\providecommand \bibinfo  [0]{\@secondoftwo}%
\providecommand \bibfield  [0]{\@secondoftwo}%
\providecommand \translation [1]{[#1]}%
\providecommand \BibitemOpen [0]{}%
\providecommand \bibitemStop [0]{}%
\providecommand \bibitemNoStop [0]{.\EOS\space}%
\providecommand \EOS [0]{\spacefactor3000\relax}%
\providecommand \BibitemShut  [1]{\csname bibitem#1\endcsname}%
\let\auto@bib@innerbib\@empty
\bibitem [{\citenamefont {Chen}\ \emph {et~al.}(2012)\citenamefont {Chen},
  \citenamefont {Gu}, \citenamefont {Liu},\ and\ \citenamefont
  {Wen}}]{XieChenScience}%
  \BibitemOpen
  \bibfield  {author} {\bibinfo {author} {\bibfnamefont {X.}~\bibnamefont
  {Chen}}, \bibinfo {author} {\bibfnamefont {Z.-C.}\ \bibnamefont {Gu}},
  \bibinfo {author} {\bibfnamefont {Z.-X.}\ \bibnamefont {Liu}}, \ and\
  \bibinfo {author} {\bibfnamefont {X.-G.}\ \bibnamefont {Wen}},\ }\href
  {https://science.sciencemag.org/content/338/6114/1604} {\bibfield  {journal}
  {\bibinfo  {journal} {Science}\ }\textbf {\bibinfo {volume} {338}},\ \bibinfo
  {pages} {1604} (\bibinfo {year} {2012})}\BibitemShut {NoStop}%
\bibitem [{\citenamefont {Chen}\ \emph
  {et~al.}(2013{\natexlab{a}})\citenamefont {Chen}, \citenamefont {Gu},
  \citenamefont {Liu},\ and\ \citenamefont {Wen}}]{Chen:2011pg}%
  \BibitemOpen
  \bibfield  {author} {\bibinfo {author} {\bibfnamefont {X.}~\bibnamefont
  {Chen}}, \bibinfo {author} {\bibfnamefont {Z.-C.}\ \bibnamefont {Gu}},
  \bibinfo {author} {\bibfnamefont {Z.-X.}\ \bibnamefont {Liu}}, \ and\
  \bibinfo {author} {\bibfnamefont {X.-G.}\ \bibnamefont {Wen}},\ }\href
  {\doibase 10.1103/PhysRevB.87.155114} {\bibfield  {journal} {\bibinfo
  {journal} {Phys. Rev. B}\ }\textbf {\bibinfo {volume} {87}},\ \bibinfo
  {pages} {155114} (\bibinfo {year} {2013}{\natexlab{a}})},\ \Eprint
  {http://arxiv.org/abs/1106.4772} {arXiv:1106.4772 [cond-mat.str-el]}
  \BibitemShut {NoStop}%
\bibitem [{\citenamefont {Chen}\ \emph
  {et~al.}(2013{\natexlab{b}})\citenamefont {Chen}, \citenamefont {Gu},
  \citenamefont {Liu},\ and\ \citenamefont {Wen}}]{cohomology}%
  \BibitemOpen
  \bibfield  {author} {\bibinfo {author} {\bibfnamefont {X.}~\bibnamefont
  {Chen}}, \bibinfo {author} {\bibfnamefont {Z.-C.}\ \bibnamefont {Gu}},
  \bibinfo {author} {\bibfnamefont {Z.-X.}\ \bibnamefont {Liu}}, \ and\
  \bibinfo {author} {\bibfnamefont {X.-G.}\ \bibnamefont {Wen}},\ }\href
  {https://journals.aps.org/prb/abstract/10.1103/PhysRevB.87.155114} {\bibfield
   {journal} {\bibinfo  {journal} {Phys. Rev. B}\ }\textbf {\bibinfo {volume}
  {87}},\ \bibinfo {pages} {155114} (\bibinfo {year}
  {2013}{\natexlab{b}})}\BibitemShut {NoStop}%
\bibitem [{\citenamefont {Gu}\ and\ \citenamefont {Wen}(2009)}]{ZCGu2009}%
  \BibitemOpen
  \bibfield  {author} {\bibinfo {author} {\bibfnamefont {Z.-C.}\ \bibnamefont
  {Gu}}\ and\ \bibinfo {author} {\bibfnamefont {X.-G.}\ \bibnamefont {Wen}},\
  }\href {https://journals.aps.org/prb/abstract/10.1103/PhysRevB.80.155131}
  {\bibfield  {journal} {\bibinfo  {journal} {Phys. Rev. B}\ }\textbf {\bibinfo
  {volume} {80}},\ \bibinfo {pages} {155131} (\bibinfo {year}
  {2009})}\BibitemShut {NoStop}%
\bibitem [{\citenamefont {Levin}\ and\ \citenamefont {Gu}(2012)}]{LevinGu}%
  \BibitemOpen
  \bibfield  {author} {\bibinfo {author} {\bibfnamefont {M.}~\bibnamefont
  {Levin}}\ and\ \bibinfo {author} {\bibfnamefont {Z.-C.}\ \bibnamefont {Gu}},\
  }\href {https://journals.aps.org/prb/abstract/10.1103/PhysRevB.86.115109}
  {\bibfield  {journal} {\bibinfo  {journal} {Phys. Rev. B}\ }\textbf {\bibinfo
  {volume} {86}},\ \bibinfo {pages} {115109} (\bibinfo {year}
  {2012})}\BibitemShut {NoStop}%
\bibitem [{\citenamefont {Vishwanath}\ and\ \citenamefont
  {Senthil}(2013)}]{Ashvin2013}%
  \BibitemOpen
  \bibfield  {author} {\bibinfo {author} {\bibfnamefont {A.}~\bibnamefont
  {Vishwanath}}\ and\ \bibinfo {author} {\bibfnamefont {T.}~\bibnamefont
  {Senthil}},\ }\href
  {https://journals.aps.org/prx/abstract/10.1103/PhysRevX.3.011016} {\bibfield
  {journal} {\bibinfo  {journal} {Phys. Rev. X}\ }\textbf {\bibinfo {volume}
  {3}},\ \bibinfo {pages} {011016} (\bibinfo {year} {2013})}\BibitemShut
  {NoStop}%
\bibitem [{\citenamefont {Chen}\ \emph {et~al.}(2014)\citenamefont {Chen},
  \citenamefont {Fidkowski},\ and\ \citenamefont {Vishwanath}}]{XieChen2014}%
  \BibitemOpen
  \bibfield  {author} {\bibinfo {author} {\bibfnamefont {X.}~\bibnamefont
  {Chen}}, \bibinfo {author} {\bibfnamefont {L.}~\bibnamefont {Fidkowski}}, \
  and\ \bibinfo {author} {\bibfnamefont {A.}~\bibnamefont {Vishwanath}},\
  }\href {https://journals.aps.org/prb/abstract/10.1103/PhysRevB.89.165132}
  {\bibfield  {journal} {\bibinfo  {journal} {Phys. Rev. B}\ }\textbf {\bibinfo
  {volume} {89}},\ \bibinfo {pages} {165132} (\bibinfo {year}
  {2014})}\BibitemShut {NoStop}%
\bibitem [{\citenamefont {Kapustin}\ \emph {et~al.}(2015)\citenamefont
  {Kapustin}, \citenamefont {Thorngren}, \citenamefont {Turzillo},\ and\
  \citenamefont {Wang}}]{Kapustin2015}%
  \BibitemOpen
  \bibfield  {author} {\bibinfo {author} {\bibfnamefont {A.}~\bibnamefont
  {Kapustin}}, \bibinfo {author} {\bibfnamefont {R.}~\bibnamefont {Thorngren}},
  \bibinfo {author} {\bibfnamefont {A.}~\bibnamefont {Turzillo}}, \ and\
  \bibinfo {author} {\bibfnamefont {Z.}~\bibnamefont {Wang}},\ }\href {\doibase
  10.1007/jhep12(2015)052} {\bibfield  {journal} {\bibinfo  {journal} {Journal
  of High Energy Physics}\ }\textbf {\bibinfo {volume} {2015}},\ \bibinfo
  {pages} {1–21} (\bibinfo {year} {2015})}\BibitemShut {NoStop}%
\bibitem [{\citenamefont {Senthil}(2015)}]{Senthil_2015}%
  \BibitemOpen
  \bibfield  {author} {\bibinfo {author} {\bibfnamefont {T.}~\bibnamefont
  {Senthil}},\ }\href {\doibase 10.1146/annurev-conmatphys-031214-014740}
  {\bibfield  {journal} {\bibinfo  {journal} {Annu. Rev. Condens. Matter
  Phys.}\ }\textbf {\bibinfo {volume} {6}},\ \bibinfo {pages} {299} (\bibinfo
  {year} {2015})}\BibitemShut {NoStop}%
\bibitem [{\citenamefont {de~Groot}\ \emph {et~al.}(2022)\citenamefont
  {de~Groot}, \citenamefont {Turzillo},\ and\ \citenamefont
  {Schuch}}]{deGroot2022}%
  \BibitemOpen
  \bibfield  {author} {\bibinfo {author} {\bibfnamefont {C.}~\bibnamefont
  {de~Groot}}, \bibinfo {author} {\bibfnamefont {A.}~\bibnamefont {Turzillo}},
  \ and\ \bibinfo {author} {\bibfnamefont {N.}~\bibnamefont {Schuch}},\ }\href
  {\doibase 10.22331/q-2022-11-10-856} {\bibfield  {journal} {\bibinfo
  {journal} {Quantum}\ }\textbf {\bibinfo {volume} {6}},\ \bibinfo {pages}
  {856} (\bibinfo {year} {2022})}\BibitemShut {NoStop}%
\bibitem [{\citenamefont {Ma}\ and\ \citenamefont {Wang}(2023)}]{MaWangASPT}%
  \BibitemOpen
  \bibfield  {author} {\bibinfo {author} {\bibfnamefont {R.}~\bibnamefont
  {Ma}}\ and\ \bibinfo {author} {\bibfnamefont {C.}~\bibnamefont {Wang}},\
  }\href {\doibase 10.1103/physrevx.13.031016} {\bibfield  {journal} {\bibinfo
  {journal} {Physical Review X}\ }\textbf {\bibinfo {volume} {13}} (\bibinfo
  {year} {2023}),\ 10.1103/physrevx.13.031016}\BibitemShut {NoStop}%
\bibitem [{\citenamefont {Lee}\ \emph {et~al.}(2023{\natexlab{a}})\citenamefont
  {Lee}, \citenamefont {Jian},\ and\ \citenamefont {Xu}}]{LeeYouXu2022}%
  \BibitemOpen
  \bibfield  {author} {\bibinfo {author} {\bibfnamefont {J.~Y.}\ \bibnamefont
  {Lee}}, \bibinfo {author} {\bibfnamefont {C.-M.}\ \bibnamefont {Jian}}, \
  and\ \bibinfo {author} {\bibfnamefont {C.}~\bibnamefont {Xu}},\ }\href@noop
  {} {\bibfield  {journal} {\bibinfo  {journal} {PRX Quantum}\ }\textbf
  {\bibinfo {volume} {4}},\ \bibinfo {pages} {030317} (\bibinfo {year}
  {2023}{\natexlab{a}})}\BibitemShut {NoStop}%
\bibitem [{\citenamefont {{Zhang}}\ \emph {et~al.}(2022)\citenamefont
  {{Zhang}}, \citenamefont {{Qi}},\ and\ \citenamefont {{Bi}}}]{ZhangQiBi2022}%
  \BibitemOpen
  \bibfield  {author} {\bibinfo {author} {\bibfnamefont {J.-H.}\ \bibnamefont
  {{Zhang}}}, \bibinfo {author} {\bibfnamefont {Y.}~\bibnamefont {{Qi}}}, \
  and\ \bibinfo {author} {\bibfnamefont {Z.}~\bibnamefont {{Bi}}},\ }\href
  {\doibase 10.48550/arXiv.2210.17485} {\bibfield  {journal} {\bibinfo
  {journal} {arXiv e-prints}\ ,\ \bibinfo {eid} {arXiv:2210.17485}} (\bibinfo
  {year} {2022})},\ \Eprint {http://arxiv.org/abs/2210.17485} {arXiv:2210.17485
  [cond-mat.str-el]} \BibitemShut {NoStop}%
\bibitem [{\citenamefont {Ma}\ \emph {et~al.}(2024)\citenamefont {Ma},
  \citenamefont {Zhang}, \citenamefont {Bi}, \citenamefont {Cheng},\ and\
  \citenamefont {Wang}}]{ma2024topological}%
  \BibitemOpen
  \bibfield  {author} {\bibinfo {author} {\bibfnamefont {R.}~\bibnamefont
  {Ma}}, \bibinfo {author} {\bibfnamefont {J.-H.}\ \bibnamefont {Zhang}},
  \bibinfo {author} {\bibfnamefont {Z.}~\bibnamefont {Bi}}, \bibinfo {author}
  {\bibfnamefont {M.}~\bibnamefont {Cheng}}, \ and\ \bibinfo {author}
  {\bibfnamefont {C.}~\bibnamefont {Wang}},\ }\href
  {https://arxiv.org/abs/2305.16399} {\enquote {\bibinfo {title} {Topological
  phases with average symmetries: the decohered, the disordered, and the
  intrinsic},}\ } (\bibinfo {year} {2024}),\ \Eprint
  {http://arxiv.org/abs/2305.16399} {arXiv:2305.16399 [cond-mat.str-el]}
  \BibitemShut {NoStop}%
\bibitem [{\citenamefont {Zhang}\ \emph
  {et~al.}(2023{\natexlab{a}})\citenamefont {Zhang}, \citenamefont {Ding},
  \citenamefont {Yang},\ and\ \citenamefont {Bi}}]{Zhang_2023}%
  \BibitemOpen
  \bibfield  {author} {\bibinfo {author} {\bibfnamefont {J.-H.}\ \bibnamefont
  {Zhang}}, \bibinfo {author} {\bibfnamefont {K.}~\bibnamefont {Ding}},
  \bibinfo {author} {\bibfnamefont {S.}~\bibnamefont {Yang}}, \ and\ \bibinfo
  {author} {\bibfnamefont {Z.}~\bibnamefont {Bi}},\ }\href {\doibase
  10.1103/PhysRevB.108.155123} {\bibfield  {journal} {\bibinfo  {journal}
  {Phys. Rev. B}\ }\textbf {\bibinfo {volume} {108}},\ \bibinfo {pages}
  {155123} (\bibinfo {year} {2023}{\natexlab{a}})}\BibitemShut {NoStop}%
\bibitem [{\citenamefont {Ma}\ and\ \citenamefont
  {Turzillo}(2024)}]{ma2024symmetry}%
  \BibitemOpen
  \bibfield  {author} {\bibinfo {author} {\bibfnamefont {R.}~\bibnamefont
  {Ma}}\ and\ \bibinfo {author} {\bibfnamefont {A.}~\bibnamefont {Turzillo}},\
  }\href {https://arxiv.org/abs/2403.13280} {\enquote {\bibinfo {title}
  {Symmetry protected topological phases of mixed states in the doubled
  space},}\ } (\bibinfo {year} {2024}),\ \Eprint
  {http://arxiv.org/abs/2403.13280} {arXiv:2403.13280 [quant-ph]} \BibitemShut
  {NoStop}%
\bibitem [{\citenamefont {Xue}\ \emph {et~al.}(2024)\citenamefont {Xue},
  \citenamefont {Lee},\ and\ \citenamefont {Bao}}]{xue2024tensor}%
  \BibitemOpen
  \bibfield  {author} {\bibinfo {author} {\bibfnamefont {H.}~\bibnamefont
  {Xue}}, \bibinfo {author} {\bibfnamefont {J.~Y.}\ \bibnamefont {Lee}}, \ and\
  \bibinfo {author} {\bibfnamefont {Y.}~\bibnamefont {Bao}},\ }\href
  {https://arxiv.org/abs/2403.17069} {\enquote {\bibinfo {title} {Tensor
  network formulation of symmetry protected topological phases in mixed
  states},}\ } (\bibinfo {year} {2024}),\ \Eprint
  {http://arxiv.org/abs/2403.17069} {arXiv:2403.17069 [cond-mat.str-el]}
  \BibitemShut {NoStop}%
\bibitem [{\citenamefont {Guo}\ \emph {et~al.}(2024{\natexlab{a}})\citenamefont
  {Guo}, \citenamefont {Zhang}, \citenamefont {Zhang}, \citenamefont {Yang},\
  and\ \citenamefont {Bi}}]{guo2024locally}%
  \BibitemOpen
  \bibfield  {author} {\bibinfo {author} {\bibfnamefont {Y.}~\bibnamefont
  {Guo}}, \bibinfo {author} {\bibfnamefont {J.-H.}\ \bibnamefont {Zhang}},
  \bibinfo {author} {\bibfnamefont {H.-R.}\ \bibnamefont {Zhang}}, \bibinfo
  {author} {\bibfnamefont {S.}~\bibnamefont {Yang}}, \ and\ \bibinfo {author}
  {\bibfnamefont {Z.}~\bibnamefont {Bi}},\ }\href@noop {} {\enquote {\bibinfo
  {title} {Locally purified density operators for symmetry-protected
  topological phases in mixed states},}\ } (\bibinfo {year}
  {2024}{\natexlab{a}}),\ \Eprint {http://arxiv.org/abs/2403.16978}
  {arXiv:2403.16978 [cond-mat.str-el]} \BibitemShut {NoStop}%
\bibitem [{\citenamefont {Chen}\ and\ \citenamefont
  {Grover}(2024{\natexlab{a}})}]{chen2024separability}%
  \BibitemOpen
  \bibfield  {author} {\bibinfo {author} {\bibfnamefont {Y.-H.}\ \bibnamefont
  {Chen}}\ and\ \bibinfo {author} {\bibfnamefont {T.}~\bibnamefont {Grover}},\
  }\href@noop {} {\enquote {\bibinfo {title} {Separability transitions in
  topological states induced by local decoherence},}\ } (\bibinfo {year}
  {2024}{\natexlab{a}}),\ \Eprint {http://arxiv.org/abs/2309.11879}
  {arXiv:2309.11879 [quant-ph]} \BibitemShut {NoStop}%
\bibitem [{\citenamefont {Chen}\ and\ \citenamefont
  {Grover}(2024{\natexlab{b}})}]{chen2024unconventional}%
  \BibitemOpen
  \bibfield  {author} {\bibinfo {author} {\bibfnamefont {Y.-H.}\ \bibnamefont
  {Chen}}\ and\ \bibinfo {author} {\bibfnamefont {T.}~\bibnamefont {Grover}},\
  }\href@noop {} {\enquote {\bibinfo {title} {Unconventional topological
  mixed-state transition and critical phase induced by self-dual coherent
  errors},}\ } (\bibinfo {year} {2024}{\natexlab{b}}),\ \Eprint
  {http://arxiv.org/abs/2403.06553} {arXiv:2403.06553 [quant-ph]} \BibitemShut
  {NoStop}%
\bibitem [{\citenamefont {Chen}\ and\ \citenamefont
  {Grover}(2023{\natexlab{a}})}]{chen2023symmetryenforced}%
  \BibitemOpen
  \bibfield  {author} {\bibinfo {author} {\bibfnamefont {Y.-H.}\ \bibnamefont
  {Chen}}\ and\ \bibinfo {author} {\bibfnamefont {T.}~\bibnamefont {Grover}},\
  }\href@noop {} {\enquote {\bibinfo {title} {Symmetry-enforced many-body
  separability transitions},}\ } (\bibinfo {year} {2023}{\natexlab{a}}),\
  \Eprint {http://arxiv.org/abs/2310.07286} {arXiv:2310.07286 [quant-ph]}
  \BibitemShut {NoStop}%
\bibitem [{\citenamefont {Lessa}\ \emph
  {et~al.}(2024{\natexlab{a}})\citenamefont {Lessa}, \citenamefont {Cheng},\
  and\ \citenamefont {Wang}}]{lessa2024mixedstate}%
  \BibitemOpen
  \bibfield  {author} {\bibinfo {author} {\bibfnamefont {L.~A.}\ \bibnamefont
  {Lessa}}, \bibinfo {author} {\bibfnamefont {M.}~\bibnamefont {Cheng}}, \ and\
  \bibinfo {author} {\bibfnamefont {C.}~\bibnamefont {Wang}},\ }\href@noop {}
  {\enquote {\bibinfo {title} {Mixed-state quantum anomaly and multipartite
  entanglement},}\ } (\bibinfo {year} {2024}{\natexlab{a}}),\ \Eprint
  {http://arxiv.org/abs/2401.17357} {arXiv:2401.17357 [cond-mat.str-el]}
  \BibitemShut {NoStop}%
\bibitem [{\citenamefont {Wang}\ and\ \citenamefont
  {Li}(2024)}]{wang2024anomaly}%
  \BibitemOpen
  \bibfield  {author} {\bibinfo {author} {\bibfnamefont {Z.}~\bibnamefont
  {Wang}}\ and\ \bibinfo {author} {\bibfnamefont {L.}~\bibnamefont {Li}},\
  }\href@noop {} {\bibfield  {journal} {\bibinfo  {journal} {arXiv preprint
  arXiv:2403.14533}\ } (\bibinfo {year} {2024})}\BibitemShut {NoStop}%
\bibitem [{\citenamefont {You}\ and\ \citenamefont {Oshikawa}(2024)}]{prelim}%
  \BibitemOpen
  \bibfield  {author} {\bibinfo {author} {\bibfnamefont {Y.}~\bibnamefont
  {You}}\ and\ \bibinfo {author} {\bibfnamefont {M.}~\bibnamefont {Oshikawa}},\
  }\href@noop {} {\bibfield  {journal} {\bibinfo  {journal} {arXiv preprint
  arXiv:2407.08786}\ } (\bibinfo {year} {2024})}\BibitemShut {NoStop}%
\bibitem [{\citenamefont {Zhang}\ \emph
  {et~al.}(2024{\natexlab{a}})\citenamefont {Zhang}, \citenamefont {Agrawal},\
  and\ \citenamefont {Vijay}}]{zhang2024quantum}%
  \BibitemOpen
  \bibfield  {author} {\bibinfo {author} {\bibfnamefont {Z.}~\bibnamefont
  {Zhang}}, \bibinfo {author} {\bibfnamefont {U.}~\bibnamefont {Agrawal}}, \
  and\ \bibinfo {author} {\bibfnamefont {S.}~\bibnamefont {Vijay}},\ }\href
  {https://arxiv.org/abs/2405.05965} {\enquote {\bibinfo {title} {Quantum
  communication and mixed-state order in decohered symmetry-protected
  topological states},}\ } (\bibinfo {year} {2024}{\natexlab{a}}),\ \Eprint
  {http://arxiv.org/abs/2405.05965} {arXiv:2405.05965 [quant-ph]} \BibitemShut
  {NoStop}%
\bibitem [{\citenamefont {Ellison}\ and\ \citenamefont
  {Cheng}(2024)}]{ellison2024towards}%
  \BibitemOpen
  \bibfield  {author} {\bibinfo {author} {\bibfnamefont {T.}~\bibnamefont
  {Ellison}}\ and\ \bibinfo {author} {\bibfnamefont {M.}~\bibnamefont
  {Cheng}},\ }\href@noop {} {\bibfield  {journal} {\bibinfo  {journal} {arXiv
  preprint arXiv:2405.02390}\ } (\bibinfo {year} {2024})}\BibitemShut {NoStop}%
\bibitem [{\citenamefont {Sohal}\ and\ \citenamefont
  {Prem}(2024)}]{sohal2024noisy}%
  \BibitemOpen
  \bibfield  {author} {\bibinfo {author} {\bibfnamefont {R.}~\bibnamefont
  {Sohal}}\ and\ \bibinfo {author} {\bibfnamefont {A.}~\bibnamefont {Prem}},\
  }\href@noop {} {\enquote {\bibinfo {title} {A noisy approach to intrinsically
  mixed-state topological order},}\ } (\bibinfo {year} {2024}),\ \Eprint
  {http://arxiv.org/abs/2403.13879} {arXiv:2403.13879 [cond-mat.str-el]}
  \BibitemShut {NoStop}%
\bibitem [{\citenamefont {Chirame}\ \emph {et~al.}(2024)\citenamefont
  {Chirame}, \citenamefont {Burnell}, \citenamefont {Gopalakrishnan},\ and\
  \citenamefont {Prem}}]{chirame2024stable}%
  \BibitemOpen
  \bibfield  {author} {\bibinfo {author} {\bibfnamefont {S.}~\bibnamefont
  {Chirame}}, \bibinfo {author} {\bibfnamefont {F.~J.}\ \bibnamefont
  {Burnell}}, \bibinfo {author} {\bibfnamefont {S.}~\bibnamefont
  {Gopalakrishnan}}, \ and\ \bibinfo {author} {\bibfnamefont {A.}~\bibnamefont
  {Prem}},\ }\href@noop {} {\enquote {\bibinfo {title} {Stable
  symmetry-protected topological phases in systems with heralded noise},}\ }
  (\bibinfo {year} {2024}),\ \Eprint {http://arxiv.org/abs/2404.16962}
  {arXiv:2404.16962 [quant-ph]} \BibitemShut {NoStop}%
\bibitem [{\citenamefont {Sala}\ \emph
  {et~al.}(2024{\natexlab{a}})\citenamefont {Sala}, \citenamefont {Alicea},\
  and\ \citenamefont {Verresen}}]{sala2024decoherence}%
  \BibitemOpen
  \bibfield  {author} {\bibinfo {author} {\bibfnamefont {P.}~\bibnamefont
  {Sala}}, \bibinfo {author} {\bibfnamefont {J.}~\bibnamefont {Alicea}}, \ and\
  \bibinfo {author} {\bibfnamefont {R.}~\bibnamefont {Verresen}},\ }\href@noop
  {} {\bibfield  {journal} {\bibinfo  {journal} {arXiv preprint
  arXiv:2409.12948}\ } (\bibinfo {year} {2024}{\natexlab{a}})}\BibitemShut
  {NoStop}%
\bibitem [{\citenamefont {Albert}\ and\ \citenamefont
  {Jiang}(2014)}]{albert2014symmetries}%
  \BibitemOpen
  \bibfield  {author} {\bibinfo {author} {\bibfnamefont {V.~V.}\ \bibnamefont
  {Albert}}\ and\ \bibinfo {author} {\bibfnamefont {L.}~\bibnamefont {Jiang}},\
  }\href@noop {} {\bibfield  {journal} {\bibinfo  {journal} {Physical Review
  A}\ }\textbf {\bibinfo {volume} {89}},\ \bibinfo {pages} {022118} (\bibinfo
  {year} {2014})}\BibitemShut {NoStop}%
\bibitem [{\citenamefont {Zhang}\ \emph
  {et~al.}(2024{\natexlab{b}})\citenamefont {Zhang}, \citenamefont {Xu},
  \citenamefont {Zhang}, \citenamefont {Xu}, \citenamefont {Bi},\ and\
  \citenamefont {Luo}}]{zhang2024strong}%
  \BibitemOpen
  \bibfield  {author} {\bibinfo {author} {\bibfnamefont {C.}~\bibnamefont
  {Zhang}}, \bibinfo {author} {\bibfnamefont {Y.}~\bibnamefont {Xu}}, \bibinfo
  {author} {\bibfnamefont {J.-H.}\ \bibnamefont {Zhang}}, \bibinfo {author}
  {\bibfnamefont {C.}~\bibnamefont {Xu}}, \bibinfo {author} {\bibfnamefont
  {Z.}~\bibnamefont {Bi}}, \ and\ \bibinfo {author} {\bibfnamefont {Z.-X.}\
  \bibnamefont {Luo}},\ }\href {https://arxiv.org/abs/2409.17530} {\enquote
  {\bibinfo {title} {Strong-to-weak spontaneous breaking of 1-form symmetry and
  intrinsically mixed topological order},}\ } (\bibinfo {year}
  {2024}{\natexlab{b}}),\ \Eprint {http://arxiv.org/abs/2409.17530}
  {arXiv:2409.17530 [quant-ph]} \BibitemShut {NoStop}%
\bibitem [{\citenamefont {Zhang}\ \emph
  {et~al.}(2024{\natexlab{c}})\citenamefont {Zhang}, \citenamefont {Xu},\ and\
  \citenamefont {Xu}}]{zhang2024fluctuation}%
  \BibitemOpen
  \bibfield  {author} {\bibinfo {author} {\bibfnamefont {J.-H.}\ \bibnamefont
  {Zhang}}, \bibinfo {author} {\bibfnamefont {C.}~\bibnamefont {Xu}}, \ and\
  \bibinfo {author} {\bibfnamefont {Y.}~\bibnamefont {Xu}},\ }\href
  {https://arxiv.org/abs/2409.18944} {\enquote {\bibinfo {title}
  {Fluctuation-dissipation theorem and information geometry in open quantum
  systems},}\ } (\bibinfo {year} {2024}{\natexlab{c}}),\ \Eprint
  {http://arxiv.org/abs/2409.18944} {arXiv:2409.18944 [quant-ph]} \BibitemShut
  {NoStop}%
\bibitem [{\citenamefont {Lee}\ \emph {et~al.}(2023{\natexlab{b}})\citenamefont
  {Lee}, \citenamefont {Jian},\ and\ \citenamefont {Xu}}]{lee2023quantum}%
  \BibitemOpen
  \bibfield  {author} {\bibinfo {author} {\bibfnamefont {J.~Y.}\ \bibnamefont
  {Lee}}, \bibinfo {author} {\bibfnamefont {C.-M.}\ \bibnamefont {Jian}}, \
  and\ \bibinfo {author} {\bibfnamefont {C.}~\bibnamefont {Xu}},\ }\href@noop
  {} {\bibfield  {journal} {\bibinfo  {journal} {arXiv preprint
  arXiv:2301.05238}\ } (\bibinfo {year} {2023}{\natexlab{b}})}\BibitemShut
  {NoStop}%
\bibitem [{\citenamefont {Su}\ \emph {et~al.}(2024)\citenamefont {Su},
  \citenamefont {Myerson-Jain}, \citenamefont {Wang}, \citenamefont {Jian},\
  and\ \citenamefont {Xu}}]{Su_2024}%
  \BibitemOpen
  \bibfield  {author} {\bibinfo {author} {\bibfnamefont {K.}~\bibnamefont
  {Su}}, \bibinfo {author} {\bibfnamefont {N.}~\bibnamefont {Myerson-Jain}},
  \bibinfo {author} {\bibfnamefont {C.}~\bibnamefont {Wang}}, \bibinfo {author}
  {\bibfnamefont {C.-M.}\ \bibnamefont {Jian}}, \ and\ \bibinfo {author}
  {\bibfnamefont {C.}~\bibnamefont {Xu}},\ }\href {\doibase
  10.1103/physrevlett.132.200402} {\bibfield  {journal} {\bibinfo  {journal}
  {Physical Review Letters}\ }\textbf {\bibinfo {volume} {132}} (\bibinfo
  {year} {2024}),\ 10.1103/physrevlett.132.200402}\BibitemShut {NoStop}%
\bibitem [{\citenamefont {Sang}\ \emph {et~al.}(2023)\citenamefont {Sang},
  \citenamefont {Zou},\ and\ \citenamefont {Hsieh}}]{sang2023mixed}%
  \BibitemOpen
  \bibfield  {author} {\bibinfo {author} {\bibfnamefont {S.}~\bibnamefont
  {Sang}}, \bibinfo {author} {\bibfnamefont {Y.}~\bibnamefont {Zou}}, \ and\
  \bibinfo {author} {\bibfnamefont {T.~H.}\ \bibnamefont {Hsieh}},\ }\href@noop
  {} {\bibfield  {journal} {\bibinfo  {journal} {arXiv preprint
  arXiv:2310.08639}\ } (\bibinfo {year} {2023})}\BibitemShut {NoStop}%
\bibitem [{\citenamefont {{Sang}}\ and\ \citenamefont
  {{Hsieh}}(2024)}]{sang2024stability}%
  \BibitemOpen
  \bibfield  {author} {\bibinfo {author} {\bibfnamefont {S.}~\bibnamefont
  {{Sang}}}\ and\ \bibinfo {author} {\bibfnamefont {T.~H.}\ \bibnamefont
  {{Hsieh}}},\ }\href {\doibase 10.48550/arXiv.2404.07251} {\bibfield
  {journal} {\bibinfo  {journal} {arXiv e-prints}\ ,\ \bibinfo {eid}
  {arXiv:2404.07251}} (\bibinfo {year} {2024})},\ \Eprint
  {http://arxiv.org/abs/2404.07251} {arXiv:2404.07251 [quant-ph]} \BibitemShut
  {NoStop}%
\bibitem [{\citenamefont {Moharramipour}\ \emph {et~al.}(2024)\citenamefont
  {Moharramipour}, \citenamefont {Lessa}, \citenamefont {Wang}, \citenamefont
  {Hsieh},\ and\ \citenamefont {Sahu}}]{moharramipour2024symmetry}%
  \BibitemOpen
  \bibfield  {author} {\bibinfo {author} {\bibfnamefont {A.}~\bibnamefont
  {Moharramipour}}, \bibinfo {author} {\bibfnamefont {L.~A.}\ \bibnamefont
  {Lessa}}, \bibinfo {author} {\bibfnamefont {C.}~\bibnamefont {Wang}},
  \bibinfo {author} {\bibfnamefont {T.~H.}\ \bibnamefont {Hsieh}}, \ and\
  \bibinfo {author} {\bibfnamefont {S.}~\bibnamefont {Sahu}},\ }\href
  {https://arxiv.org/abs/2406.08542} {\enquote {\bibinfo {title} {Symmetry
  enforced entanglement in maximally mixed states},}\ } (\bibinfo {year}
  {2024}),\ \Eprint {http://arxiv.org/abs/2406.08542} {arXiv:2406.08542
  [quant-ph]} \BibitemShut {NoStop}%
\bibitem [{\citenamefont {Sang}\ \emph {et~al.}(2024)\citenamefont {Sang},
  \citenamefont {Hsieh},\ and\ \citenamefont {Zou}}]{sang2024approximate}%
  \BibitemOpen
  \bibfield  {author} {\bibinfo {author} {\bibfnamefont {S.}~\bibnamefont
  {Sang}}, \bibinfo {author} {\bibfnamefont {T.~H.}\ \bibnamefont {Hsieh}}, \
  and\ \bibinfo {author} {\bibfnamefont {Y.}~\bibnamefont {Zou}},\ }\href
  {https://arxiv.org/abs/2406.09555} {\enquote {\bibinfo {title} {Approximate
  quantum error correcting codes from conformal field theory},}\ } (\bibinfo
  {year} {2024}),\ \Eprint {http://arxiv.org/abs/2406.09555} {arXiv:2406.09555
  [quant-ph]} \BibitemShut {NoStop}%
\bibitem [{\citenamefont {Lessa}\ \emph
  {et~al.}(2024{\natexlab{b}})\citenamefont {Lessa}, \citenamefont {Ma},
  \citenamefont {Zhang}, \citenamefont {Bi}, \citenamefont {Cheng},\ and\
  \citenamefont {Wang}}]{lessa2024strong}%
  \BibitemOpen
  \bibfield  {author} {\bibinfo {author} {\bibfnamefont {L.~A.}\ \bibnamefont
  {Lessa}}, \bibinfo {author} {\bibfnamefont {R.}~\bibnamefont {Ma}}, \bibinfo
  {author} {\bibfnamefont {J.-H.}\ \bibnamefont {Zhang}}, \bibinfo {author}
  {\bibfnamefont {Z.}~\bibnamefont {Bi}}, \bibinfo {author} {\bibfnamefont
  {M.}~\bibnamefont {Cheng}}, \ and\ \bibinfo {author} {\bibfnamefont
  {C.}~\bibnamefont {Wang}},\ }\href {https://arxiv.org/abs/2405.03639}
  {\enquote {\bibinfo {title} {Strong-to-weak spontaneous symmetry breaking in
  mixed quantum states},}\ } (\bibinfo {year} {2024}{\natexlab{b}}),\ \Eprint
  {http://arxiv.org/abs/2405.03639} {arXiv:2405.03639 [quant-ph]} \BibitemShut
  {NoStop}%
\bibitem [{\citenamefont {Gu}\ \emph {et~al.}(2024)\citenamefont {Gu},
  \citenamefont {Wang},\ and\ \citenamefont {Wang}}]{gu2024spontaneous}%
  \BibitemOpen
  \bibfield  {author} {\bibinfo {author} {\bibfnamefont {D.}~\bibnamefont
  {Gu}}, \bibinfo {author} {\bibfnamefont {Z.}~\bibnamefont {Wang}}, \ and\
  \bibinfo {author} {\bibfnamefont {Z.}~\bibnamefont {Wang}},\ }\href
  {https://arxiv.org/abs/2406.19381} {\enquote {\bibinfo {title} {Spontaneous
  symmetry breaking in open quantum systems: strong, weak, and
  strong-to-weak},}\ } (\bibinfo {year} {2024}),\ \Eprint
  {http://arxiv.org/abs/2406.19381} {arXiv:2406.19381 [quant-ph]} \BibitemShut
  {NoStop}%
\bibitem [{\citenamefont {Sala}\ \emph
  {et~al.}(2024{\natexlab{b}})\citenamefont {Sala}, \citenamefont
  {Gopalakrishnan}, \citenamefont {Oshikawa},\ and\ \citenamefont
  {You}}]{sala2024spontaneous}%
  \BibitemOpen
  \bibfield  {author} {\bibinfo {author} {\bibfnamefont {P.}~\bibnamefont
  {Sala}}, \bibinfo {author} {\bibfnamefont {S.}~\bibnamefont
  {Gopalakrishnan}}, \bibinfo {author} {\bibfnamefont {M.}~\bibnamefont
  {Oshikawa}}, \ and\ \bibinfo {author} {\bibfnamefont {Y.}~\bibnamefont
  {You}},\ }\href {https://arxiv.org/abs/2405.02402} {\enquote {\bibinfo
  {title} {Spontaneous strong symmetry breaking in open systems: Purification
  perspective},}\ } (\bibinfo {year} {2024}{\natexlab{b}}),\ \Eprint
  {http://arxiv.org/abs/2405.02402} {arXiv:2405.02402 [quant-ph]} \BibitemShut
  {NoStop}%
\bibitem [{\citenamefont {Huang}\ \emph {et~al.}(2024)\citenamefont {Huang},
  \citenamefont {Qi}, \citenamefont {Zhang},\ and\ \citenamefont
  {Lucas}}]{huang2024hydro}%
  \BibitemOpen
  \bibfield  {author} {\bibinfo {author} {\bibfnamefont {X.}~\bibnamefont
  {Huang}}, \bibinfo {author} {\bibfnamefont {M.}~\bibnamefont {Qi}}, \bibinfo
  {author} {\bibfnamefont {J.-H.}\ \bibnamefont {Zhang}}, \ and\ \bibinfo
  {author} {\bibfnamefont {A.}~\bibnamefont {Lucas}},\ }\href
  {https://arxiv.org/abs/2407.08760} {\enquote {\bibinfo {title} {Hydrodynamics
  as the effective field theory of strong-to-weak spontaneous symmetry
  breaking},}\ } (\bibinfo {year} {2024}),\ \Eprint
  {http://arxiv.org/abs/2407.08760} {arXiv:2407.08760 [cond-mat.str-el]}
  \BibitemShut {NoStop}%
\bibitem [{\citenamefont {Guo}\ \emph {et~al.}(2024{\natexlab{b}})\citenamefont
  {Guo}, \citenamefont {Ding},\ and\ \citenamefont {Yang}}]{guo2024new}%
  \BibitemOpen
  \bibfield  {author} {\bibinfo {author} {\bibfnamefont {Y.}~\bibnamefont
  {Guo}}, \bibinfo {author} {\bibfnamefont {K.}~\bibnamefont {Ding}}, \ and\
  \bibinfo {author} {\bibfnamefont {S.}~\bibnamefont {Yang}},\ }\href
  {https://arxiv.org/abs/2408.03239} {\enquote {\bibinfo {title} {A new
  framework for quantum phases in open systems: Steady state of imaginary-time
  lindbladian evolution},}\ } (\bibinfo {year} {2024}{\natexlab{b}}),\ \Eprint
  {http://arxiv.org/abs/2408.03239} {arXiv:2408.03239 [quant-ph]} \BibitemShut
  {NoStop}%
\bibitem [{\citenamefont {Vijay}\ \emph {et~al.}(2015)\citenamefont {Vijay},
  \citenamefont {Haah},\ and\ \citenamefont {Fu}}]{Vijay2015-jj}%
  \BibitemOpen
  \bibfield  {author} {\bibinfo {author} {\bibfnamefont {S.}~\bibnamefont
  {Vijay}}, \bibinfo {author} {\bibfnamefont {J.}~\bibnamefont {Haah}}, \ and\
  \bibinfo {author} {\bibfnamefont {L.}~\bibnamefont {Fu}},\ }\href@noop {}
  {\bibfield  {journal} {\bibinfo  {journal} {Phys. Rev. B Condens. Matter}\
  }\textbf {\bibinfo {volume} {92}},\ \bibinfo {pages} {235136} (\bibinfo
  {year} {2015})}\BibitemShut {NoStop}%
\bibitem [{\citenamefont {Haah}(2011)}]{Haah2011-ny}%
  \BibitemOpen
  \bibfield  {author} {\bibinfo {author} {\bibfnamefont {J.}~\bibnamefont
  {Haah}},\ }\href@noop {} {\bibfield  {journal} {\bibinfo  {journal} {Phys.
  Rev. A}\ }\textbf {\bibinfo {volume} {83}},\ \bibinfo {pages} {042330}
  (\bibinfo {year} {2011})}\BibitemShut {NoStop}%
\bibitem [{\citenamefont {Chamon}(2005)}]{Chamon2005-fc}%
  \BibitemOpen
  \bibfield  {author} {\bibinfo {author} {\bibfnamefont {C.}~\bibnamefont
  {Chamon}},\ }\href@noop {} {\bibfield  {journal} {\bibinfo  {journal} {Phys.
  Rev. Lett.}\ }\textbf {\bibinfo {volume} {94}},\ \bibinfo {pages} {040402}
  (\bibinfo {year} {2005})}\BibitemShut {NoStop}%
\bibitem [{\citenamefont {Bravyi}\ \emph {et~al.}(2010)\citenamefont {Bravyi},
  \citenamefont {Hastings},\ and\ \citenamefont
  {Michalakis}}]{bravyi2010topological}%
  \BibitemOpen
  \bibfield  {author} {\bibinfo {author} {\bibfnamefont {S.}~\bibnamefont
  {Bravyi}}, \bibinfo {author} {\bibfnamefont {M.~B.}\ \bibnamefont
  {Hastings}}, \ and\ \bibinfo {author} {\bibfnamefont {S.}~\bibnamefont
  {Michalakis}},\ }\href@noop {} {\bibfield  {journal} {\bibinfo  {journal}
  {Journal of mathematical physics}\ }\textbf {\bibinfo {volume} {51}},\
  \bibinfo {pages} {093512} (\bibinfo {year} {2010})}\BibitemShut {NoStop}%
\bibitem [{\citenamefont {Vijay}\ \emph {et~al.}(2016)\citenamefont {Vijay},
  \citenamefont {Haah},\ and\ \citenamefont {Fu}}]{Vijay2016-dr}%
  \BibitemOpen
  \bibfield  {author} {\bibinfo {author} {\bibfnamefont {S.}~\bibnamefont
  {Vijay}}, \bibinfo {author} {\bibfnamefont {J.}~\bibnamefont {Haah}}, \ and\
  \bibinfo {author} {\bibfnamefont {L.}~\bibnamefont {Fu}},\ }\href@noop {}
  {\bibfield  {journal} {\bibinfo  {journal} {Phys. Rev. B Condens. Matter}\
  }\textbf {\bibinfo {volume} {94}},\ \bibinfo {pages} {235157} (\bibinfo
  {year} {2016})}\BibitemShut {NoStop}%
\bibitem [{\citenamefont {Yoshida}(2013)}]{yoshida2013exotic}%
  \BibitemOpen
  \bibfield  {author} {\bibinfo {author} {\bibfnamefont {B.}~\bibnamefont
  {Yoshida}},\ }\href@noop {} {\bibfield  {journal} {\bibinfo  {journal}
  {Physical Review B}\ }\textbf {\bibinfo {volume} {88}},\ \bibinfo {pages}
  {125122} (\bibinfo {year} {2013})}\BibitemShut {NoStop}%
\bibitem [{\citenamefont {Pretko}\ \emph {et~al.}(2020)\citenamefont {Pretko},
  \citenamefont {Chen},\ and\ \citenamefont {You}}]{pretko2020fracton}%
  \BibitemOpen
  \bibfield  {author} {\bibinfo {author} {\bibfnamefont {M.}~\bibnamefont
  {Pretko}}, \bibinfo {author} {\bibfnamefont {X.}~\bibnamefont {Chen}}, \ and\
  \bibinfo {author} {\bibfnamefont {Y.}~\bibnamefont {You}},\ }\href@noop {}
  {\bibfield  {journal} {\bibinfo  {journal} {International Journal of Modern
  Physics A}\ }\textbf {\bibinfo {volume} {35}},\ \bibinfo {pages} {2030003}
  (\bibinfo {year} {2020})}\BibitemShut {NoStop}%
\bibitem [{\citenamefont {Yoshida}\ \emph {et~al.}(2015)\citenamefont
  {Yoshida}, \citenamefont {Morimoto},\ and\ \citenamefont
  {Furusaki}}]{yoshida2015bosonic}%
  \BibitemOpen
  \bibfield  {author} {\bibinfo {author} {\bibfnamefont {T.}~\bibnamefont
  {Yoshida}}, \bibinfo {author} {\bibfnamefont {T.}~\bibnamefont {Morimoto}}, \
  and\ \bibinfo {author} {\bibfnamefont {A.}~\bibnamefont {Furusaki}},\
  }\href@noop {} {\bibfield  {journal} {\bibinfo  {journal} {Physical Review
  B}\ }\textbf {\bibinfo {volume} {92}},\ \bibinfo {pages} {245122} (\bibinfo
  {year} {2015})}\BibitemShut {NoStop}%
\bibitem [{\citenamefont {Ma}\ \emph {et~al.}(2018)\citenamefont {Ma},
  \citenamefont {Hermele},\ and\ \citenamefont {Chen}}]{ma2018fracton}%
  \BibitemOpen
  \bibfield  {author} {\bibinfo {author} {\bibfnamefont {H.}~\bibnamefont
  {Ma}}, \bibinfo {author} {\bibfnamefont {M.}~\bibnamefont {Hermele}}, \ and\
  \bibinfo {author} {\bibfnamefont {X.}~\bibnamefont {Chen}},\ }\href {\doibase
  10.1103/PhysRevB.98.035111} {\bibfield  {journal} {\bibinfo  {journal} {Phys.
  Rev. B}\ }\textbf {\bibinfo {volume} {98}},\ \bibinfo {pages} {035111}
  (\bibinfo {year} {2018})}\BibitemShut {NoStop}%
\bibitem [{\citenamefont {You}\ \emph {et~al.}(2020)\citenamefont {You},
  \citenamefont {Bi},\ and\ \citenamefont {Pretko}}]{you2019emergent}%
  \BibitemOpen
  \bibfield  {author} {\bibinfo {author} {\bibfnamefont {Y.}~\bibnamefont
  {You}}, \bibinfo {author} {\bibfnamefont {Z.}~\bibnamefont {Bi}}, \ and\
  \bibinfo {author} {\bibfnamefont {M.}~\bibnamefont {Pretko}},\ }\href
  {\doibase 10.1103/PhysRevResearch.2.013162} {\bibfield  {journal} {\bibinfo
  {journal} {Phys. Rev. Res.}\ }\textbf {\bibinfo {volume} {2}},\ \bibinfo
  {pages} {013162} (\bibinfo {year} {2020})}\BibitemShut {NoStop}%
\bibitem [{\citenamefont {Xu}\ and\ \citenamefont {Fisher}(2007)}]{xu2007bond}%
  \BibitemOpen
  \bibfield  {author} {\bibinfo {author} {\bibfnamefont {C.}~\bibnamefont
  {Xu}}\ and\ \bibinfo {author} {\bibfnamefont {M.~P.}\ \bibnamefont
  {Fisher}},\ }\href@noop {} {\bibfield  {journal} {\bibinfo  {journal}
  {Physical Review B}\ }\textbf {\bibinfo {volume} {75}},\ \bibinfo {pages}
  {104428} (\bibinfo {year} {2007})}\BibitemShut {NoStop}%
\bibitem [{\citenamefont {Paramekanti}\ \emph {et~al.}(2002)\citenamefont
  {Paramekanti}, \citenamefont {Balents},\ and\ \citenamefont
  {Fisher}}]{paramekanti2002ring}%
  \BibitemOpen
  \bibfield  {author} {\bibinfo {author} {\bibfnamefont {A.}~\bibnamefont
  {Paramekanti}}, \bibinfo {author} {\bibfnamefont {L.}~\bibnamefont
  {Balents}}, \ and\ \bibinfo {author} {\bibfnamefont {M.~P.}\ \bibnamefont
  {Fisher}},\ }\href@noop {} {\bibfield  {journal} {\bibinfo  {journal}
  {Physical Review B}\ }\textbf {\bibinfo {volume} {66}},\ \bibinfo {pages}
  {054526} (\bibinfo {year} {2002})}\BibitemShut {NoStop}%
\bibitem [{\citenamefont {Tay}\ and\ \citenamefont
  {Motrunich}(2010)}]{tay2010possible}%
  \BibitemOpen
  \bibfield  {author} {\bibinfo {author} {\bibfnamefont {T.}~\bibnamefont
  {Tay}}\ and\ \bibinfo {author} {\bibfnamefont {O.~I.}\ \bibnamefont
  {Motrunich}},\ }\href@noop {} {\bibfield  {journal} {\bibinfo  {journal}
  {Physical review letters}\ }\textbf {\bibinfo {volume} {105}},\ \bibinfo
  {pages} {187202} (\bibinfo {year} {2010})}\BibitemShut {NoStop}%
\bibitem [{\citenamefont {Seiberg}\ and\ \citenamefont
  {Shao}(2020)}]{seiberg2020exotic}%
  \BibitemOpen
  \bibfield  {author} {\bibinfo {author} {\bibfnamefont {N.}~\bibnamefont
  {Seiberg}}\ and\ \bibinfo {author} {\bibfnamefont {S.-H.}\ \bibnamefont
  {Shao}},\ }\href@noop {} {\bibfield  {journal} {\bibinfo  {journal} {SciPost
  Physics}\ }\textbf {\bibinfo {volume} {9}},\ \bibinfo {pages} {046} (\bibinfo
  {year} {2020})}\BibitemShut {NoStop}%
\bibitem [{\citenamefont {You}(2024{\natexlab{a}})}]{you2024quantum}%
  \BibitemOpen
  \bibfield  {author} {\bibinfo {author} {\bibfnamefont {Y.}~\bibnamefont
  {You}},\ }\href@noop {} {\bibfield  {journal} {\bibinfo  {journal} {Annual
  Review of Condensed Matter Physics}\ }\textbf {\bibinfo {volume} {16}}
  (\bibinfo {year} {2024}{\natexlab{a}})}\BibitemShut {NoStop}%
\bibitem [{\citenamefont {Nandkishore}\ and\ \citenamefont
  {Hermele}(2019)}]{nandkishore2019fractons}%
  \BibitemOpen
  \bibfield  {author} {\bibinfo {author} {\bibfnamefont {R.~M.}\ \bibnamefont
  {Nandkishore}}\ and\ \bibinfo {author} {\bibfnamefont {M.}~\bibnamefont
  {Hermele}},\ }\href@noop {} {\bibfield  {journal} {\bibinfo  {journal}
  {Annual Review of Condensed Matter Physics}\ }\textbf {\bibinfo {volume}
  {10}},\ \bibinfo {pages} {295} (\bibinfo {year} {2019})}\BibitemShut
  {NoStop}%
\bibitem [{\citenamefont {Burnell}\ \emph {et~al.}(2022)\citenamefont
  {Burnell}, \citenamefont {Devakul}, \citenamefont {Gorantla}, \citenamefont
  {Lam},\ and\ \citenamefont {Shao}}]{burnell2022anomaly}%
  \BibitemOpen
  \bibfield  {author} {\bibinfo {author} {\bibfnamefont {F.~J.}\ \bibnamefont
  {Burnell}}, \bibinfo {author} {\bibfnamefont {T.}~\bibnamefont {Devakul}},
  \bibinfo {author} {\bibfnamefont {P.}~\bibnamefont {Gorantla}}, \bibinfo
  {author} {\bibfnamefont {H.~T.}\ \bibnamefont {Lam}}, \ and\ \bibinfo
  {author} {\bibfnamefont {S.-H.}\ \bibnamefont {Shao}},\ }\href@noop {}
  {\bibfield  {journal} {\bibinfo  {journal} {Physical Review B}\ }\textbf
  {\bibinfo {volume} {106}},\ \bibinfo {pages} {085113} (\bibinfo {year}
  {2022})}\BibitemShut {NoStop}%
\bibitem [{\citenamefont {You}\ \emph {et~al.}(2018{\natexlab{a}})\citenamefont
  {You}, \citenamefont {Devakul}, \citenamefont {Burnell},\ and\ \citenamefont
  {Sondhi}}]{you2018subsystem}%
  \BibitemOpen
  \bibfield  {author} {\bibinfo {author} {\bibfnamefont {Y.}~\bibnamefont
  {You}}, \bibinfo {author} {\bibfnamefont {T.}~\bibnamefont {Devakul}},
  \bibinfo {author} {\bibfnamefont {F.~J.}\ \bibnamefont {Burnell}}, \ and\
  \bibinfo {author} {\bibfnamefont {S.~L.}\ \bibnamefont {Sondhi}},\
  }\href@noop {} {\bibfield  {journal} {\bibinfo  {journal} {Physical Review
  B}\ }\textbf {\bibinfo {volume} {98}},\ \bibinfo {pages} {035112} (\bibinfo
  {year} {2018}{\natexlab{a}})}\BibitemShut {NoStop}%
\bibitem [{\citenamefont {You}\ \emph {et~al.}(2018{\natexlab{b}})\citenamefont
  {You}, \citenamefont {Devakul}, \citenamefont {Burnell},\ and\ \citenamefont
  {Sondhi}}]{you2018symmetric}%
  \BibitemOpen
  \bibfield  {author} {\bibinfo {author} {\bibfnamefont {Y.}~\bibnamefont
  {You}}, \bibinfo {author} {\bibfnamefont {T.}~\bibnamefont {Devakul}},
  \bibinfo {author} {\bibfnamefont {F.}~\bibnamefont {Burnell}}, \ and\
  \bibinfo {author} {\bibfnamefont {S.}~\bibnamefont {Sondhi}},\ }\href@noop {}
  {\bibfield  {journal} {\bibinfo  {journal} {arXiv preprint arXiv:1805.09800}\
  } (\bibinfo {year} {2018}{\natexlab{b}})}\BibitemShut {NoStop}%
\bibitem [{\citenamefont {You}\ \emph {et~al.}(2019)\citenamefont {You},
  \citenamefont {Devakul}, \citenamefont {Sondhi},\ and\ \citenamefont
  {Burnell}}]{you2019fractonic}%
  \BibitemOpen
  \bibfield  {author} {\bibinfo {author} {\bibfnamefont {Y.}~\bibnamefont
  {You}}, \bibinfo {author} {\bibfnamefont {T.}~\bibnamefont {Devakul}},
  \bibinfo {author} {\bibfnamefont {S.}~\bibnamefont {Sondhi}}, \ and\ \bibinfo
  {author} {\bibfnamefont {F.}~\bibnamefont {Burnell}},\ }\href@noop {}
  {\bibfield  {journal} {\bibinfo  {journal} {arXiv preprint arXiv:1904.11530}\
  } (\bibinfo {year} {2019})}\BibitemShut {NoStop}%
\bibitem [{\citenamefont {You}\ \emph {et~al.}(2021)\citenamefont {You},
  \citenamefont {Burnell},\ and\ \citenamefont {Hughes}}]{you2019multipolar}%
  \BibitemOpen
  \bibfield  {author} {\bibinfo {author} {\bibfnamefont {Y.}~\bibnamefont
  {You}}, \bibinfo {author} {\bibfnamefont {F.~J.}\ \bibnamefont {Burnell}}, \
  and\ \bibinfo {author} {\bibfnamefont {T.~L.}\ \bibnamefont {Hughes}},\
  }\href {\doibase 10.1103/PhysRevB.103.245128} {\bibfield  {journal} {\bibinfo
   {journal} {Phys. Rev. B}\ }\textbf {\bibinfo {volume} {103}},\ \bibinfo
  {pages} {245128} (\bibinfo {year} {2021})}\BibitemShut {NoStop}%
\bibitem [{\citenamefont {Devakul}\ \emph {et~al.}(2019)\citenamefont
  {Devakul}, \citenamefont {You}, \citenamefont {Burnell},\ and\ \citenamefont
  {Sondhi}}]{devakul2018fractal}%
  \BibitemOpen
  \bibfield  {author} {\bibinfo {author} {\bibfnamefont {T.}~\bibnamefont
  {Devakul}}, \bibinfo {author} {\bibfnamefont {Y.}~\bibnamefont {You}},
  \bibinfo {author} {\bibfnamefont {F.~J.}\ \bibnamefont {Burnell}}, \ and\
  \bibinfo {author} {\bibfnamefont {S.~L.}\ \bibnamefont {Sondhi}},\ }\href
  {\doibase 10.21468/SciPostPhys.6.1.007} {\bibfield  {journal} {\bibinfo
  {journal} {SciPost Phys.}\ }\textbf {\bibinfo {volume} {6}},\ \bibinfo
  {pages} {007} (\bibinfo {year} {2019})}\BibitemShut {NoStop}%
\bibitem [{\citenamefont {Devakul}\ \emph
  {et~al.}(2018{\natexlab{a}})\citenamefont {Devakul}, \citenamefont
  {Williamson},\ and\ \citenamefont {You}}]{devakul2018strong}%
  \BibitemOpen
  \bibfield  {author} {\bibinfo {author} {\bibfnamefont {T.}~\bibnamefont
  {Devakul}}, \bibinfo {author} {\bibfnamefont {D.~J.}\ \bibnamefont
  {Williamson}}, \ and\ \bibinfo {author} {\bibfnamefont {Y.}~\bibnamefont
  {You}},\ }\href@noop {} {\bibfield  {journal} {\bibinfo  {journal} {arXiv
  preprint arXiv:1808.05300}\ } (\bibinfo {year}
  {2018}{\natexlab{a}})}\BibitemShut {NoStop}%
\bibitem [{\citenamefont {May-Mann}\ \emph {et~al.}(2022)\citenamefont
  {May-Mann}, \citenamefont {You}, \citenamefont {Hughes},\ and\ \citenamefont
  {Bi}}]{may2022interaction}%
  \BibitemOpen
  \bibfield  {author} {\bibinfo {author} {\bibfnamefont {J.}~\bibnamefont
  {May-Mann}}, \bibinfo {author} {\bibfnamefont {Y.}~\bibnamefont {You}},
  \bibinfo {author} {\bibfnamefont {T.~L.}\ \bibnamefont {Hughes}}, \ and\
  \bibinfo {author} {\bibfnamefont {Z.}~\bibnamefont {Bi}},\ }\href@noop {}
  {\bibfield  {journal} {\bibinfo  {journal} {Physical Review B}\ }\textbf
  {\bibinfo {volume} {105}},\ \bibinfo {pages} {245122} (\bibinfo {year}
  {2022})}\BibitemShut {NoStop}%
\bibitem [{\citenamefont {Wang}\ and\ \citenamefont
  {Gu}(2018)}]{Wang_2018Classification}%
  \BibitemOpen
  \bibfield  {author} {\bibinfo {author} {\bibfnamefont {Q.-R.}\ \bibnamefont
  {Wang}}\ and\ \bibinfo {author} {\bibfnamefont {Z.-C.}\ \bibnamefont {Gu}},\
  }\href {\doibase 10.1103/physrevx.8.011055} {\bibfield  {journal} {\bibinfo
  {journal} {Physical Review X}\ }\textbf {\bibinfo {volume} {8}} (\bibinfo
  {year} {2018}),\ 10.1103/physrevx.8.011055}\BibitemShut {NoStop}%
\bibitem [{\citenamefont {Wang}\ and\ \citenamefont {Gu}(2020)}]{Wang_2020}%
  \BibitemOpen
  \bibfield  {author} {\bibinfo {author} {\bibfnamefont {Q.-R.}\ \bibnamefont
  {Wang}}\ and\ \bibinfo {author} {\bibfnamefont {Z.-C.}\ \bibnamefont {Gu}},\
  }\href {\doibase 10.1103/physrevx.10.031055} {\bibfield  {journal} {\bibinfo
  {journal} {Physical Review X}\ }\textbf {\bibinfo {volume} {10}} (\bibinfo
  {year} {2020}),\ 10.1103/physrevx.10.031055}\BibitemShut {NoStop}%
\bibitem [{\citenamefont {Wang}\ \emph {et~al.}(2021)\citenamefont {Wang},
  \citenamefont {Ning},\ and\ \citenamefont {Cheng}}]{SpectralSequence}%
  \BibitemOpen
  \bibfield  {author} {\bibinfo {author} {\bibfnamefont {Q.-R.}\ \bibnamefont
  {Wang}}, \bibinfo {author} {\bibfnamefont {S.-Q.}\ \bibnamefont {Ning}}, \
  and\ \bibinfo {author} {\bibfnamefont {M.}~\bibnamefont {Cheng}},\
  }\href@noop {} {\  (\bibinfo {year} {2021})},\ \Eprint
  {http://arxiv.org/abs/2104.13233} {arXiv:2104.13233 [cond-mat.str-el]}
  \BibitemShut {NoStop}%
\bibitem [{\citenamefont {Zhang}\ \emph
  {et~al.}(2023{\natexlab{b}})\citenamefont {Zhang}, \citenamefont {Cheng},\
  and\ \citenamefont {Bi}}]{zhang2023classification}%
  \BibitemOpen
  \bibfield  {author} {\bibinfo {author} {\bibfnamefont {J.-H.}\ \bibnamefont
  {Zhang}}, \bibinfo {author} {\bibfnamefont {M.}~\bibnamefont {Cheng}}, \ and\
  \bibinfo {author} {\bibfnamefont {Z.}~\bibnamefont {Bi}},\ }\href@noop {}
  {\bibfield  {journal} {\bibinfo  {journal} {Physical Review B}\ }\textbf
  {\bibinfo {volume} {108}},\ \bibinfo {pages} {045133} (\bibinfo {year}
  {2023}{\natexlab{b}})}\BibitemShut {NoStop}%
\bibitem [{\citenamefont {You}(2024{\natexlab{b}})}]{you2024higher}%
  \BibitemOpen
  \bibfield  {author} {\bibinfo {author} {\bibfnamefont {Y.}~\bibnamefont
  {You}},\ }\href@noop {} {\bibfield  {journal} {\bibinfo  {journal} {New
  Journal of Physics}\ }\textbf {\bibinfo {volume} {26}},\ \bibinfo {pages}
  {093028} (\bibinfo {year} {2024}{\natexlab{b}})}\BibitemShut {NoStop}%
\bibitem [{\citenamefont {You}\ \emph {et~al.}(2014)\citenamefont {You},
  \citenamefont {Bi}, \citenamefont {Rasmussen}, \citenamefont {Slagle},\ and\
  \citenamefont {Xu}}]{you2014wave}%
  \BibitemOpen
  \bibfield  {author} {\bibinfo {author} {\bibfnamefont {Y.-Z.}\ \bibnamefont
  {You}}, \bibinfo {author} {\bibfnamefont {Z.}~\bibnamefont {Bi}}, \bibinfo
  {author} {\bibfnamefont {A.}~\bibnamefont {Rasmussen}}, \bibinfo {author}
  {\bibfnamefont {K.}~\bibnamefont {Slagle}}, \ and\ \bibinfo {author}
  {\bibfnamefont {C.}~\bibnamefont {Xu}},\ }\href@noop {} {\bibfield  {journal}
  {\bibinfo  {journal} {Physical review letters}\ }\textbf {\bibinfo {volume}
  {112}},\ \bibinfo {pages} {247202} (\bibinfo {year} {2014})}\BibitemShut
  {NoStop}%
\bibitem [{\citenamefont {Lepori}\ \emph {et~al.}(2023)\citenamefont {Lepori},
  \citenamefont {Burrello}, \citenamefont {Trombettoni},\ and\ \citenamefont
  {Paganelli}}]{lepori2023strange}%
  \BibitemOpen
  \bibfield  {author} {\bibinfo {author} {\bibfnamefont {L.}~\bibnamefont
  {Lepori}}, \bibinfo {author} {\bibfnamefont {M.}~\bibnamefont {Burrello}},
  \bibinfo {author} {\bibfnamefont {A.}~\bibnamefont {Trombettoni}}, \ and\
  \bibinfo {author} {\bibfnamefont {S.}~\bibnamefont {Paganelli}},\ }\href@noop
  {} {\bibfield  {journal} {\bibinfo  {journal} {Physical Review B}\ }\textbf
  {\bibinfo {volume} {108}},\ \bibinfo {pages} {035110} (\bibinfo {year}
  {2023})}\BibitemShut {NoStop}%
\bibitem [{\citenamefont {Scaffidi}\ and\ \citenamefont
  {Ringel}(2016)}]{Scaffidi16}%
  \BibitemOpen
  \bibfield  {author} {\bibinfo {author} {\bibfnamefont {T.}~\bibnamefont
  {Scaffidi}}\ and\ \bibinfo {author} {\bibfnamefont {Z.}~\bibnamefont
  {Ringel}},\ }\href {\doibase 10.1103/PhysRevB.93.115105} {\bibfield
  {journal} {\bibinfo  {journal} {Phys. Rev. B}\ }\textbf {\bibinfo {volume}
  {93}},\ \bibinfo {pages} {115105} (\bibinfo {year} {2016})}\BibitemShut
  {NoStop}%
\bibitem [{\citenamefont {Uhlmann}(1976)}]{uhlmann1976transition}%
  \BibitemOpen
  \bibfield  {author} {\bibinfo {author} {\bibfnamefont {A.}~\bibnamefont
  {Uhlmann}},\ }\href@noop {} {\bibfield  {journal} {\bibinfo  {journal}
  {Reports on Mathematical Physics}\ }\textbf {\bibinfo {volume} {9}},\
  \bibinfo {pages} {273} (\bibinfo {year} {1976})}\BibitemShut {NoStop}%
\bibitem [{\citenamefont {Briegel}\ \emph {et~al.}(2009)\citenamefont
  {Briegel}, \citenamefont {Browne}, \citenamefont {D{\"u}r}, \citenamefont
  {Raussendorf},\ and\ \citenamefont {Van~den Nest}}]{briegel2009measurement}%
  \BibitemOpen
  \bibfield  {author} {\bibinfo {author} {\bibfnamefont {H.~J.}\ \bibnamefont
  {Briegel}}, \bibinfo {author} {\bibfnamefont {D.~E.}\ \bibnamefont {Browne}},
  \bibinfo {author} {\bibfnamefont {W.}~\bibnamefont {D{\"u}r}}, \bibinfo
  {author} {\bibfnamefont {R.}~\bibnamefont {Raussendorf}}, \ and\ \bibinfo
  {author} {\bibfnamefont {M.}~\bibnamefont {Van~den Nest}},\ }\href@noop {}
  {\bibfield  {journal} {\bibinfo  {journal} {Nature Physics}\ }\textbf
  {\bibinfo {volume} {5}},\ \bibinfo {pages} {19} (\bibinfo {year}
  {2009})}\BibitemShut {NoStop}%
\bibitem [{\citenamefont {Raussendorf}\ \emph {et~al.}(2003)\citenamefont
  {Raussendorf}, \citenamefont {Browne},\ and\ \citenamefont
  {Briegel}}]{raussendorf2003measurement}%
  \BibitemOpen
  \bibfield  {author} {\bibinfo {author} {\bibfnamefont {R.}~\bibnamefont
  {Raussendorf}}, \bibinfo {author} {\bibfnamefont {D.~E.}\ \bibnamefont
  {Browne}}, \ and\ \bibinfo {author} {\bibfnamefont {H.~J.}\ \bibnamefont
  {Briegel}},\ }\href@noop {} {\bibfield  {journal} {\bibinfo  {journal}
  {Physical review A}\ }\textbf {\bibinfo {volume} {68}},\ \bibinfo {pages}
  {022312} (\bibinfo {year} {2003})}\BibitemShut {NoStop}%
\bibitem [{\citenamefont {Wen}\ and\ \citenamefont
  {Potter}(2023)}]{wen2023bulk}%
  \BibitemOpen
  \bibfield  {author} {\bibinfo {author} {\bibfnamefont {R.}~\bibnamefont
  {Wen}}\ and\ \bibinfo {author} {\bibfnamefont {A.~C.}\ \bibnamefont
  {Potter}},\ }\href@noop {} {\bibfield  {journal} {\bibinfo  {journal}
  {Physical Review B}\ }\textbf {\bibinfo {volume} {107}},\ \bibinfo {pages}
  {245127} (\bibinfo {year} {2023})}\BibitemShut {NoStop}%
\bibitem [{\citenamefont {Thorngren}\ \emph
  {et~al.}(2021{\natexlab{a}})\citenamefont {Thorngren}, \citenamefont
  {Vishwanath},\ and\ \citenamefont {Verresen}}]{thorngren2021intrinsically}%
  \BibitemOpen
  \bibfield  {author} {\bibinfo {author} {\bibfnamefont {R.}~\bibnamefont
  {Thorngren}}, \bibinfo {author} {\bibfnamefont {A.}~\bibnamefont
  {Vishwanath}}, \ and\ \bibinfo {author} {\bibfnamefont {R.}~\bibnamefont
  {Verresen}},\ }\href@noop {} {\bibfield  {journal} {\bibinfo  {journal}
  {Physical Review B}\ }\textbf {\bibinfo {volume} {104}},\ \bibinfo {pages}
  {075132} (\bibinfo {year} {2021}{\natexlab{a}})}\BibitemShut {NoStop}%
\bibitem [{\citenamefont {Li}\ \emph {et~al.}(2023)\citenamefont {Li},
  \citenamefont {Oshikawa},\ and\ \citenamefont {Zheng}}]{li2023intrinsically}%
  \BibitemOpen
  \bibfield  {author} {\bibinfo {author} {\bibfnamefont {L.}~\bibnamefont
  {Li}}, \bibinfo {author} {\bibfnamefont {M.}~\bibnamefont {Oshikawa}}, \ and\
  \bibinfo {author} {\bibfnamefont {Y.}~\bibnamefont {Zheng}},\ }\href@noop {}
  {\bibfield  {journal} {\bibinfo  {journal} {arXiv preprint arXiv:2307.04788}\
  } (\bibinfo {year} {2023})}\BibitemShut {NoStop}%
\bibitem [{\citenamefont {Thorngren}\ \emph
  {et~al.}(2021{\natexlab{b}})\citenamefont {Thorngren}, \citenamefont
  {Vishwanath},\ and\ \citenamefont {Verresen}}]{gaplessSPT}%
  \BibitemOpen
  \bibfield  {author} {\bibinfo {author} {\bibfnamefont {R.}~\bibnamefont
  {Thorngren}}, \bibinfo {author} {\bibfnamefont {A.}~\bibnamefont
  {Vishwanath}}, \ and\ \bibinfo {author} {\bibfnamefont {R.}~\bibnamefont
  {Verresen}},\ }\href {\doibase 10.1103/PhysRevB.104.075132} {\bibfield
  {journal} {\bibinfo  {journal} {Physical Review B}\ }\textbf {\bibinfo
  {volume} {104}},\ \bibinfo {pages} {075132} (\bibinfo {year}
  {2021}{\natexlab{b}})}\BibitemShut {NoStop}%
\bibitem [{\citenamefont {Scaffidi}\ \emph {et~al.}(2017)\citenamefont
  {Scaffidi}, \citenamefont {Parker},\ and\ \citenamefont
  {Vasseur}}]{scaffidi2017gapless}%
  \BibitemOpen
  \bibfield  {author} {\bibinfo {author} {\bibfnamefont {T.}~\bibnamefont
  {Scaffidi}}, \bibinfo {author} {\bibfnamefont {D.~E.}\ \bibnamefont
  {Parker}}, \ and\ \bibinfo {author} {\bibfnamefont {R.}~\bibnamefont
  {Vasseur}},\ }\href@noop {} {\bibfield  {journal} {\bibinfo  {journal}
  {Physical Review X}\ }\textbf {\bibinfo {volume} {7}},\ \bibinfo {pages}
  {041048} (\bibinfo {year} {2017})}\BibitemShut {NoStop}%
\bibitem [{\citenamefont {Bi}\ \emph {et~al.}(2015)\citenamefont {Bi},
  \citenamefont {Rasmussen}, \citenamefont {Slagle},\ and\ \citenamefont
  {Xu}}]{bi2015classification}%
  \BibitemOpen
  \bibfield  {author} {\bibinfo {author} {\bibfnamefont {Z.}~\bibnamefont
  {Bi}}, \bibinfo {author} {\bibfnamefont {A.}~\bibnamefont {Rasmussen}},
  \bibinfo {author} {\bibfnamefont {K.}~\bibnamefont {Slagle}}, \ and\ \bibinfo
  {author} {\bibfnamefont {C.}~\bibnamefont {Xu}},\ }\href@noop {} {\bibfield
  {journal} {\bibinfo  {journal} {Physical Review B}\ }\textbf {\bibinfo
  {volume} {91}},\ \bibinfo {pages} {134404} (\bibinfo {year}
  {2015})}\BibitemShut {NoStop}%
\bibitem [{\citenamefont {You}\ and\ \citenamefont
  {Xu}(2014)}]{you2014symmetry}%
  \BibitemOpen
  \bibfield  {author} {\bibinfo {author} {\bibfnamefont {Y.-Z.}\ \bibnamefont
  {You}}\ and\ \bibinfo {author} {\bibfnamefont {C.}~\bibnamefont {Xu}},\
  }\href@noop {} {\bibfield  {journal} {\bibinfo  {journal} {Physical Review
  B}\ }\textbf {\bibinfo {volume} {90}},\ \bibinfo {pages} {245120} (\bibinfo
  {year} {2014})}\BibitemShut {NoStop}%
\bibitem [{\citenamefont {Xu}\ and\ \citenamefont
  {Senthil}(2013)}]{xu2013wave}%
  \BibitemOpen
  \bibfield  {author} {\bibinfo {author} {\bibfnamefont {C.}~\bibnamefont
  {Xu}}\ and\ \bibinfo {author} {\bibfnamefont {T.}~\bibnamefont {Senthil}},\
  }\href@noop {} {\bibfield  {journal} {\bibinfo  {journal} {Physical Review
  B}\ }\textbf {\bibinfo {volume} {87}},\ \bibinfo {pages} {174412} (\bibinfo
  {year} {2013})}\BibitemShut {NoStop}%
\bibitem [{\citenamefont {You}\ \emph {et~al.}(2015)\citenamefont {You},
  \citenamefont {Bi}, \citenamefont {Rasmussen}, \citenamefont {Cheng},\ and\
  \citenamefont {Xu}}]{you2015bridging}%
  \BibitemOpen
  \bibfield  {author} {\bibinfo {author} {\bibfnamefont {Y.-Z.}\ \bibnamefont
  {You}}, \bibinfo {author} {\bibfnamefont {Z.}~\bibnamefont {Bi}}, \bibinfo
  {author} {\bibfnamefont {A.}~\bibnamefont {Rasmussen}}, \bibinfo {author}
  {\bibfnamefont {M.}~\bibnamefont {Cheng}}, \ and\ \bibinfo {author}
  {\bibfnamefont {C.}~\bibnamefont {Xu}},\ }\href@noop {} {\bibfield  {journal}
  {\bibinfo  {journal} {New Journal of Physics}\ }\textbf {\bibinfo {volume}
  {17}},\ \bibinfo {pages} {075010} (\bibinfo {year} {2015})}\BibitemShut
  {NoStop}%
\bibitem [{\citenamefont {You}\ and\ \citenamefont
  {You}(2016)}]{you2016stripe}%
  \BibitemOpen
  \bibfield  {author} {\bibinfo {author} {\bibfnamefont {Y.}~\bibnamefont
  {You}}\ and\ \bibinfo {author} {\bibfnamefont {Y.-Z.}\ \bibnamefont {You}},\
  }\href@noop {} {\bibfield  {journal} {\bibinfo  {journal} {Physical Review
  B}\ }\textbf {\bibinfo {volume} {93}},\ \bibinfo {pages} {195141} (\bibinfo
  {year} {2016})}\BibitemShut {NoStop}%
\bibitem [{\citenamefont {Zhang}\ \emph {et~al.}(2022)\citenamefont {Zhang},
  \citenamefont {Qi},\ and\ \citenamefont {Bi}}]{zhang2022strange}%
  \BibitemOpen
  \bibfield  {author} {\bibinfo {author} {\bibfnamefont {J.-H.}\ \bibnamefont
  {Zhang}}, \bibinfo {author} {\bibfnamefont {Y.}~\bibnamefont {Qi}}, \ and\
  \bibinfo {author} {\bibfnamefont {Z.}~\bibnamefont {Bi}},\ }\href@noop {}
  {\bibfield  {journal} {\bibinfo  {journal} {arXiv preprint arXiv:2210.17485}\
  } (\bibinfo {year} {2022})}\BibitemShut {NoStop}%
\bibitem [{\citenamefont {Devakul}\ \emph
  {et~al.}(2018{\natexlab{b}})\citenamefont {Devakul}, \citenamefont
  {Williamson},\ and\ \citenamefont {You}}]{devakul2018classification}%
  \BibitemOpen
  \bibfield  {author} {\bibinfo {author} {\bibfnamefont {T.}~\bibnamefont
  {Devakul}}, \bibinfo {author} {\bibfnamefont {D.~J.}\ \bibnamefont
  {Williamson}}, \ and\ \bibinfo {author} {\bibfnamefont {Y.}~\bibnamefont
  {You}},\ }\href@noop {} {\bibfield  {journal} {\bibinfo  {journal} {Physical
  Review B}\ }\textbf {\bibinfo {volume} {98}},\ \bibinfo {pages} {235121}
  (\bibinfo {year} {2018}{\natexlab{b}})}\BibitemShut {NoStop}%
\bibitem [{\citenamefont {Devakul}\ \emph {et~al.}(2020)\citenamefont
  {Devakul}, \citenamefont {Shirley},\ and\ \citenamefont
  {Wang}}]{devakul2020strong}%
  \BibitemOpen
  \bibfield  {author} {\bibinfo {author} {\bibfnamefont {T.}~\bibnamefont
  {Devakul}}, \bibinfo {author} {\bibfnamefont {W.}~\bibnamefont {Shirley}}, \
  and\ \bibinfo {author} {\bibfnamefont {J.}~\bibnamefont {Wang}},\ }\href@noop
  {} {\bibfield  {journal} {\bibinfo  {journal} {Physical Review Research}\
  }\textbf {\bibinfo {volume} {2}},\ \bibinfo {pages} {012059} (\bibinfo {year}
  {2020})}\BibitemShut {NoStop}%
\bibitem [{\citenamefont {Tantivasadakarn}\ and\ \citenamefont
  {Vijay}(2020)}]{tantivasadakarn2020searching}%
  \BibitemOpen
  \bibfield  {author} {\bibinfo {author} {\bibfnamefont {N.}~\bibnamefont
  {Tantivasadakarn}}\ and\ \bibinfo {author} {\bibfnamefont {S.}~\bibnamefont
  {Vijay}},\ }\href@noop {} {\bibfield  {journal} {\bibinfo  {journal}
  {Physical Review B}\ }\textbf {\bibinfo {volume} {101}},\ \bibinfo {pages}
  {165143} (\bibinfo {year} {2020})}\BibitemShut {NoStop}%
\bibitem [{\citenamefont {Chitambar}\ and\ \citenamefont
  {Gour}(2019)}]{QRT_RMP}%
  \BibitemOpen
  \bibfield  {author} {\bibinfo {author} {\bibfnamefont {E.}~\bibnamefont
  {Chitambar}}\ and\ \bibinfo {author} {\bibfnamefont {G.}~\bibnamefont
  {Gour}},\ }\href {\doibase 10.1103/RevModPhys.91.025001} {\bibfield
  {journal} {\bibinfo  {journal} {Rev. Mod. Phys.}\ }\textbf {\bibinfo {volume}
  {91}},\ \bibinfo {pages} {025001} (\bibinfo {year} {2019})}\BibitemShut
  {NoStop}%
\bibitem [{\citenamefont {Hastings}\ and\ \citenamefont
  {Wen}(2005)}]{hastings2005quasiadiabatic}%
  \BibitemOpen
  \bibfield  {author} {\bibinfo {author} {\bibfnamefont {M.~B.}\ \bibnamefont
  {Hastings}}\ and\ \bibinfo {author} {\bibfnamefont {X.-G.}\ \bibnamefont
  {Wen}},\ }\href@noop {} {\bibfield  {journal} {\bibinfo  {journal} {Physical
  review b}\ }\textbf {\bibinfo {volume} {72}},\ \bibinfo {pages} {045141}
  (\bibinfo {year} {2005})}\BibitemShut {NoStop}%
\bibitem [{\citenamefont {Stephen}\ \emph {et~al.}(2024)\citenamefont
  {Stephen}, \citenamefont {Nandkishore},\ and\ \citenamefont
  {Zhang}}]{stephen2024manybody}%
  \BibitemOpen
  \bibfield  {author} {\bibinfo {author} {\bibfnamefont {D.~T.}\ \bibnamefont
  {Stephen}}, \bibinfo {author} {\bibfnamefont {R.}~\bibnamefont
  {Nandkishore}}, \ and\ \bibinfo {author} {\bibfnamefont {J.-H.}\ \bibnamefont
  {Zhang}},\ }\href {https://arxiv.org/abs/2410.23354} {\enquote {\bibinfo
  {title} {Many-body quantum catalysts for transforming between phases of
  matter},}\ } (\bibinfo {year} {2024}),\ \Eprint
  {http://arxiv.org/abs/2410.23354} {arXiv:2410.23354 [quant-ph]} \BibitemShut
  {NoStop}%
\bibitem [{\citenamefont {Gopalakrishnan}(2023)}]{gopalakrishnan2023push}%
  \BibitemOpen
  \bibfield  {author} {\bibinfo {author} {\bibfnamefont {S.}~\bibnamefont
  {Gopalakrishnan}},\ }\href@noop {} {\bibfield  {journal} {\bibinfo  {journal}
  {Journal of Physics A: Mathematical and Theoretical}\ } (\bibinfo {year}
  {2023})}\BibitemShut {NoStop}%
\bibitem [{\citenamefont {Chen}\ \emph
  {et~al.}(2024{\natexlab{a}})\citenamefont {Chen}, \citenamefont {Dua},
  \citenamefont {Hermele}, \citenamefont {Stephen}, \citenamefont
  {Tantivasadakarn}, \citenamefont {Vanhove},\ and\ \citenamefont
  {Zhao}}]{chen2023sequential}%
  \BibitemOpen
  \bibfield  {author} {\bibinfo {author} {\bibfnamefont {X.}~\bibnamefont
  {Chen}}, \bibinfo {author} {\bibfnamefont {A.}~\bibnamefont {Dua}}, \bibinfo
  {author} {\bibfnamefont {M.}~\bibnamefont {Hermele}}, \bibinfo {author}
  {\bibfnamefont {D.~T.}\ \bibnamefont {Stephen}}, \bibinfo {author}
  {\bibfnamefont {N.}~\bibnamefont {Tantivasadakarn}}, \bibinfo {author}
  {\bibfnamefont {R.}~\bibnamefont {Vanhove}}, \ and\ \bibinfo {author}
  {\bibfnamefont {J.-Y.}\ \bibnamefont {Zhao}},\ }\href@noop {} {\bibfield
  {journal} {\bibinfo  {journal} {Physical Review B}\ }\textbf {\bibinfo
  {volume} {109}},\ \bibinfo {pages} {075116} (\bibinfo {year}
  {2024}{\natexlab{a}})}\BibitemShut {NoStop}%
\bibitem [{\citenamefont {Chen}\ \emph
  {et~al.}(2024{\natexlab{b}})\citenamefont {Chen}, \citenamefont {Hermele},\
  and\ \citenamefont {Stephen}}]{chen2024sequential}%
  \BibitemOpen
  \bibfield  {author} {\bibinfo {author} {\bibfnamefont {X.}~\bibnamefont
  {Chen}}, \bibinfo {author} {\bibfnamefont {M.}~\bibnamefont {Hermele}}, \
  and\ \bibinfo {author} {\bibfnamefont {D.~T.}\ \bibnamefont {Stephen}},\
  }\href@noop {} {\bibfield  {journal} {\bibinfo  {journal} {arXiv preprint
  arXiv:2402.03433}\ } (\bibinfo {year} {2024}{\natexlab{b}})}\BibitemShut
  {NoStop}%
\bibitem [{\citenamefont {Huang}\ \emph {et~al.}(2020)\citenamefont {Huang},
  \citenamefont {Kueng},\ and\ \citenamefont {Preskill}}]{Huang_2020}%
  \BibitemOpen
  \bibfield  {author} {\bibinfo {author} {\bibfnamefont {H.-Y.}\ \bibnamefont
  {Huang}}, \bibinfo {author} {\bibfnamefont {R.}~\bibnamefont {Kueng}}, \ and\
  \bibinfo {author} {\bibfnamefont {J.}~\bibnamefont {Preskill}},\ }\href
  {\doibase 10.1038/s41567-020-0932-7} {\bibfield  {journal} {\bibinfo
  {journal} {Nature Physics}\ }\textbf {\bibinfo {volume} {16}},\ \bibinfo
  {pages} {1050–1057} (\bibinfo {year} {2020})}\BibitemShut {NoStop}%
\bibitem [{\citenamefont {Rath}\ \emph {et~al.}(2021)\citenamefont {Rath},
  \citenamefont {Branciard}, \citenamefont {Minguzzi},\ and\ \citenamefont
  {Vermersch}}]{Rath_2021}%
  \BibitemOpen
  \bibfield  {author} {\bibinfo {author} {\bibfnamefont {A.}~\bibnamefont
  {Rath}}, \bibinfo {author} {\bibfnamefont {C.}~\bibnamefont {Branciard}},
  \bibinfo {author} {\bibfnamefont {A.}~\bibnamefont {Minguzzi}}, \ and\
  \bibinfo {author} {\bibfnamefont {B.}~\bibnamefont {Vermersch}},\ }\href
  {\doibase 10.1103/physrevlett.127.260501} {\bibfield  {journal} {\bibinfo
  {journal} {Physical Review Letters}\ }\textbf {\bibinfo {volume} {127}}
  (\bibinfo {year} {2021}),\ 10.1103/physrevlett.127.260501}\BibitemShut
  {NoStop}%
\bibitem [{\citenamefont {Chen}\ and\ \citenamefont
  {Grover}(2023{\natexlab{b}})}]{chen2023separability}%
  \BibitemOpen
  \bibfield  {author} {\bibinfo {author} {\bibfnamefont {Y.-H.}\ \bibnamefont
  {Chen}}\ and\ \bibinfo {author} {\bibfnamefont {T.}~\bibnamefont {Grover}},\
  }\href@noop {} {\bibfield  {journal} {\bibinfo  {journal} {arXiv preprint
  arXiv:2309.11879}\ } (\bibinfo {year} {2023}{\natexlab{b}})}\BibitemShut
  {NoStop}%
\bibitem [{\citenamefont {Heinrich}\ and\ \citenamefont
  {Levin}(2018)}]{Heinrich_2018}%
  \BibitemOpen
  \bibfield  {author} {\bibinfo {author} {\bibfnamefont {C.}~\bibnamefont
  {Heinrich}}\ and\ \bibinfo {author} {\bibfnamefont {M.}~\bibnamefont
  {Levin}},\ }\href {\doibase 10.1103/physrevb.98.035101} {\bibfield  {journal}
  {\bibinfo  {journal} {Physical Review B}\ }\textbf {\bibinfo {volume} {98}}
  (\bibinfo {year} {2018}),\ 10.1103/physrevb.98.035101}\BibitemShut {NoStop}%
\bibitem [{\citenamefont {Lam}\ \emph {et~al.}(2024)\citenamefont {Lam},
  \citenamefont {Han},\ and\ \citenamefont {You}}]{lam2024topological}%
  \BibitemOpen
  \bibfield  {author} {\bibinfo {author} {\bibfnamefont {H.~T.}\ \bibnamefont
  {Lam}}, \bibinfo {author} {\bibfnamefont {J.~H.}\ \bibnamefont {Han}}, \ and\
  \bibinfo {author} {\bibfnamefont {Y.}~\bibnamefont {You}},\ }\href@noop {}
  {\bibfield  {journal} {\bibinfo  {journal} {arXiv preprint arXiv:2403.13880}\
  } (\bibinfo {year} {2024})}\BibitemShut {NoStop}%
\bibitem [{\citenamefont {Lam}(2024)}]{lam2024classification}%
  \BibitemOpen
  \bibfield  {author} {\bibinfo {author} {\bibfnamefont {H.~T.}\ \bibnamefont
  {Lam}},\ }\href@noop {} {\bibfield  {journal} {\bibinfo  {journal} {Physical
  Review B}\ }\textbf {\bibinfo {volume} {109}},\ \bibinfo {pages} {115142}
  (\bibinfo {year} {2024})}\BibitemShut {NoStop}%
\bibitem [{\citenamefont {Han}\ \emph {et~al.}(2024)\citenamefont {Han},
  \citenamefont {Lake}, \citenamefont {Lam}, \citenamefont {Verresen},\ and\
  \citenamefont {You}}]{han2024topological}%
  \BibitemOpen
  \bibfield  {author} {\bibinfo {author} {\bibfnamefont {J.~H.}\ \bibnamefont
  {Han}}, \bibinfo {author} {\bibfnamefont {E.}~\bibnamefont {Lake}}, \bibinfo
  {author} {\bibfnamefont {H.~T.}\ \bibnamefont {Lam}}, \bibinfo {author}
  {\bibfnamefont {R.}~\bibnamefont {Verresen}}, \ and\ \bibinfo {author}
  {\bibfnamefont {Y.}~\bibnamefont {You}},\ }\href@noop {} {\bibfield
  {journal} {\bibinfo  {journal} {Physical Review B}\ }\textbf {\bibinfo
  {volume} {109}},\ \bibinfo {pages} {125121} (\bibinfo {year}
  {2024})}\BibitemShut {NoStop}%
\bibitem [{\citenamefont {Xu}\ and\ \citenamefont
  {Jian}(2024)}]{xu2024average}%
  \BibitemOpen
  \bibfield  {author} {\bibinfo {author} {\bibfnamefont {Y.}~\bibnamefont
  {Xu}}\ and\ \bibinfo {author} {\bibfnamefont {C.-M.}\ \bibnamefont {Jian}},\
  }\href {https://arxiv.org/abs/2406.07417} {\enquote {\bibinfo {title}
  {Average-exact mixed anomalies and compatible phases},}\ } (\bibinfo {year}
  {2024}),\ \Eprint {http://arxiv.org/abs/2406.07417} {arXiv:2406.07417
  [cond-mat.str-el]} \BibitemShut {NoStop}%
\bibitem [{\citenamefont {Sala}\ \emph
  {et~al.}(2024{\natexlab{c}})\citenamefont {Sala}, \citenamefont {You},
  \citenamefont {Hauschild},\ and\ \citenamefont {Motrunich}}]{sala2024exotic}%
  \BibitemOpen
  \bibfield  {author} {\bibinfo {author} {\bibfnamefont {P.}~\bibnamefont
  {Sala}}, \bibinfo {author} {\bibfnamefont {Y.}~\bibnamefont {You}}, \bibinfo
  {author} {\bibfnamefont {J.}~\bibnamefont {Hauschild}}, \ and\ \bibinfo
  {author} {\bibfnamefont {O.}~\bibnamefont {Motrunich}},\ }\href@noop {}
  {\bibfield  {journal} {\bibinfo  {journal} {Physical Review B}\ }\textbf
  {\bibinfo {volume} {109}},\ \bibinfo {pages} {014406} (\bibinfo {year}
  {2024}{\natexlab{c}})}\BibitemShut {NoStop}%
\end{thebibliography}%

 \end{document}